\documentclass[
 aip, 
 amsmath,amssymb,
 reprint, superscriptaddress,
 citeautoscript,  
]{revtex4-1}
\usepackage[utf8]{inputenc}
\usepackage{graphicx}
\usepackage{float}
\usepackage{enumitem}
\usepackage{bm}
\usepackage{bbold}
\usepackage{cases}
\usepackage[usenames, dvipsnames]{color}
\usepackage[colorlinks=true, citecolor=Periwinkle, linkcolor=MidnightBlue, urlcolor=Orchid]{hyperref}
\usepackage{hyperref}
\usepackage{braket}
\usepackage[normalem]{ulem}
\usepackage{textgreek}
\usepackage{gensymb}
\usepackage{dcolumn}
\usepackage{array}

\newcommand*{\citen}[1]{%
  \begingroup
    \romannumeral-`\x % remove space at the beginning of \setcitestyle
    \setcitestyle{numbers}%
    \cite{#1}%
  \endgroup   
}

\makeatletter
\def\@email#1#2{%
 \endgroup
 \patchcmd{\titleblock@produce}
  {\frontmatter@RRAPformat}
  {\frontmatter@RRAPformat{\produce@RRAP{*#1\href{mailto:#2}{#2}}}\frontmatter@RRAPformat}
  {}{}
}%
\makeatother

\begin{document}

\title{How to improve the accuracy of semiclassical and quasiclassical dynamics with and without generalized quantum master equations}

\author{Matthew R. Laskowski}
\affiliation{Department of Chemistry, University of Colorado Boulder, Boulder, CO 80309, USA \looseness=-1}

\author{Srijan Bhattacharyya}
\affiliation{Department of Chemistry, University of Colorado Boulder, Boulder, CO 80309, USA  \looseness=-1}

\author{Andr\'{e}s Montoya-Castillo}
\homepage{Andres.MontoyaCastillo@colorado.edu}
\affiliation{Department of Chemistry, University of Colorado Boulder, Boulder, CO 80309, USA  \looseness=-1} 

\date{\today}

\begin{abstract}
Semi- and quasi-classical (SC) theories can handle anharmonic interactions and are thus well-suited to predict atomistic quantum dynamics in condensed phases that encode energy and charge transport, spectroscopic responses, and chemical reactivity. However, SC theories can be computationally expensive and inaccurate. When combined with generalized quantum master equations (GQMEs), the resulting SC-GQMEs can enhance the efficiency and accuracy of SC dynamics. Yet, while the origin of improved efficiency is clear, the mechanism that improves accuracy remains elusive. Even worse, SC-GQMEs can yield unphysical dynamics in challenging parameter regimes---a shortcoming that might be avoided if the mechanism of accuracy improvement were understood. Here, we uncover this mechanism. We leverage short-time analyses to prove that exact, ``left-handed'' time-derivatives delay the onset of SC inaccuracy, even \textit{without} the GQME. However, these derivatives are a double-edged sword: while offering greater short-time accuracy, they become unphysical in challenging parameter regimes. Because short-lived SC-GQME kernels combine short-time accuracy with long-time stability, we develop a protocol to unambiguously determine the memory kernel cutoff, even in challenging cases where previous treatments had failed. Our protocol employs only SC calculations and combines self-consistency with mixed-accuracy auxiliary kernels to triangulate a propitious kernel cutoff, yielding SC-GQMEs with greater accuracy than SC theory alone, while remaining physical and accurate over arbitrary times. Our insights into accuracy improvement, identification of when the SC-GQME is advantageous, and kernel cutoff protocol are general and can be expected to apply to complex systems that go beyond simple models.
\end{abstract}

\maketitle
 
Accurate quantum dynamics is essential for describing fundamental processes such as charge transfer in solution, energy transport in light-harvesting complexes, and the optical spectra of molecules. When a problem can be written in terms of an open quantum system coupled to a Gaussian environment \cite{breuer2002theory}, one can use a variety of numerically exact methods \cite{Tanimura1989, Meyer1990, Makri1995, Thoss2001, Wang2003, Ishizaki2005, shi2009efficient, ishizaki2009theoretical, Suess2014, makri2014blip, Tamascelli2019, makri2020small, kundu2023pathsum, Fux2024}. However, many physical systems exhibit anharmonic interactions that can engender non-Gaussian statistics, including spin systems \cite{Packwood2011,norris2016qubit, Degen2017, sung2019non}, photosynthetic complexes \cite{golub2018rigid, Cho2025}, electrochemical interfaces \cite{Small2003, martin2012non, willard2014molecular}, and technologically relevant semiconductors\cite{Xu2023, Xie2017, ranalli2024electron, Jasrasaria2025}. Hence, approximate methods become necessary. Among these, the semi- and quasi-classical hierarchy\cite{herman1994dynamics, muller1998consistent, wang1998semiclassical, Sun1998, jang1999path, thoss1999mapping, wang1999semiclassical, ben2002ab,thoss2004semiclassical, kay2005semiclassical, ananth2007semiclassical, richardson2013communication, ananth2013mapping, kapral2015quantum, crespo2018recent, curchod2018ab} stands out due to its compatibility with generally anharmonic \textit{ab initio} forces and ability to treat non-Gaussian fluctuations. Yet, despite their promise, semiclassics involve uncontrolled approximations that can lead to inaccurate dynamics. 

Semi- and quasi-classical (SC) theories treat some degrees of freedom quantum mechanically and others classically, offering a means to approximate quantum dynamics in systems where a fully quantum description is intractable. This family of methods has been instrumental in understanding superconductivity \cite{Wang2021, Kim2024}, modeling charge transport from first principles\cite{Bernardi2016, Zheng2023, Yao2025}, describing complex interactions in spin systems \cite{Davidson2015, Schachenmayer2015, Zhu2019}, and simulating linear and non-linear spectroscopies \cite{heller1981semiclassical,  walton1996new, mukamel1995principles, noid2003optical, shi2005comparison, Hanna2009, wehrle2014fly, Provazza2018, Gao2020, provazza2021analytic, loring2022calculating, Atsango2023, lieberherr2025two, zeng2025quantumness}. However, SC methods are beset by problems, such as violating detailed balance\cite{bowman1989method, guo1996analysis, stock1999flow, Parandekar2006} and creating artificial resonances or shifts in spectroscopic peaks\cite{cao1994formulation, Habershon2008, Witt2009, Provazza2018, althorpe2021path, althorpe2024path}. Researchers have adopted various strategies to address these challenges. For example, one can incorporate more quantum effects using the path integral formalism \cite{pechukas1969time, miller2001semiclassical, shi2003relationship, craig2005chemical, bonella2005land,  dunkel2008iterative, lambert2012quantum} or the quantum-classical Liouville equation \cite{kim2008quantum, hsieh2012nonadiabatic, kelly2011mixed,  kelly2012mapping, kim2014improving}, but this generally entails a trade-off between higher accuracy and increased computational cost. Another way is to partition a system's degrees of freedom into two different categories and treat one set with numerically exact or perturbative approaches and the other set semiclassically \cite{wang2001systematic, berkelbach2012reduced, berkelbach2012reduced2, montoya2015extending, fetherolf2017linear, schile2019simulating}, but this generally requires Gaussian environments or exactly solvable Hamiltonian partitions. Finally, one can combine SC theory with generalized quantum master equations (GQME). These SC-GQMEs extract the non-Markovian generator of the dynamics, i.e., the memory kernel, using SC theory, enabling the subsequent solution of the GQME. SC-GQMEs have been shown to provide significant boosts in accuracy and efficiency to SC methods, and have been demonstrated both in \cite{montoya2017approximate} and out\cite{shi2004semiclassical, shi2004derivation, kelly2013efficient, Kelly2015,  noneq1, whencanonewin, Mulvihill2019, mulvihill2019combining, mulvihill2021road, Mulvihill2022, Amati2022, liu2024combining} of equilibrium, for atomistic systems\cite{pfalzgraff2015nonadiabatic}, multi-state problems \cite{pfalzgraff2019efficient, Mulvihill2021}, and even multitime correlation functions for nonlinear spectroscopies \cite{sayer2024generalized}. However, fundamental questions remain regarding the source of their improved accuracy, when they are advantageous, and why they sometimes fail. We consider these questions here. 

How and when do GQMEs improve SC dynamics? Early work posited that it was the confluence of short-lived memory kernels and the short-time accuracy of SC methods that endowed SC-GQMEs with improved accuracy \cite{Kelly2015}. However, later work  showed that long-lived memory kernels could still improve the accuracy of SC dynamics, and proposed that this improvement arises from exact sampling of bath correlations at time $t=0$ used to construct the SC-GQME kernel \cite{noneq1}. Building on these insights, Ref.~\citen{whencanonewin} established mathematical requirements for when one should expect the self-consistent extraction of the memory kernel in the SC-GQME to provide either the same or different (not necessarily better) level of accuracy compared to the original SC dynamics. Indeed, it was later shown that the SC-GQME can generate worse dynamics than the original SC approximation, even with unphysical, negative populations, in systems with large energy biases and system-bath coupling\cite{Amati2022}. This is a worrying situation because, to date, no diagnostic exists to determine if and when the SC-GQME should be expected to \textit{improve} SC dynamics. This presents a problem when describing complex processes where exact quantum dynamics become challenging, including energy flow in photosynthetic complexes or charge transport in transition metal oxides, which exhibit strong electronic-nuclear coupling and significant energetic disorder \cite{ Rtsep2011, dai2025comparison, dai2025polarons}.

Here, we resolve these issues by addressing the following questions: 
\phantomsection
\label{qs}
\begin{enumerate}
\renewcommand{\labelenumi}{\textbf{Q\arabic{enumi}}:}
    \item What is the mechanism of accuracy improvement in the SC-GQME?
    \item What is the role of self-consistency in the SC-GQME?
    \item How can one reliably improve SC dynamics with and without the SC-GQME?
\end{enumerate}
In particular, we leverage a short-time expansion analysis to demonstrate that improved accuracy arises from \textit{exact} time derivatives on the initial condition that delay the onset of error arising from the SC approximation. This enables us to show that simple numerical integration of these time derivatives yields improved accuracy, even in the absence of the GQME. However, we find that, while offering greater short-time accuracy, these time derivatives can become inaccurate, even unphysical, at long times in challenging parameter regimes. Returning to the SC-GQME, our analysis shows that the self-consistent structure of the SC-GQME is not strictly necessary to improve the accuracy of SC dynamics. Instead, we establish that it can help ensure sum rules, such as population conservation in the calculated nonequilibrium averages of reduced density matrices. By understanding static sampling errors in SC correlation functions, we show that one can impose the same conservation laws with a predictable uniform shift of the time derivatives. Noting, then, that SC-GQMEs are most beneficial when they exhibit short-lived kernels that capitalize on short-time accuracy while circumventing long-time instability of exact time derivative-rotated initial conditions, we propose a new, unambiguous protocol to truncate the memory kernel, even in challenging parameter regimes where previous protocols had failed. Our kernel truncation metric is self-contained and only requires access to the SC dynamics used to construct the SC-GQME kernel. Importantly, our protocol is compatible with advances in sampling initial conditions for atomistic systems\cite{shi2003semiclassicala, poulsen2003practical, liu2009simple, montoya2017path, bose2019coherent, ple2019sampling}, and may be expected to allow for physically meaningful and accurate dynamics that surpass those of SC methods alone. We observe that, with this protocol, the resulting SC-GQME dynamics are \textit{always} as accurate or more than the direct SC approximation. 

\section{SC Method and Illustrative Models}

While our conclusions in this work are independent of the choice of SC theory and model, we illustrate our arguments by employing the linearized semiclassical initial value representation (LSC) method \cite{Sun1998} as our choice of SC theory to simulate the reduced (spin) density matrix dynamics of the spin-boson model \cite{leggett1987dynamics}. Specifically, we quantify the accuracy of SC theory and our modifications to it by benchmarking against the population dynamics, $\langle \sigma_z(t)\rangle$, obtained using the numerically exact, path integral-based time-evolving matrix product operator (TEMPO) approach \cite{strathearn2018efficient}. In the Supplementary Information (SI)~Sec.~I, we further test the performance of our ideas with respect to coherence dynamics and other SC methods (see SI~Sec.~I A and B).

We choose the spin-boson model because it is the paradigmatic open quantum system that illustrates decoherence, dissipation, and thermalization, and can be used to model electron, charge, and spin transfer in the condensed phase\cite{leggett1987dynamics, weiss1992quantum}. However, our conclusions are broadly applicable to systems in contact with thermal reservoirs, including, e.g., the Frenkel exciton model (see SI~Sec.~I C). The spin-boson model consists of a central spin connected to a Gaussian thermal environment,
\begin{equation}
    \hat{H} = (\varepsilon + \hat{V}_B) \hat{\sigma}_z + \Delta \hat{\sigma}_x + \frac{1}{2}\sum_n \big( \hat{p}_n^2 + \omega_n^2 \hat{x}_n^2\big),
\end{equation}
where $\{ \hat{\sigma}_i \}$ are Pauli matrices, $\varepsilon$ is spin's bias, $\Delta$ is its diabatic coupling, $\omega_n$, $\hat{p}_n$, and $\hat{x}_n$, are the frequency and mass-weighted momenta and positions of the $n^{\rm th}$ oscillator in the thermal bath, and $\hat{V}_B = \sum_n c_n\hat{x}_n$ is the bath part of the system-bath coupling, with $c_n$ being the coupling to the $n^{\rm th}$ oscillator. The spectral density,
\begin{equation}
    J(\omega) = \frac{\pi}{2} \sum_n \frac{c_n^2}{\omega_n} \delta\left(\omega - \omega_n\right)
\end{equation}
determines the coupling between the system and bath. For simplicity, we employ an Ohmic spectral density with an exponential cutoff $J(\omega) = \frac{\pi}{2}\eta \omega e^{-\omega/\omega_c}$, where $\frac{\pi}{2}\eta =  \pi\lambda/\omega_c$, is the Kondo parameter, and $\lambda = \frac{1}{\pi}\int_0^{\infty}{\rm d}\omega\ J(\omega)/\omega$ is the reorganization energy quantifying the strength of system-bath coupling and $\omega_c$ is the cutoff frequency that determines how quickly the thermal bath dissipates energy. 

Arguably the simplest method in the SC hierarchy, the LSC method \cite{Sun1998, shi2003relationship} is equivalent to truncated Wigner approximation\cite{Schachenmayer2015, czischek2020neural, hosseinabadi2025user} and offers accuracy comparable to mean-field methods such as Ehrenfest theory \cite{mclachlan1964variational, ehrenstock1995semiclassical, tully1998mixed, grunwald2009quantum} .
Because LSC contains the least quantum treatment, it offers a hard test for our analysis. LSC encodes quantum mechanics only in the initial and final conditions for measurement and employs classical dynamics to evolve all variables. In Wigner phase space, LSC approximates a quantum correlation function thus:
\begin{equation}
\begin{split}
\label{lsc-ivr-approx}
    C_{AB}(t) &= \mathrm{Tr}\{\hat{\rho}_B \hat{A}e^{i\mathcal{L}t}\hat{B}\}\\
    &\approx (2\pi)^{-f} \int {\rm d}\boldsymbol{\Gamma}\ (\hat{\rho}_B \hat{A})^W_{ \boldsymbol{\Gamma}}e^{i\mathcal{L}^Wt}B^W_{\boldsymbol{\Gamma}}\\
    &= (2\pi)^{-f} \int {\rm d}\boldsymbol{\Gamma}\ (\hat{\rho}_B \hat{A})^W_{ \boldsymbol{\Gamma}}B^W_{\boldsymbol{\Gamma}}(t).
\end{split}
\end{equation}
Here, $f$ is the total number of degrees of freedom (electronic and nuclear), $\Gamma = \{\mathbf{X}, \mathbf{P}, \mathbf{x}, \mathbf{p} \}$ constitutes the phase space for the subsystem and bath, and $\mathcal{L}^W$ is the Wigner-transformed \cite{imre1967wigner, hillery1984distribution} quantum Liouvillian, which is equivalent to the classical Poisson bracket, with $e^{i\mathcal{L}^W t}$ becoming the classical Hamiltonian propagator. To obtain a continuous, Cartesian representation of the spin's outer product states, $\{ \ket{j}\bra{k}\} \mapsto \{ \mathbf{X}, \mathbf{P}\}$, the LSC treatment of such open quantum systems employ the Meyer-Miller-Stock-Thoss mapping \cite{miller1979classical, stock1997semiclassical}. We refer the reader to the SI Secs.~II A and B for details on exact simulations and LSC simulations, which we propagate using a split-operator algorithm \cite{strang1968construction}. 

\section{Self-consistent GQMEs}
We start by delineating the anatomy of GQMEs and SC-GQMEs, which necessitates introducing new notation. Here, we provide a table summarizing our notation (see Table~\ref{table:table1}).
\begin{table}[h]
\renewcommand{\arraystretch}{1.5}
\begin{ruledtabular}
\begin{tabular}{c | c}
\parbox[t]{23.5mm}{$\mathcal{C}(t)$} &
\parbox[t]{55mm}{Correlation function} \\ 
\hline
\parbox[c]{23.5mm}{$\mathcal{K}(t)$}  &
\parbox[c]{55mm}{Memory kernel} \\
\hline
\parbox[c]{23.5mm}{$\big[\mathcal{C}(t)\big]_{\rm LSC}$} &
\parbox[c]{55mm}{LSC approximated $\mathcal{C}(t)$}  \\
\hline
\parbox[c]{23.5mm}{$\partial_t^n\big[\mathcal{C}(t)\big]_{\rm LSC}$}&
\parbox[c]{55mm}{$n^{\rm th}$ numerical time derivative of  $\big[\mathcal{C}(t)\big]_{\rm LSC}$}  \\
\hline
\parbox[c]{23.5mm}{$\big[\dot{\mathcal{C}}^{(nL/nR)}(t)\big]_{\rm LSC}$} &
\parbox[c]{55mm}{LSC approximated $n^{\rm th}$ left- or right-handed derivative(s) of $\mathcal{C}(t)$} \\
\hline
\parbox[c]{23.5mm}{$\big[\mathcal{C}^{(nL/nR)}(t)\big]_{\rm LSC}$} &
\parbox[c]{55mm}{Integrated $\big[\dot{\mathcal{C}}^{(nL/nR)}(t)\big]_{\rm LSC}$} \\
\hline
\parbox[c]{23.5mm}{$\big[\dot{\bar{\mathcal{C}}}^{(nL/nR)}(t)\big]_{\rm LSC}$} &
\parbox[c]{55mm}{Statically shifted $\big[\dot{\bar{\mathcal{C}}}^{(nL/nR)}(t)\big]_{\rm LSC}$ } \\
\hline
\parbox[c]{23.5mm}{$\big[\bar{\mathcal{C}}^{(nL/nR)}(t)\big]_{\rm LSC}$} &
\parbox[t]{55mm}{Integrated $\big[\dot{\bar{\mathcal{C}}}^{(nL/nR)}(t)\big]_{\rm LSC}$} \\
\hline
\parbox[c]{23.5mm}{$\mathcal{K}^{(x)}(t)$} & 
\parbox[t]{55mm}{Auxiliary kernel, $x\in \{ 1, 3b, 3f \}$} \\
\hline
\raisebox{-0.5\height}{
\parbox[c]{23.5mm}{$\mathcal{K}_{\rm LSC}^{(nL)}(t)$}} & 
\parbox[t]{55mm}{Memory kernel constructed from $\big[\bar{\mathcal{C}}^{nL}(t)\big]_{\rm LSC}$ and its time derivatives}
\end{tabular}
\caption{\label{table:table1} Summary of notation.}
\end{ruledtabular}

\end{table}

GQMEs\cite{nakajima1958quantum, zwanzig1960ensemble, mori1965transport} are low-dimensional equations of motion for a few ``interesting'' degrees of freedom, say electronic excitations in a molecular aggregate \cite{sayer2023compact}, transition dipoles in a spectroscopy experiment \cite{Mulvihill2021, Sayer2024, liu2025memory}, density fluctuations in a liquid \cite{breuer1993master}, or electric and thermal currents in materials\cite{bhattacharyya2024mori}. One can derive a GQME using projection operator techniques \cite{grabert1982projection, fick1990quantum}, and the choice of projection operator, $\mathcal{P} = |\textbf{A})(\textbf{A}|$, determines which degrees of freedom one is interested in tracking, $\{A_j \}$. This procedure yields a Volterra equation for $\mathcal{C}(t)$, the correlation function or nonequilibrium average of interest:
\begin{equation}
\label{mnz}
    \dot{\mathcal{C}}(t) = \mathcal{C}(t)\dot{\mathcal{C}}(0) - \int^t_0 {\rm d}s \; \mathcal{C}(t-s)\mathcal{K}(s),
\end{equation}
where $\mathcal{C}(t) = (\textbf{A}|e^{i\mathcal{L}t}|\textbf{A})$, $\mathcal{L}$ is the Liouvillian for the entire system, and $\mathcal{K}(t)$ is the memory kernel, which describes the influence of all degrees of freedom not included in $\{ A_j \}$. The definition of the inner product, $(\mathbf{A}|\hat{\mathcal{O}}|\mathbf{B})$ depends on the definition of the projection operator and should be chosen to ensure the idempotency of the projector, $\mathcal{P}^2 = \mathcal{P}$, and its complement $\mathcal{Q} = 1 - \mathcal{P}$. Here, we illustrate our ideas using a definition of the inner product that yields nonequilibrium averages in the electronic subspace of an open quantum system linked to a bosonic reservoir: $(A_j|\mathcal{O} |A_k) \equiv \mathrm{Tr}[\hat{\rho}_B A_j^{\dagger}\mathcal{O}A_k]$, with $\rho_B A^{\dagger}_k$ encoding the initial condition. In particular, we consider the Argyres-Kelley projector \cite{argyres1964theory}, for which $A_j \in \{\ket{j}\bra{k} \}$ spans the coherences and populations of the discrete subspace, but note that the same conclusions arise when using the population projector\cite{sparpaglione1988dielectric}, for which $A_j \in \{ \ket{j}\bra{j}\}$ spans only the populations.
Nevertheless, our conclusions are general and apply both in and out of equilibrium and for other definitions of the inner product. 

The projected propagator, $e^{i\mathcal{Q}\mathcal{L}t}$, in the memory kernel, $\mathcal{K}(t) = (\textbf{A}|\left(\mathcal{LQ}\right)e^{i\mathcal{QL}t}\left(\mathcal{QL}\right)|\textbf{A})$ has historically made this quantity difficult to calculate. However, one may invoke the Dyson identity to obtain a self-consistent expansion of the memory kernel \cite{shi2003new, noneq1},
\begin{equation}
\label{ksc1}
    \mathcal{K}(t) = \mathcal{K}^{(1)}(t) + \int^t_0 {\rm d}s \; \mathcal{K}^{(3b)}(t-s)\mathcal{K}(s),
\end{equation}
into two auxiliary kernels
\begin{subequations}
\begin{align}
    \label{k1}
   \mathcal{K}^{(1)}(t) &= (\textbf{A}|\left(\mathcal{LQ}\right) e^{i\mathcal{L}t} \left(\mathcal{QL}\right)|\textbf{A}), \\
   \label{k3b}
   \mathcal{K}^{(3b)}(t) &= i(\textbf{A}|\left(\mathcal{LQ}\right) e^{i\mathcal{L}t} |\textbf{A}),
\end{align}
\end{subequations}
that no longer use the projected propagator. This enables one to use any dynamics solver to obtain $\mathcal{K}^{(1)}(t)$ and $\mathcal{K}^{(3b)}(t)$. Evaluating $\mathcal{K}^{(1)}(t)$ and $\mathcal{K}^{(3b)}(t)$ using SC dynamics is what constitutes the SC-GQME. 

%%%%%%%%%%%%%%%%%%%%Figure  1  begin %%%%%%%%%%%%%%%%%%%%%%%
\begin{figure*}[t!]
\centering
\includegraphics[width=\linewidth]{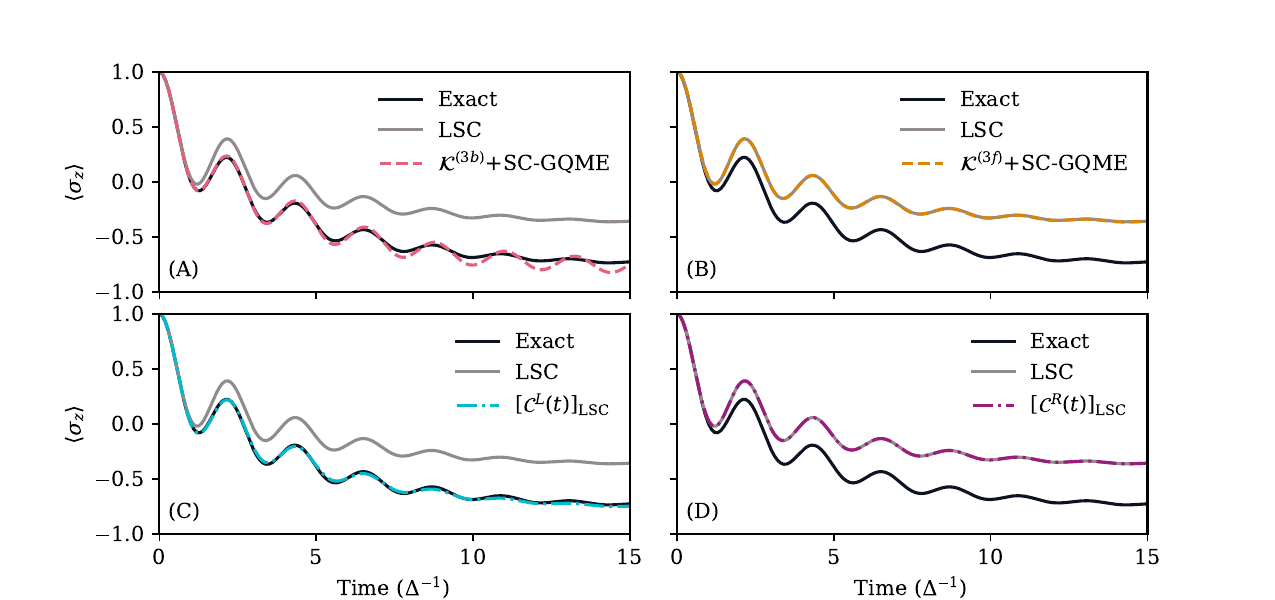}
\caption{\label{setup}  \textbf{Accuracy improvement in the SC-GQME and its relation to left- and right-handed derivatives.} Nonequilibrium population dynamics in the spin–boson model subject to initial condition $\rho(0) = \rho_B\ket{1}\bra{1}$ with parameters $\epsilon = \Delta$, $\beta = 5.0\Delta^{-1}$, $\omega_c = 2.0\Delta$, $\eta = 0.2\Delta$ with Ohmic spectral density. Panels compare exact (black) and LSC dynamics (gray) with (A) $\mathcal{K}^{(3b)}(t)$ as the generator of the SC-GQME, (B) $\mathcal{K}^{(3f)}(t)$ as the generator of the SC-GQME. Panels (C) and (D)  show the dynamics obtained by integrating $\big[\dot{\mathcal{C}}^L(t)\big]_{\rm LSC}$ and $\big[\dot{\mathcal{C}}^R(t)\big]_{\rm LSC}$, respectively.  }
\label{fig:fig1}
\end{figure*}
%%%%%%%%%%%%%%%%%%%%Figure 1 end %%%%%%%%%%%%%%%%%%

When employing exact quantum dynamics, one can alternatively derive an equally valid self-consistent expansion of the memory kernel \cite{noneq1},
\begin{equation}
\label{ksc2}
    \mathcal{K}(t) = \mathcal{K}^{(1)}(t) + \int^t_0 {\rm d}s \; \mathcal{K}(t-s)\mathcal{K}^{(3f)}(s),
\end{equation}
where, 
\begin{equation}
    \label{k3f}
    \mathcal{K}^{(3f)}(t) = i(\textbf{A}| e^{i\mathcal{L}t} \left(\mathcal{QL}\right)|\textbf{A}).
\end{equation}
Equations~\eqref{ksc1} and~\eqref{ksc2} are equally valid when using numerically exact quantum dynamics to evaluate the auxiliary kernels and yield the same memory kernel and GQME dynamics. However, this equivalence does not generally hold when using approximate dynamics \cite{noneq1}. For example, noting that one can rewrite $\mathcal{K}^{(1)}(t)$ as a time derivative of $\mathcal{K}^{(3b)}(t)$ and $\mathcal{K}^{(3f)}(t)$, i.e., 
\begin{subequations}
    \begin{align}
        \mathcal{K}^{(1)}(t) &= \dot{\mathcal{K}}^{(3b)}(t) - \mathcal{K}^{(3b)}(t)\dot{\mathcal{C}}(0), \\
        \mathcal{K}^{(1)}(t) &= \dot{\mathcal{K}}^{(3f)}(t) - \dot{\mathcal{C}}(0)\mathcal{K}^{(3f)}(t),
    \end{align}
\end{subequations}
Ref.~\citen{noneq1} showed that adopting $\mathcal{K}^{(3b)}(t)$ as the generator of the memory kernel yielded a \textit{backward SC-GQME} with improved dynamics but using $\mathcal{K}^{(3f)}(t)$ as the generator of the memory kernel resulted in a \textit{forward SC-GQME}, whose dynamics were as inaccurate as the original SC dynamics (see Fig.~\ref{fig:fig1} (A) and (B)). We provide expressions for all auxiliary kernels in SI Sec.~III C.

\section{Mechanism for accuracy improvement}

The difference in accuracy improvement afforded by $\mathcal{K}^{(3b)}(t)$ and $\mathcal{K}^{(3f)}(t)$ continues to be particularly surprising since the only difference between these auxiliary kernels is the direction in which one applies the Liouvillian operator as the generator of the time derivative, either on the initial condition in Eq.~\eqref{k3b} or the final measurement in Eq.~\eqref{k3f}. Hence, there exists an asymmetry when taking the analytical derivative on the initial versus final conditions from a semiclassical perspective. 

To highlight this asymmetry, we rewrite the auxiliary kernels by explicitly noting whether they invoke right- or left-handed derivatives of $\mathcal{C}(t)$---$\dot{\mathcal{C}}^R(t)$ and $\dot{\mathcal{C}}^L(t)$, respectively. To obtain such expressions, one substitutes the definition of $\mathcal{Q}$ into the form of each auxiliary kernel,
\begin{subequations}\label{aux_all}
\begin{align}
    \mathcal{K}^{(1)}(t) = &-\ddot{\mathcal{C}}^{LR}(t) + \dot{\mathcal{C}}(0)\dot{\mathcal{C}}^R(t) + \dot{\mathcal{C}}^L(t)\dot{\mathcal{C}}(0) \nonumber \label{eq:K1-mixed-accuracy} \\
    \; &- \dot{\mathcal{C}}(0)\mathcal{C}(t)\dot{\mathcal{C}}(0) \\
    \mathcal{K}^{(3b)}(t) = &-\dot{\mathcal{C}}^L(t) + \dot{\mathcal{C}}(0)\mathcal{C}(t) \\
    \mathcal{K}^{(3f)}(t) = &-\dot{\mathcal{C}}^R(t) + \mathcal{C}(t)\dot{\mathcal{C}}(0) 
\end{align}
\end{subequations}
where,
\begin{subequations}
    \begin{align}
    \dot{\mathcal{C}}^L(t) &= -i(\mathcal{L}\textbf{A}|e^{i\mathcal{L}t}|\textbf{A}), \\
    \dot{\mathcal{C}}^R(t) &= i(\textbf{A}|e^{i\mathcal{L}t}|\mathcal{L}\textbf{A}), \\
    \ddot{\mathcal{C}}^{LR}(t) &= (\mathcal{L}\textbf{A}|e^{i\mathcal{L}t}|\mathcal{L}\textbf{A}).
    \end{align}
\end{subequations}
Because performing a numerical time derivative of $\mathcal{C}(t)$ is equivalent to taking a single right-handed derivative \cite{whencanonewin}, it appears that the SC-GQME relies on this asymmetry to obtain accuracy improvements. While true for the first right-handed derivative, we show more broadly where this ostensible equivalence holds in Eqs.~\eqref{eq:lsc-time-derivative}-\eqref{eq:lsc-equivalence-of-rhd-numerical-derivative}. 

\subsection{Is self-consistency necessary?} 

Given that the only difference in constructing the backward and forward SC-GQMEs is their use of left- versus right-handed time derivatives, respectively, can one obtain similar improvements by \textit{directly integrating} $\dot{\mathcal{C}}^L(t)$ versus $\dot{\mathcal{C}}^R(t)$ to recover $\mathcal{C}(t)$? Specifically, one could calculate $\mathcal{C}(t)$ by integrating the time-derivative, bypassing the self-consistent structure of the SC-GQME altogether,
\begin{equation}
\label{integration}
    \mathcal{C}^{(L/R)}(t) \equiv \mathcal{C}(0) - \int^t_0 {\rm d}s \; \dot{\mathcal{C}}^{(L/R)}(s). 
\end{equation}
To pursue this question, we return to the population dynamics in Fig.~\ref{fig:fig1} (A) and (B). Panels (C) and (D) show the results of numerically integrating the LSC approximations to $\dot{\mathcal{C}}^L(t)$ and $\dot{\mathcal{C}}^R(t)$ that we used to construct the backward and forward SC-GQMEs, respectively. We provide expressions for these derivatives in SI Sec.~III D. The results show that integrating $\big[\dot{\mathcal{C}}^L(t)\big]_{\rm LSC}$ yields dynamics that closely match the numerically exact dynamics and the improved dynamics afforded by the backward SC-GQME that uses $\mathcal{K}^{(3b)}(t)$, while integrating $\big[\dot{\mathcal{C}}^R(t)\big]_{\rm LSC}$ reproduces the original and inaccurate LSC dynamics for $\langle \sigma_z(t)\rangle $ and the forward SC-GQME dynamics obtained from $\mathcal{K}^{(3f)}(t)$. Similar improvement extends to observables governed by electronic coherences, such as $\langle \sigma_x \rangle$ and $\langle \sigma_y \rangle$, where integrating $\big[\dot{\mathcal{C}}^L(t)\big]_{\rm LSC}$ consistently maintains agreement with the exact dynamics for a longer period of time compared to the LSC dynamics in line with trends reported for the SC-GQME\cite{liu2024combining} (see SI Sec.~I A) . We thus conclude that \textit{self-consistency is not} strictly \textit{necessary to improve the accuracy of SC dynamics}. 

Figure~\ref{fig:fig1} raises two intriguing questions. 
First, how does integrating $\big[\dot{\mathcal{C}}^L(t)\big]_{\rm LSC}$ produce any accuracy improvement whereas $\big[\dot{\mathcal{C}}^R(t)\big]_{\rm LSC}$ does not? To answer \textbf{\hyperref[qs]{Q1}}, we must understand this asymmetry. The following observation also motivates a second question: while both the backward SC-GQME and the numerical integration of $\big[\dot{\mathcal{C}}^L(t)\big]_{\rm LSC}$ yield predictions with improved accuracy, the two do not exactly agree with each other. So, why do they differ? To apply the SC-GQME in a controlled manner, we must resolve these questions. 

\subsection{Why does $\mathcal{C}^R(t)$ not improve accuracy?}

We start with the null result by explaining why $\big[\dot{\mathcal{C}}_{jk}^R(t)\big]_{\rm LSC}$ yields no improvement. This is because the time-derivative of the LSC-approximated $\mathcal{C}_{jk}(t)$ equals the LSC-approximated correlation function where the rotation caused by the time derivative (i.e., the action of the Liouvillian on $A_k$) is done \textit{quantum mechanically}, before application of the LSC approximation. This latter expression is the definition of the right-handed derivative,
\begin{equation}\label{eq:lsc-time-derivative}
\begin{split}
    \partial_t[\mathcal{C}_{jk}(t)]_{\rm LSC} &= \int {\rm d}\boldsymbol{\Gamma} \; \left(\hat{\rho}_B A_j\right)^W e^{i\mathcal{L}^Wt} (i\mathcal{L}^W) {A_k}^W, \\
    &= \int {\rm d}\boldsymbol{\Gamma} \; \left(\hat{\rho}_B A_j\right)^W e^{i\mathcal{L}^Wt} \left(i\mathcal{L}A_k\right)^W\\
    &= [\dot{\mathcal{C}}^{R}_{jk}(t)]_{\rm LSC},
\end{split}
\end{equation}
where $\partial_t$ denotes the \textit{numerical} time derivative. Hence, since $\big[\dot{\mathcal{C}}^R(t)\big]_{\rm LSC} = \partial_t[\mathcal{C}_{jk}(t)]_{\rm LSC}$, its numerical integral cannot produce any accuracy improvement.

We note, however, that right-handed and numerical derivatives are not equivalent to all others---if they were, then the LSC approximation would recover the full quantum dynamics. In fact, there is a critical value $n^*$---which depends on the model and the observable---beyond which the (quantum mechanical) right-handed and numerical (SC) derivatives acting on the measured variable cease to agree,
\begin{equation} \label{cond2}
  \left((i\mathcal{L})^nA_k\right)^W  \neq ((i\mathcal{L})^W)^n {A_k}^W,\ \ \ {\rm for\ } n \geq n^*.
\end{equation}
The explanation for this is simple: the two flavors of derivative agree when the application of the Wigner transform on the resulting Liouvillian-rotated operator have no correction terms arising from the Moyal product (see SI Sec.~IV A). In the case of the spin-boson (and related energy and charge transfer models), this equivalence holds for the first few derivatives of subsystem operators, $\{\ket{j}\bra{k} \}$ (see SI Sec.~IV A). Hence, one can conclude,
\begin{equation}\label{eq:lsc-equivalence-of-rhd-numerical-derivative}
    \partial_t[\mathcal{C}_{jk}(t)]_{\rm LSC} = [\dot{\mathcal{C}}^{R}_{jk}(t)]_{\rm LSC}.
\end{equation}
This equivalence breaks down at different orders for different operators. For example, in the spin-boson model, $n^* = 4$ for $\sigma_z$, while $n^* = 3$ for $\sigma_x$ (see SI Sec.~IV B). While one could expect discrepancies to arise in the evolution of these operators at orders beyond $n^{*}$, the order of accuracy of their \textit{correlation functions} can be higher, as closure against the initial condition can fortuitously cancel some of the erroneous terms (see SI Sec.~IV D). 

We now turn to $\dot{\mathcal{C}}_{jk}^L(t)$. Here, the Liouvillian acts on the initial condition, $\hat{\rho}_BA_j^{\dagger}$, where we demonstrate that the order of applying the Wigner transformation matters even at the level of the first derivative (see SI Sec.~IV C), 
\begin{equation}
    (i\mathcal{L} \hat{\rho}_BA_j)^W \neq i\mathcal{L}^W \left( \hat{\rho}_BA_j\right)^W.
\end{equation}
This implies that, 
\begin{equation}
   \partial_t[\mathcal{C}_{jk}(t)]_{\rm LSC} \neq \dot{\mathcal{C}}^L_{jk}(t). 
\end{equation}
This inequality emerges because the Wigner transformation of an operator product does not generally equal the product of the Wigner-transformed operators. Accounting for this distinction has been essential for accurately evaluating quantum correlation functions within semiclassical approaches\cite{hillery1984distribution,  ka2005vibrational, whencanonewin, saller2022accurate}.

Hence, upon integrating $\dot{\mathcal{C}}^L_{jk}(t)$, one should not expect to recover the original LSC approximation to $\mathcal{C}_{jk}(t)$. This condition is compatible with earlier work delineating conditions for when the SC-GQME can be expected to give the same level of accuracy as the original SC dynamics\cite{whencanonewin}. Yet, while this analysis shows \textit{why} $\big[\mathcal{C}^{R}(t)\big]_{\rm LSC} = \big[\mathcal{C}(t)\big]_{\rm LSC}$ and $\big[\mathcal{C}^{L}(t)\big]_{\rm LSC} \neq \big[\mathcal{C}(t)\big]_{\rm LSC}$, it does not explain \textit{how} employing $\dot{\mathcal{C}}^L(t)$ leads to improved dynamics. 

    %%%%%%%%%%%%%%%%%%%%Figure  2  begin %%%%%%%%%%%%%%%%%%%%%%%
\begin{figure*}[t]
\centering
\includegraphics[width=\linewidth]{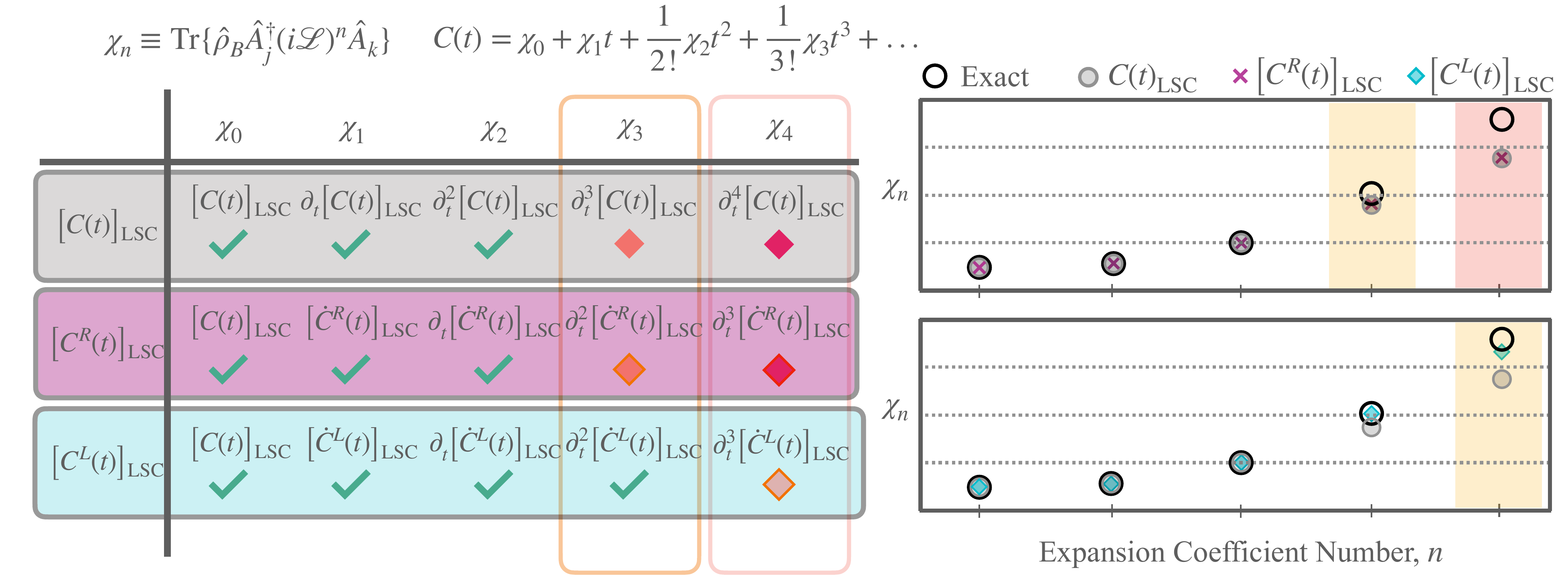}
\caption{\label{setup} \textbf{Schematic of how left-handed derivatives can delay the onset of inaccuracy of semiclassical dynamics.} \textit{Left:} Taylor expansion of a representative correlation function, $\mathcal{C}(t)$, where $\chi_n$ represents the $n^{\text{th}}$ static coefficient. Green checks indicate agreement with the exact static coefficient while an orange (red) diamond suggests minor (major) disagreement with the exact static coefficient. \textit{Right:} graphical representation of the static coefficients of the same representative correlation function, $\mathcal{C}(t)$. \textit{Top right:} the static moments of $\big[\mathcal{C}^R(t)\big]$ are identical to $\mathcal{C}(t)$ and have minor disagreement from exact dynamics at $4^{\rm th}$ order (yellow region). Larger disagreements arise at higher orders (red region). \textit{Bottom right:} the static moments of $\big[\mathcal{C}^L(t)\big]$ match the exact quantum dynamics through $4^{\rm th}$ order, but minor disagreements arise at  $5^{\rm th}$ order. 
 }
\label{fig:fig2}
\end{figure*}
%%%%%%%%%%%%%%%%%%%%Figure 2 end %%%%%%%%%%%%%%%%%%%%%%%

\subsection{Why and how does $\mathcal{C}^L(t)$ improve accuracy?}

We answer this question by turning to short-time analysis based on Taylor expansions. We distinguish between two types of Taylor expansions, one on a Heisenberg-evolved operator:
\begin{equation}
    \label{operator_exp}
    A_k(t) = \sum_{n=0}^{\infty} i^n\frac{\mathcal{L}^nA_k}{n!} t^n,
\end{equation}
and one on the correlation function itself:
\begin{equation}
    \label{corr_exp}
    \mathcal{C}_{jk}(t) = \sum_{n=0}^{\infty} i^n\frac{\mathcal{C}^{(n)}_{jk}(0)}{n!} t^n,
\end{equation}
where $\mathcal{C}^{(n)}_{jk}(0) \equiv \frac{{\rm d}^n}{{\rm d}t^n}\mathcal{C}_{jk}(t)\big|_{t=0}$. The critical concept here is how the accuracy of these short-time expansions changes when one applies the SC approximation directly to $\mathcal{C}(t)$ versus to versions of the correlation functions where one applies left-handed derivatives. 

We consider the Taylor expansion of the SC-approximated $A_k(t)$ and $\mathcal{C}(t)$. The LSC approximation to an operator's evolution takes the form:
\begin{equation}
    [A_k(t)]_{\rm LSC} = e^{i\mathcal{L}^W t} A_k^W = \sum_{n=0}^{\infty} i^n\frac{(\mathcal{L}^W)^nA_k^W}{n!} t^n.
\end{equation}
In contrast, to recover the \textit{exact} quantum dynamics of this operator in phase space, one would instead need to apply the Wigner transform to both sides of Eq.~\eqref{operator_exp}, meaning that one instead requires $(\mathcal{L}^nA_k)^W$. While we may expect $(\mathcal{L}^nA_k)^W = (\mathcal{L}^W)^nA_k^W$ for low $n$, this equality breaks down at a threshold order that depends on the form of $A_k$ \cite{golosov2001classical}. For example, in the case of $A_k = \sigma_z$, we have already shown the order at which these start to differ is $n=4$ (see Eq.~\eqref{cond2} and SI Sec.~IV B). Naively, this suggests that the LSC approximation to the quantum dynamics of $\langle \sigma_z(t) \rangle$ is only accurate to $\mathcal{O}(t^4)$ in a Taylor series in time. However, a serendipitous error cancellation via closure against the initial condition leads some erroneous terms to average out to zero. In the case of population dynamics subject to an initial excitation on site $1$, the deviation between the exact and LSC dynamics only starts at $n=6$ order (see SI Sec.~IV D).

We now posit that modifying the sequence of exact left-handed versus right-handed derivatives consistent with the LSC approximation in Eq.~\eqref{corr_exp} allows one to obtain different predictions for $\mathcal{C}(t)$ while still using SC dynamics. Specifically, we hypothesize that incorporating a left-handed derivative into $\mathcal{C}(t)$, before applying the LSC approximation, can delay the onset of inaccuracy in the LSC dynamics. Indeed, the Taylor expansion for the left-handed version of Eq.~\eqref{integration} takes the form:
\begin{equation}
    \label{modified_exp}
    \mathcal{C}^{L}(t) = \mathcal{C}(0)+ \sum_{n=1}^{\infty} i^{n}\frac{\mathcal{C}^{(1|n-1)}(0)}{n!} t^{n},
\end{equation}
where $\mathcal{C}^{(1|n-1)}_{jk}(0) = \partial_t^{(n-1)}\dot{\mathcal{C}}^{L}(t)\big|_{t=0}$ denotes the quantity obtained by taking an exact quantum mechanical derivative on the initial condition followed by Wigner transformation and all other derivatives as the outcome of the repeated application of $\mathcal{L}^{W}$ on the final measurement. This modification of derivatives does not change the order at which the operator's LSC evolution differs from the exact quantum result (see Eq.~\eqref{cond2}). However, it alters the initial condition against which the operator is closed, leading one to conclude that the order of accuracy for the resulting $[\mathcal{C}^{L}(t)]_{\text{LSC}}$ can be changed relative to $[\mathcal{C}(t)]_{\text{LSC}}$ (see schematic in Fig.~\ref{fig:fig2}). In the case of population dynamics subject to an initial excitation on site $1$, this new closure yields greater error cancellation and the deviation between exact and LSC dynamics described by Eq.~\eqref{modified_exp} starts beyond  $n=6$ (see SI Sec.~IV D). Hence, our hypothesis is correct: by taking an exact left-handed derivative before Wigner transformation, we have delayed the onset of error from the LSC approximation in the short-time expansion of the correlation function of interest. 

 %%%%%%%%%%%%%%%%%%%%Figure  3  begin %%%%%%%%%%%%%%%%%%%%%
\begin{figure*}[ht]
\centering
\includegraphics[width=\linewidth]{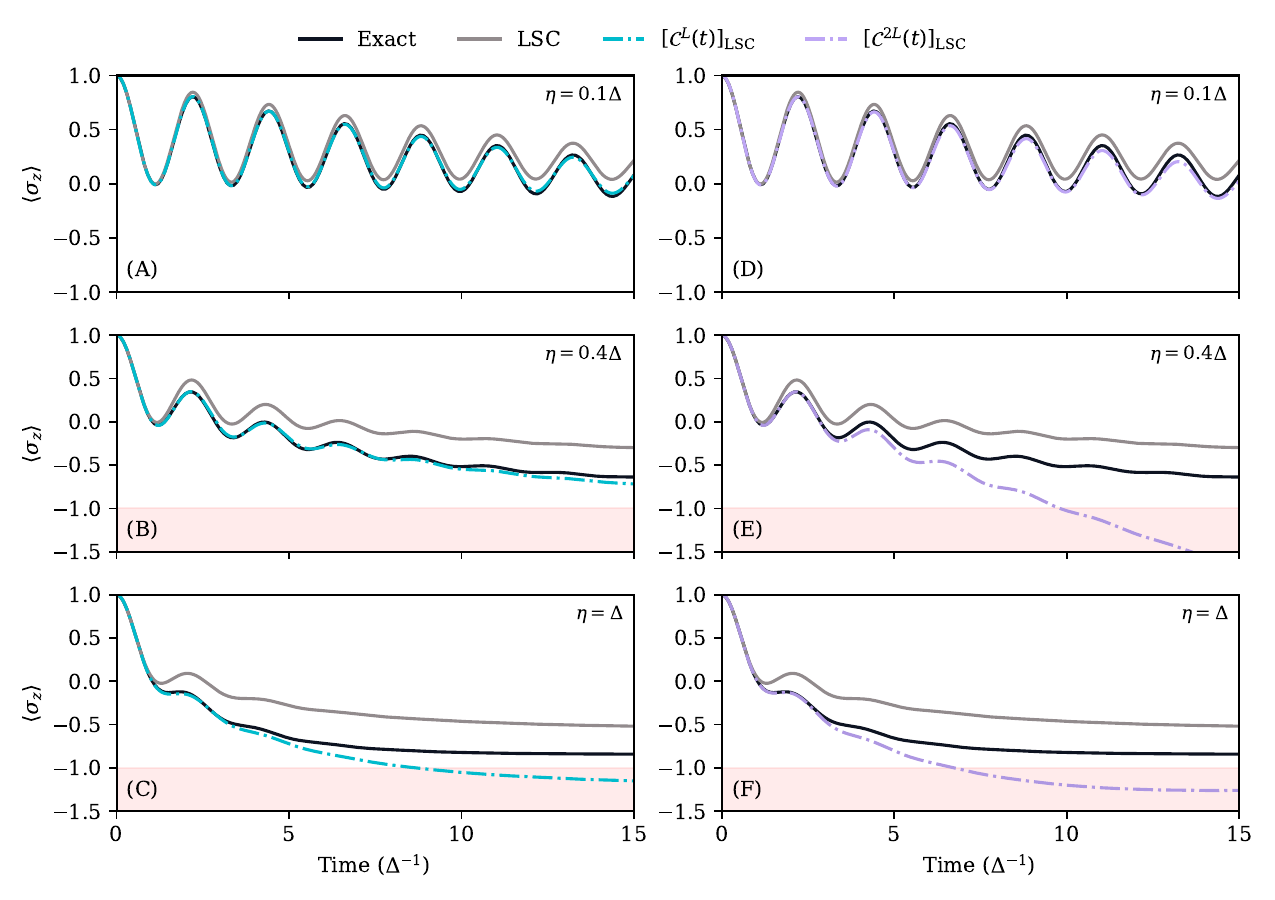}
\caption{\label{setup} \textbf{Extent to which left-handed derivatives can improve the accuracy of semiclassical dynamics.} Comparison of exact (black) and  LSC (gray) population dynamics to integrating $\big[\mathcal{C}^L(t)\big]_{\rm LSC}$ and $\big[\mathcal{C}^{2L}(t)\big]$. All panels correspond to the spin–boson model with $\epsilon = \Delta$, $\beta = 5.0\Delta^{-1}$, $\omega_c = \Delta$, an Ohmic spectral density, and varying system–bath coupling, $\eta$, subject to initial condition $\rho(0) = \rho_B\ket{1}\bra{1}$. Red shaded regions denote unphysical negative populations.}

\label{fig:fig3}
\end{figure*}
%%%%%%%%%%%%%%%%%%%%Figure 3 end %%%%%%%%%%%%%%%%%%%%%%%

\subsection{When does $\mathcal{C}^L(t)$ struggle?}

Just as the SC-GQME stops improving SC dynamics in challenging parameter regimes \cite{Amati2022}, one might expect there to be cases where $\mathcal{C}^L(t)$ is not sufficient to improve the accuracy of LSC dynamics. These parameter regimes include instances of large bias, which can be expected to exacerbate the difficulties SC methods face in capturing detailed balance, and strong system-bath coupling regimes where the mean-field approximation in LSC is known to fail. For example, in Fig.~\ref{fig:fig3} panels (A), (B), and (C) we show that $[\dot{\mathcal{C}}^{L}(t)]_{\text{LSC}}$ ceases to provide an advantage over $[\mathcal{C}(t)]_{\text{LSC}}$ as a function of increasing system-bath coupling, $\eta$. Most worrying, however, is the unphysical dynamics of $\big[\mathcal{C}^{L}(t)\big]_{\rm LSC}$, exhibiting negative populations in the case with the highest coupling, $\eta = 1$, in panel (C). In such parameter regimes where the one left-handed derivative proves not only insufficient but instead catastrophic, one might wonder if taking additional left-handed derivatives can help. This expectation aligns with our hypothesis that adopting exact left-handed derivatives delays the onset of inaccuracy for the correlation function. Yet, as we show below, one must strike a balance between the delayed onset of inaccuracy in time and the complexity of the initial condition to be sampled, as generated by the repeated action of the exact Liouvillian. 

We illustrate this complexity of the initial condition by considering a  correlation function starting from the initial condition $\hat{\rho}_B\ket{1}\bra{1}$, 
\begin{equation} \label{eq:noderiv-lsc}
    [\mathcal{C}_{1B}(t)]_{\rm LSC} = \int {\rm d} \boldsymbol{\Gamma}\ \rho_B^W [\ket{1}\bra{1}]^W B^W_{\boldsymbol{\Gamma}}(t).
\end{equation}
Upon taking the first left-handed derivative, we obtain,
\begin{equation}\label{eq:1LderivLSC}
\begin{split}
    &[\dot{\mathcal{C}}_{1B}^{L}(t)]_{\rm LSC} = -i\int {\rm d} \boldsymbol{\Gamma} \; (\mathcal{L}\rho_B^W \ket{1}\bra{1})^W B^W_{\boldsymbol{\Gamma}}(t)\\
    &\quad =  \int {\rm d} \boldsymbol{\Gamma}\  \rho_B^W( - \Delta \sigma_y^W  +2\xi^W[\ket{1}\bra{1}]^W )B^W_{\boldsymbol{\Gamma}}(t),\\
\end{split}
\end{equation}
where $\xi^W = \sum_n \frac{-c_n \tanh\left(\frac{\beta \omega_n}{2}\right)p_n}{\omega_n}$. The complexity of Eq.~\eqref{eq:1LderivLSC} is greater than that of Eq.~\eqref{eq:noderiv-lsc} as one must evaluate correlation functions where the initial condition contains contributions from both coherences ($\sigma_y^W$) and populations $(\ket{1}\bra{1})^W$ dressed by bath operators ($\xi^W$). Leveraging the intuition that Monte Carlo integration of increasingly complex operators becomes more difficult and computationally demanding, one might imagine that converging these correlation functions subject to these more complex initial conditions takes significantly more sampling than one would require for Eq.~\ref{eq:noderiv-lsc}. In turn, one can expect that incomplete cancellation of error from finite sampling at the level of Eq.~\ref{eq:1LderivLSC} gets amplified in the integrative construction of $\mathcal{C}^L(t)$ in Eq.~\eqref{integration}. Indeed, $\sim10^5$ trajectories suffice to converge Eq.~\eqref{eq:noderiv-lsc}, while converging Eq.~\eqref{eq:1LderivLSC} $\sim~10^6$ trajectories to achieve similar accuracy when using Eq.~\eqref{integration}. 

This complexity only rises further when we pursue the second left-handed derivative,
\begin{widetext}
\begin{equation}\label{eq:2LderivLSC}
    \begin{split}
       [\ddot{\mathcal{C}}_{1B}^{2L}(t)]_{\rm LSC} &= -\int {\rm d} \boldsymbol{\Gamma}\ (\mathcal{L}^2\rho_B^W \ket{1}\bra{1})^W B^W_{\boldsymbol{\Gamma}}(t) \\
       &= -\int {\rm d} \boldsymbol{\Gamma}\  \rho_B^W \left[ 2\Delta^2 \sigma_z^W - 2\Delta (\varepsilon  +  V_B^W) \sigma_x^W  + 2\Delta\xi^W \sigma_y^W + 2(\varphi^W - \zeta^W)(\ket{1}\bra{1})^W \right] B^W_{\boldsymbol{\Gamma}}(t).
    \end{split}
\end{equation}
\end{widetext}

\noindent Here $\ddot{\mathcal{C}}^{2L}(t) = -(\mathcal{L}^2\mathbf{A}|e^{i\mathcal{L}t}|\mathbf{A}) $, $V_B^W = \sum_n c_n x_n$, $ \varphi^W = \sum_n c_n \omega_n \tanh(\beta\omega_n/2) x_n$, and $\zeta^W = 2 (\xi^W)^2 - \sum_n c_n^2 \frac{ \tanh\left(\beta \omega_n/2\right)}{\omega_n}$. Equation~\eqref{eq:2LderivLSC} poses a monumental challenge to converge, requiring $\sim 10^7-10^8$ trajectories. Recovering $\mathcal{C}(t)$ from its second time derivative simply requires double integration, 
\begin{equation}
   \label{double-int}
    \mathcal{C}^{2L}(t) = \mathcal{C}(0) + \int_0^t {\rm d} s_1 \; \int_0^{s_1} {\rm d}s_2 \; \ddot{\mathcal{C}}^{2L}(s_2).
\end{equation}
Starting with an easy parameter regime, Fig.~\ref{fig:fig3} (A) and (D) show that $\big[\mathcal{C}^{L}(t)\big]_{\rm LSC}$ in Eq.~\eqref{integration} yields a more accurate result at long times than $\big[\mathcal{C}^{2L}(t)\big]_{\rm LSC}$ in Eq.~\eqref{double-int}---despite our expectation that the second left-handed derivative should provide greater accuracy. This small but significant disagreement illustrates the difficulty of striking a balance between increased accuracy at short times and the much higher computational cost of converging the second left-handed derivative. 

Despite the greater complexity of the second left-handed derivative, one might envision parameter regimes where the first derivative struggles but the second derivative can delay the onset of inaccuracy. Figure~\ref{fig:fig3} (E) and (F) focus on stronger system-bath coupling, which is challenging to SC theories. Consistent with the mechanism of short-time accuracy that we have outlined, $[\mathcal{C}^{L}(t)]_{\rm LSC}$ in (C) offers a short-time improvement over $[\mathcal{C}(t)]_{\rm LSC}$, but deviates from the exact result, even turning unphysical at long times. As hypothesized, $\big[\mathcal{C}^{2L}(t)\big]_{\rm LSC}$ can offer greater short-time accuracy than $\big[\mathcal{C}^{L}(t)\big]_{\rm LSC}$ in (F), but ultimately falls prey to long-time instability in both (E) and (F).
These results suggest that \textit{although first and second left-handed derivatives progressively improve short-time accuracy---with diminishing gains---both approaches ultimately exhibit long-time instability in challenging parameter regimes}.

\subsection{Summary}

Our above analysis establishes a framework that elucidates how to improve the accuracy of SC dynamics via left-handed derivatives. We show that exact (left-handed) derivatives acting on the initial condition before transforming into SC phase space delay the onset of inaccuracy in short-time SC dynamics, providing a concrete answer to \textbf{\hyperref[qs]{Q1}}, even if long-time inaccuracy can become worse. While one can, in principle, employ any number of analytical derivatives on the initial condition to further delay inaccuracies, this approach is analytically demanding, generally inaccessible unless one has access to an explicit Liouvillian, and poses a significant computational challenge. However, SC-GQMEs may provide an advantage over direct integration of these left-handed derivatives. This is because a GQME's memory kernel may decay on fast timescales---perhaps faster than the onset of the instability in the left-handed derivatives. In such cases, the SC-GQME can benefit from the enhanced short-time accuracy afforded by the left-handed derivatives while offering a path to construct well-behaved and improved dynamics over \textit{all} time. Hence, we turn back to the SC-GQME to uncover how it operates and determine whether it works at all differently from simply integrating left-handed derivatives. 

\section{Apparent differences, conservation laws, \& sampling errors}

While it appears that the SC-GQME improves the accuracy of LSC dynamics by leveraging the left-handed derivative $\dot{\mathcal{C}}^{L}(t)$ in $\mathcal{K}^{(1)}(t)$ and $\mathcal{K}^{(3b)}(t)$ (see Eq.~\eqref{k1} and ~\eqref{k3b}), Fig.~\ref{fig:fig1} (A) and (C) demonstrate that predictions from the SC-GQME do not \textit{exactly} match predictions from $[\mathcal{C}^L(t)]_{\rm LSC}$ , especially at long times, despite both yielding improved accuracy. This discrepancy motivates \textbf{Q2} and implies that the SC-GQME, perhaps through its self-consistent construction of the memory kernel, is doing something different than simply integrating $\big[\dot{\mathcal{C}}^L(t)\big]_{\rm LSC}$. Motivated by this insight, we ask how the SC-GQME and $\big[\mathcal{C}^{L}(t)\big]_{\rm LSC}$ differ for alternative metrics, including conservation laws. 

To address \textbf{\hyperref[qs]{Q2}}, we consider the extent to which $\big[\mathcal{C}^{L}(t)\big]_{\rm LSC}$ and the SC-GQME conserve probability, $h$, in the electronic subsystem. To illustrate this point, we describe the time-dependence of $h$ using matrix elements of $\dot{\mathcal{C}}^L(t)$. Specifically, we consider the matrix elements of $\dot{\mathcal{C}}^L(t)$ that correspond to initializing an excitation in state $\ket{1}\bra{1}$ and measuring the excitation in site $\ket{1}\bra{1}$ and $\ket{2}\bra{2}$,
\begin{equation}
    \frac{{\rm d}h^L}{{\rm d}t} = \dot{\mathcal{C}}^L_{11}(t) + \dot{\mathcal{C}}^L_{12}(t),
\end{equation}
In the spin-boson model, exact quantum dynamics trivially satisfy this conservation law,
\begin{align}
\begin{split}
    \frac{{\rm d}h^L}{{\rm d}t} &= -\Delta\mathrm{Tr}\left\{\hat{\rho}_B \hat{\sigma}_y\right\}   -i\mathrm{Tr}\{[\hat{V}_B, \hat{\rho}_B] \ket{1}\bra{1}\} = 0.
\end{split}
\end{align}
However, under the LSC approximation, finite sampling of the initial conditions can cause this condition to no longer be strictly satisfied, 
\begin{align}
\begin{split}
\label{conserve}
    \frac{{\rm d}h^{L}_{\rm LSC}}{{\rm d}t} &= -\Delta\int {\rm d}\boldsymbol{\Gamma} \; \rho_B^W \sigma_y^W 
    + 2 \int {\rm d}\boldsymbol{\Gamma} \; \xi^W \rho_B^W [\ket{1}\bra{1}]^W\approx 0.
\end{split}
\end{align}
As the right-hand side of Eq.~\eqref{conserve} is calculated from sampling of initial conditions and is independent of the generator of dynamics, it remains a constant through time. Therefore, one would expect the total population to change linearly when integrating $\big[\dot{\mathcal{C}}^L(t)\big]_{\rm LSC}$. As such, when one samples $\big[\dot{\mathcal{C}}^L(t)\big]_{\rm LSC}$ and integrates it using Eq.~\eqref{integration}, one should not expect to conserve the total electronic population to high precision, i.e., it should not obey the probability conservation law. 

\begin{figure}[h]
\centering
\includegraphics[width=\linewidth]{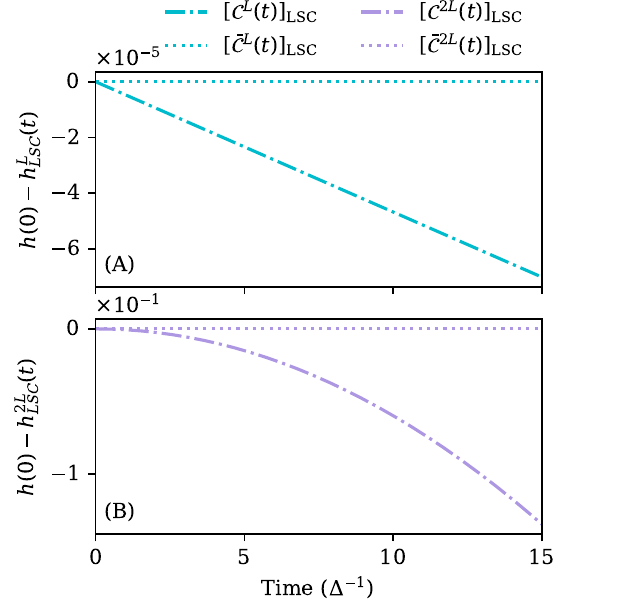}
\caption{\label{setup} \textbf{A simple shift allows left-handed derivatives to conserve population.} Conservation of total population, $h$, in (A) $\big[\mathcal{C}^L(t)\big]_{\rm LSC}$ versus $\big[\bar{\mathcal{C}}^L(t)\big]_{\rm LSC}$ and (B) $\big[\mathcal{C}^{2L}(t)\big]_{\rm LSC}$ versus $\big[\bar{\mathcal{C}}^{2L}(t)\big]_{\rm LSC}$ for a spin-boson model with parameters $\epsilon = \Delta$, $\beta = 5.0\Delta^{-1}$, $\omega_c = 2.0\Delta$, $\eta = 0.2\Delta$ subject to initial condition $\rho(0) = \rho_B\ket{1}\bra{1}$. }
\label{fig:figa1}
\end{figure}

This static error biases the entire LSC correlation function and is magnified upon numerical integration. If this insight is correct, by removing the zero-time bias uniformly across time and instead replacing the zero-time value by either the analytical answer or a highly converged static calculation, 
\begin{subequations} \label{reference}
\begin{align}
    \dot{\bar{\mathcal{C}}}^{L}(t) &= \dot{\mathcal{C}}^{L}(t) - \dot{\mathcal{C}}^{L}(0) + \dot{\mathcal{C}}^{\rm ref}(0),\\
    \ddot{\bar{\mathcal{C}}}^{2L}(t) &= \ddot{\mathcal{C}}^{2L}(t) - \ddot{\mathcal{C}}^{2L}(0) + \ddot{\mathcal{C}}^{\rm ref}(0),
\end{align}
\end{subequations}
one should conserve population upon numerical integration of the left-handed derivatives. Figure~\ref{fig:figa1} demonstrates this to be true. Thus, the numerically integrated $\big[\mathcal{C}^L(t)\big]_{\rm LSC}$ fails to conserve population because finite sampling of its initial conditions leave a residual constant bias in the correlation function---a problem that is easy to resolve with the simple shift in Eq.~\eqref{reference}---rather than from any deficiency in the underlying SC methodology.

%%%%%%%%%%%%%%%%%%%%Figure 4 start %%%%%%%%%%%%%%%%%%%%%%
%%%%%%%%%%%%%%%%%%%%Figure 4 end %%%%%%%%%%%%%%%%%%%%%%

\begin{figure}[h]
\centering
\includegraphics[width=\linewidth]{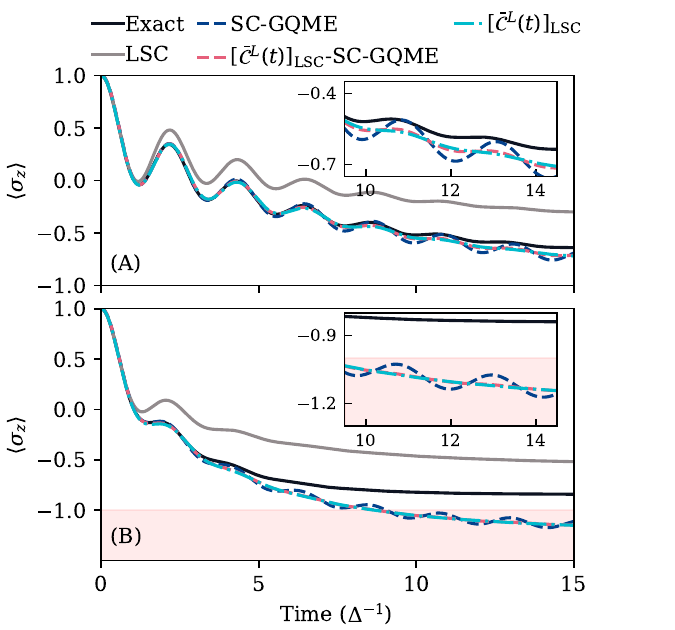}
\caption{\label{setup} \textbf{Resolving apparent discrepancies between the SC-GQME and the left-handed derivative.} Nonequilibrium population dynamics for the spin-boson model with $\epsilon = \Delta$, $\beta = 5.0\Delta^{-1}$, $\omega_c =\Delta$, an Ohmic spectral density, and  subject to initial condition $\rho(0) = \rho_B\ket{1}\bra{1}$. (A) $\eta = 0.4\Delta$. (B) $\eta = \Delta$. Red shaded regions denote nonphysical negative populations. Note that the dynamics obtained from the single-accuracy SC-GQME built on the shifted left-handed derivative, $\big[\bar{\mathcal{C}}^L(t)\big]_{{\rm LSC}}$-SC-GQME (dashed cyan line), and the direct integration of $\big[\bar{\mathcal{C}}^L(t)\big]_{{\rm LSC}}$ (dashed red line) agree, in contrast to the more strongly oscillatory dynamics predicted with the traditional SC-GQME (dashed dark blue line), which is built on mixed-accuracy inputs to the auxiliary kernels. Use of single-accuracy constructions validates the use of equivalence proofs\cite{noneq1, whencanonewin} for GQME dynamics.}
\label{fig:fig4}
\end{figure}

Interestingly, the current formulation of the SC-GQME preserves the total population, $h$, for the Argyres-Kelley and population-based projectors in spin-boson model-like problems. However, this is a fortuitous accident arising from the structure of $\dot{\mathcal{C}}(0)$ and its action in the auxiliary kernels for these projectors, respectively (see SI Sec.~V). While we encounter conservation of total population for the chosen projectors in this work, by choosing a different projector---for example, one that projects onto a single population---one no longer guarantees conservation of total population, which can lead to inaccurate descriptions of the dynamics \cite{Mulvihill2022}. Nevertheless, applying the same shift of the left-handed derivative can minimize sampling error and ensure total population conserving dynamics (see SI Sec.~V E). Moreover, as previous studies \cite{whencanonewin} have proven, if one constructs a memory using a correlation function and its (commensurably accurate) time derivatives, one must predict the same correlation function from the SC-GQME. Thus, if one builds the memory kernel using only the corrected $\big[\bar{\mathcal{C}}^L(t)\big]_{\rm LSC}$, which preserves population, and its time derivatives, the SC-GQME must return the same population-conserving dynamics observed in $\big[\bar{\mathcal{C}}^L(t)\big]_{\rm LSC}$. Figure~\ref{fig:fig4} verifies this fact, with $\big[\bar{\mathcal{C}}^L(t)\big]_{\rm LSC}$ matching the dynamics of the SC-GQME whose memory kernel we built using $\big[\bar{\mathcal{C}}^L(t)\big]_{\rm LSC}$ and its time derivatives within $0.01\%$ error. Figure~\ref{fig:fig4} also reveals that the traditional SC-GQME (labeled `SC-GQME'), which uses mixed (0$^{\rm th}$ and $1^{\rm st}$) left-handed derivatives to construct the auxiliary kernels, mixes two levels of accuracy and thus yields similar (albeit not the same) dynamics as the SC-GQME built consistently from the left-handed derivative $\big[\bar{\mathcal{C}}^{L}(t)\big]_{\rm LSC}$ and its time-derivatives.

    %%%%%%%%%%%%%%%%%%%%Figure    begin %%%%%%%%%%%%%%%%%%%%%%%
%%%%%%%%%%%%%%%%%%%%Figure 5 end %%%%%%%%%%%%%%%%%%%%%%

    %%%%%%%%%%%%%%%%%%%%Figure  6  begin %%%%%%%%%%%%%%%%%%%%%%%
\begin{figure*}[t]
\centering
\includegraphics[width=\linewidth]{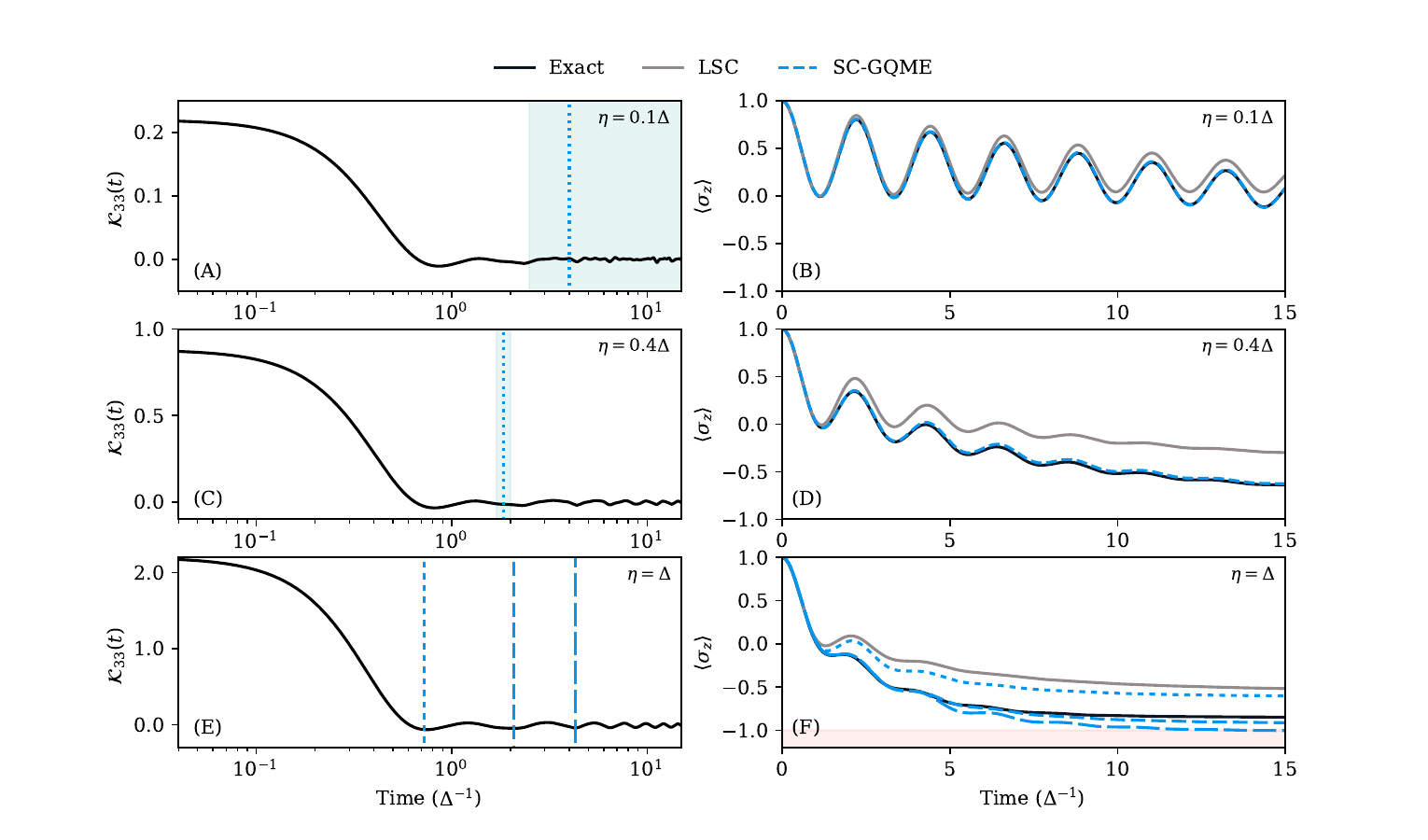}
\caption{\label{setup} \textbf{Vanishing ``plateaus of stability" complicate the choice of the memory kernel truncation time, $\boldsymbol{\tau_{\mathcal{K}}}$.} Memory kernel elements (\textit{left}) and nonequilibrium population dynamics (\textit{right}) for the spin-boson model with $\epsilon = \Delta$, $\beta = 5.0\Delta^{-1}$, $\omega_c = \Delta$, as a function of increasing system-bath coupling, $\eta$, with an Ohmic spectral density, and subject to initial condition $\rho(0) = \rho_B\ket{1}\bra{1}$. Green shaded regions indicate the plateau of stability in (A) and (C), and vertical dotted lines correspond to proposed memory kernel cutoffs, $\tau_{\mathcal{K}}$. Memory cutoffs chosen within this region yield accurate population dynamics shown in (B) and (D). For strong system–bath coupling, the plateau of stability vanishes. Truncation at different times (blue dashed lines in (E)) generate population dynamics (F) with differing long-time limits. Red shaded regions denote unphysical, negative populations.}
\label{fig:fig5}
\end{figure*}
%%%%%%%%%%%%%%%%%%%%Figure 6 end %%%%%%%%%%%%%%%%%%%%%%%

In summary, we demonstrate that dynamics predicted from integrating $\big[\dot{\mathcal{C}}^L(t)\big]_{\rm LSC}$ and the SC-GQME differ due to the preservation of conservation laws. We find that the source of this error was a time-independent sampling bias that numerical integration amplifies when obtaining $\big[\mathcal{C}^{L}(t)\big]_{\rm LSC}$ from $\big[\dot{\mathcal{C}}^{L}(t)\big]_{\rm LSC}$ or $\big[\mathcal{C}^{2L}(t)\big]_{\rm LSC}$ from $\big[\ddot{\mathcal{C}}^{2L}(t)\big]_{\rm LSC}$. We show that we could fix this problem easily by shifting $\big[\dot{\mathcal{C}}^L(t)\big]_{\rm LSC}$ and $\big[\ddot{\mathcal{C}}^{2L}(t)\big]_{\rm LSC}$ by the appropriate reference constants, as given by Eqs.~\eqref{reference}. Use of these corrected left-handed derivatives, $\big[\dot{\bar{\mathcal{C}}}^L(t)\big]_{\rm LSC}$ and $\big[\ddot{\bar{\mathcal{C}}}^{2L}(t)\big]_{\rm LSC}$, leads to internally consistent integrated and SC-GQME dynamics that conserve population. Finally, our work reveals that employing the same level of left-handed derivatives in the construction of the auxiliary kernels---in contrast to the mixed-accuracy objects relying on different numbers of left-handed derivatives employed in previous versions of the SC-GQME---ensures the internal consistency of the SC-GQME dynamics. Together, these observations allow one to answer \textbf{\hyperref[qs]{Q2}} and clarify the role of self-consistency in SC-GQME. The self-consistent structure of the SC-GQME preserves conservation laws when constructed with mixed-accuracy or biased auxiliary kernels. However, the self-consistency neither improves nor impairs the SC-GQME relative to integrating a shifted left-handed derivative (which obeys conservation of population), provided that the auxiliary kernels are constructed from the same number of correspondingly shifted left-handed derivatives.

\section{When is the SC-GQME advantageous?}
%%%%%%%%%%%%%%%%%%%%Figure  7  begin %%%%%%%%%%%%%%%%%%%%%%%
\begin{figure*}[t]
\centering
\includegraphics[width=\linewidth]{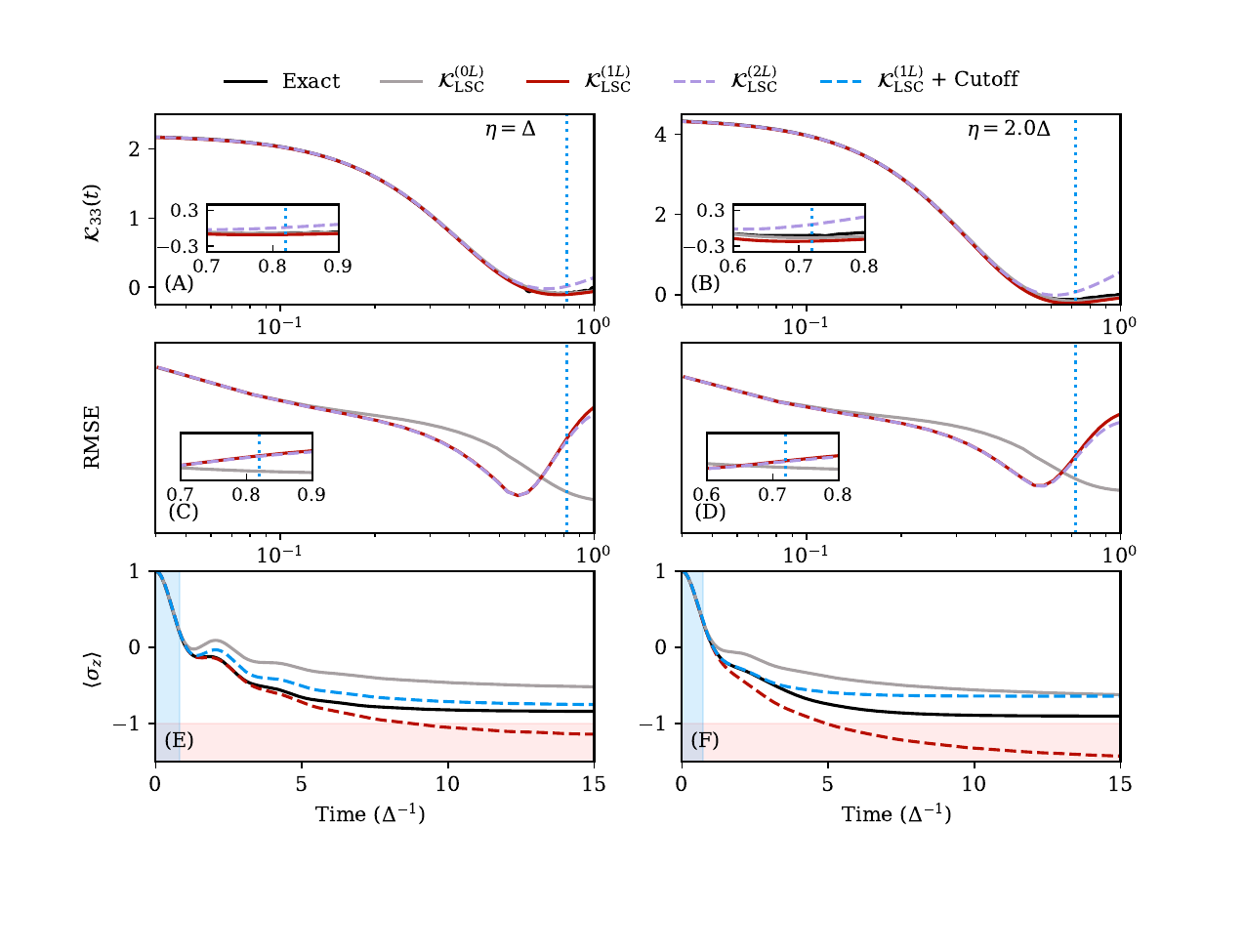}
\vskip-55pt
\caption{\label{setup} \textbf{Our RMSE protocol determines optimal cutoff times $\tau_M$, even in challenging parameter regimes.} \textit{Single-accuracy} memory kernel elements constructed from various numbers of left-handed derivatives (A-B), RMSE of the SC-GQME dynamics with respect to the bare LSC dynamics as a function of proposed kernel truncation time $\tau_M$ (C-D), and nonequilibrium population dynamics (E-F) for the spin-boson model with $\epsilon = \Delta$, $\beta = 5.0\Delta^{-1}$, $\omega_c = \Delta$, as a function of increasing system-bath coupling, $\eta$, with an Ohmic spectral density, and subject to initial condition $\rho(0) = \rho_B\ket{1}\bra{1}$. We choose $\tau_M$ as the time when the RMSE of $\mathcal{K}_{\rm{LSC}}^{(1L)}$ (solid red line) and $\mathcal{K}_{\rm{LSC}}^{(2L)}$ (dashed purple line) start to deviate (dashed blue line). Memory truncation times: $\tau_M = 0.83$ in (C) and $\tau_M = 0.72$ (D). (Insets): Enlarged view of the memory kernels (A-B) and RMSE (C-D) around $\tau_M$. (E-F) Note that the resulting population dynamics from $\mathcal{K}^{\rm{LSC}}_{1L}$ with memory truncation (dashed blue) are always better than the direct LSC approximation (gray) while avoiding the unphysical behavior evident at long times when using the SC-GQME without (dashed red) memory truncation. The shaded blue region highlights the time up to $\tau_M$. Red shaded regions denote nonphysical negative populations.}
\label{fig:fig7}
\end{figure*}

Until now, we have seen that the SC-GQME appears to offer the same level of accuracy as $\bar{\mathcal{C}}^{L}(t)$. However, we have not yet exploited one of the central benefits of the GQME: when appropriately designed, a GQME's $\mathcal{K}(t)$ decays faster (at $\tau_{\mathcal{K}}$) than $\mathcal{C}(t)$ (at $\tau_C$). In such cases, one only needs to simulate $\mathcal{C}(t)$ for short times to construct $\mathcal{K}(t)$, which then enables one to generate GQME dynamics over arbitrary times at comparatively trivial computational cost. When $\tau_{\mathcal{K}} \ll \tau_C$, the reduction in the computational cost can reach multiple orders of magnitude \cite{ bhattacharyya2024anomalous, bhattacharyya2025space, sayer2025efficient}. Beyond this efficiency boost, a short-lived memory kernel can have a second benefit in SC-GQMEs: that $\big[\bar{\mathcal{C}}^{L}\big]_{\rm LSC}$ and $\big[\bar{\mathcal{C}}^{2L}\big]_{\rm LSC}$ may be able to offer improved accuracy over $t \sim \tau_{\mathcal{K}}$ before their accuracy {degrades for $t > \tau_\mathcal{K}$, allowing one to avoid the long-time instabilities evident in Fig~\ref{fig:fig3} (E) and (F). Under these conditions, the SC-GQME may offer improved accuracy, even for parameter regimes where the single and double integration of $\big[\dot{\bar{\mathcal{C}}}^{L}(t)\big]_{\rm LSC}$ and $\big[\ddot{\bar{\mathcal{C}}}^{2L}(t)\big]_{\rm LSC}$ would be expected to fail. We demonstrate this below. 

\subsection{Difficulties of choosing $\tau_M$} 

In well-behaved dissipative problems, the memory kernel decays to $0$, allowing one to set an upper bound on the convolution integral in Eq.~\eqref{mnz} equal to the memory lifetime, $\tau_M$, for $t \geq \tau_M$. This protocol accurately recovers the correlation function $\mathcal{C}(t)$. Early implementations of the GQME that employed exact solvers for impurity models routinely showed that their GQME dynamics converged with increasing cutoff time, $\tau_M$ \cite{ cohen2011memory, Cohen2013(2), cohen2013numerically}. This type of analysis also allowed researchers to identify $\tau_M$ in SC-GQMEs \cite{kelly2013efficient, Kelly2015}. However, subsequent work showed that memory kernels exhibit high-frequency oscillations when the spectral density of the system spans a wide energetic breadth, making it difficult to identify its decay timescale \cite{noneq1}. Further, the limited accuracy of SC methods can result in memory kernels that do not decay fully \cite{noneq1, Mulvihill2019}. To resolve this ambiguity, Ref.~\citen{noneq1} proposed the idea of a ``plateau of stability'', corresponding to the range of time during which truncating the memory kernel ceases to influence the SC-GQME dynamics. In practice, one determines $\tau_M$ as any point within this plateau of stability. Unfortunately, these same studies demonstrated that the size of this plateau of stability depends on the parameter regime \cite{noneq1, montoya2017approximate}. Even worse, in some parameter regimes, this plateau of stability may not even exist \cite{noneq1, montoya2017approximate, Mulvihill2019, Amati2022}. Thus, in challenging parameter regimes, choosing $\tau_M$ via a disappearing plateau of stability becomes ambiguous at best and impossible at worst. 

Figure~\ref{fig:fig5} illustrates how the plateau of stability becomes narrower and even disappears as the parameter regime becomes more challenging to LSC---in this case, as $\eta$ increases. Despite this difficulty, the smoothness of the SC kernels in Fig.~\ref{fig:fig5} (E) enables us to serendipitously choose a good $\tau_M$ value (i.e., one that leads to SC-GQME dynamics that closely match the exact benchmark) by picking the second dip of the memory kernel. But, why not pick any point after this? Doing so yields suboptimal---even unphysical---dynamics. The situation worsens when the physical problem involves broader spectral densities, e.g., of a Debye form. In such cases, the resulting SC memory kernels display high-frequency noise \cite{noneq1} that, while globally innocuous, makes it difficult to \textit{visually} identify when the memory kernel has decayed, revealing the need for an automated method, such as that of plateau of stability identification. Yet, this plateau disappears as the coupling grows, making it clear that a more general method is required.

%%%%%%%%%%%%%%%%%%%%Figure  8  begin %%%%%%%%%%%%%%%%%%%%%%%
\begin{figure*}[t]
\centering
\includegraphics[width=\linewidth]{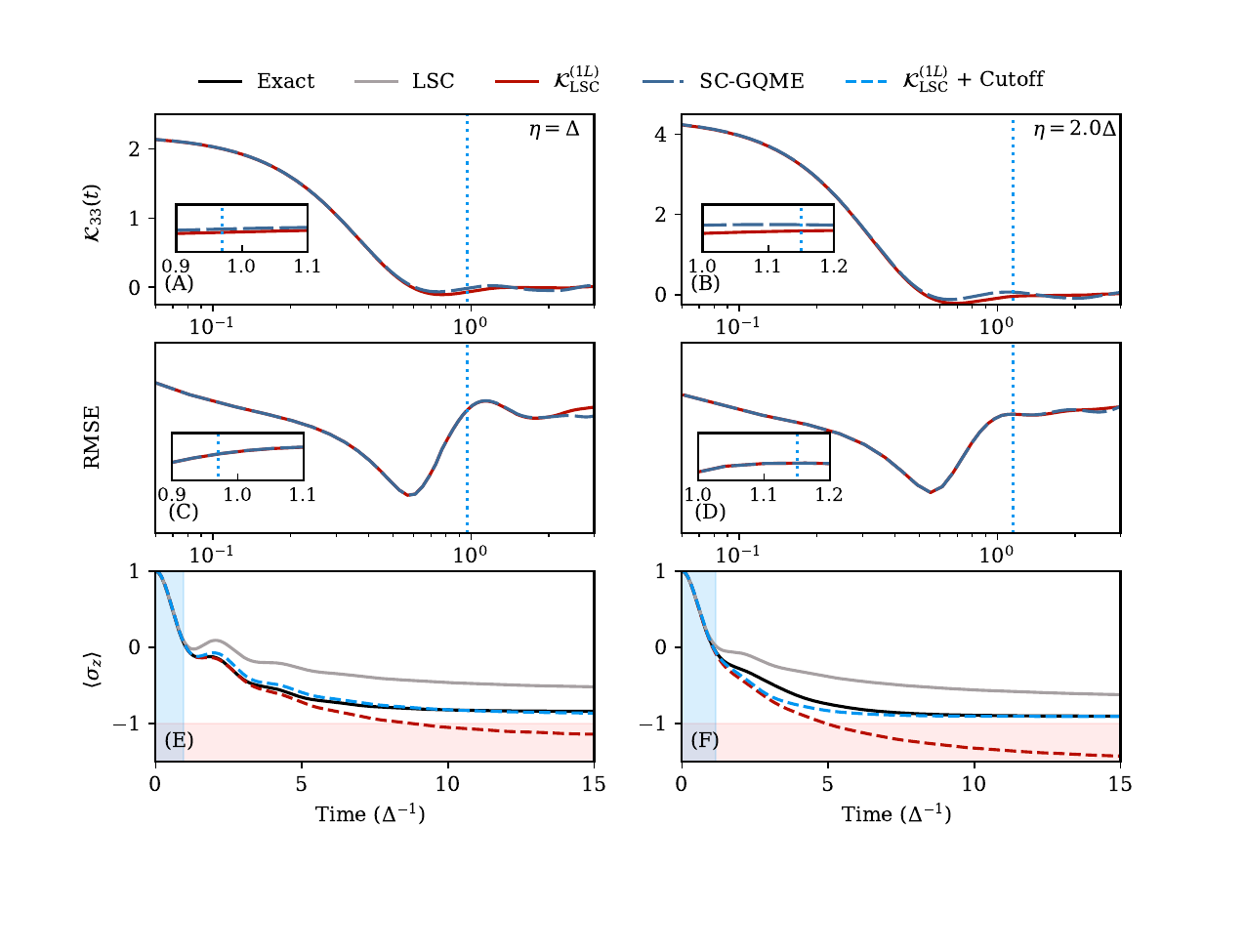}
\vskip-55pt
\caption{\label{setup} \textbf{Our RMSE protocol affords even greater accuracy when combined with mixed-accuracy kernels.} Here, we report analogous results to Fig.~\ref{fig:fig7} with the main difference being that we compare \textit{single-accuracy} and \textit{mixed-accuracy} constructions of the memory kernels. That is, we construct $\mathcal{K}^{(1L)}_{\rm LSC}(t)$ from single-accuracy auxiliary kernels and $\mathcal{K}(t)$ from the SC-GQME using mixed-accuracy auxiliary kernels defined in Eq.~\eqref{aux_all}. Memory truncation times: $\tau_M = 0.97$ in (C) and $\tau_M = 1.15$ (D). Particularly noteworthy in (E-F) is the closer agreement to the numerically exact population dynamics (black line) of $\mathcal{K}^{(1L)}_{\rm LSC}(t)$ coupled with our proposed RMSE-based memory truncation protocol.}
\label{fig:fig8}
\end{figure*}

\subsection{Triangulating accuracy limits}

Figure~\ref{fig:fig5} reveals a major challenge for the SC-GQME: when the plateau of stability disappears, how does one choose a favorable $\tau_M$ without prior knowledge of the exact dynamics? To answer this question, previous work \cite{mulvihill2021road} has attempted to address this challenge by proposing closures that combine auxiliary kernels of mixed accuracy to stabilize the long-time behavior of the memory kernel and avoid the ambiguities illustrated in Fig.~\ref{fig:fig5}. While these closures consistently exhibit a plateau of stability and provide an unambiguous truncation time, memory truncation within this framework can still produce unphysical long-time dynamics under strong system-bath coupling (see SI Sec.~VI C). Here, we take a fundamentally different approach to address \textbf{\hyperref[qs]{Q3}}: rather than modifying the construction of the memory kernel, we leverage our understanding of the short-time dynamics to identity the optimal cutoff time for a memory kernel, even one without a plateau of stability, allowing us to optimize accuracy while always providing physically meaningful dynamics. We know, for example, how to ensure that the SC-GQME retains the same accuracy as the dynamics from which we build its memory kernel. That is, if one uses $[\mathcal{C}(t)]_{\rm LSC}$ and its numerical time derivatives, $\partial_t[\mathcal{C}(t)]_{\rm LSC}$ and $\partial_t^{2}[\mathcal{C}(t)]_{\rm LSC}$, to construct $\mathcal{K}(t) \equiv \mathcal{K}_{\rm LSC}^{(0L)}(t)$, the resulting SC-GQME dynamics are identical to $[\mathcal{C}(t)]_{\rm LSC}$ \cite{noneq1, whencanonewin}. Our analysis above also reveals that if one uses dynamics obtained from the single-accuracy first left-handed derivative, $[\bar{\mathcal{C}}^{L}(t)]_{\rm LSC}$ and $[\dot{\bar{\mathcal{C}}}^{L}(t)]_{\rm LSC}$, and its numerical time derivative, $\partial_t[\dot{\bar{\mathcal{C}}}^{L}(t)]_{\rm LSC}$, to construct $\mathcal{K}(t) \equiv \mathcal{K}_{\rm LSC}^{(1L)}(t)$, the resulting SC-GQME dynamics are identical to $[\bar{\mathcal{C}}^{L}(t)]_{\rm LSC}$. Hence, if one were to use $\big[\bar{\mathcal{C}}^{2L}(t)\big]_{\rm LSC}$, $\big[\dot{\bar{\mathcal{C}}}^{2L}(t)\big]_{\rm LSC}$, and $\big[\ddot{\bar{\mathcal{C}}}^{2L}(t)\big]_{\rm LSC}$ to construct $\mathcal{K} \equiv \mathcal{K}_{\rm LSC}^{(2L)}(t)$, one should recover dynamics consistent with $\big[\bar{{\mathcal{C}}}^{2L}(t)\big]_{\rm LSC}$. Since increasing the number of left-handed derivatives progressively extends the short-time accuracy of the LSC dynamics, we propose that the memory kernels obtained from each of these protocols should offer short-time accuracy for increasingly longer times before long-time inaccuracy, or even instability, becomes dominant. 

To test this hypothesis, we construct $\{ \mathcal{K}_{\rm LSC}^{(0L)}(t), \mathcal{K}_{\rm LSC}^{(1L)}(t), \mathcal{K}_{\rm LSC}^{(2L)}(t)\}$---memory kernels with $0$, $1$, and $2$ left-handed derivatives. Given the progressively longer short-time accuracy of these memory kernels, we expect all three kernels to agree over a short time. When the the inaccuracy in $\mathcal{K}_{\rm LSC}^{(0L)}(t)$, which is the memory kernel built from correlation functions that remain accurate to the lowest order, sets in at time $\tau_{0,1}$, we expect $\mathcal{K}_{\rm LSC}^{(0L)}(t)$ to start deviating from $\mathcal{K}_{\rm LSC}^{(1L)}(t)$ and $\mathcal{K}_{\rm LSC}^{(2L)}(t)$. Then, at $\tau_{1,2}$ we expect $\mathcal{K}_{\rm LSC}^{(1L)}(t)$ to start differing from $\mathcal{K}_{\rm LSC}^{(2L)}(t)$, which continues to be accurate for some time after this point. Figures~\ref{fig:fig7} (A) and (B) confirm the validity of this expectation. However, in the absence of exact dynamics, one cannot determine how long after $\tau_{1,2}$ this short-time accuracy in $\mathcal{K}_{\rm LSC}^{(2L)}(t)$ lasts. Conservatively, one might truncate $\mathcal{K}_{\rm LSC}^{(1L)}(t)$ or $\mathcal{K}_{\rm LSC}^{(2L)}(t)$ at $\tau_{1,2}$. In addition, anticipating our results, one might also worry about the higher $n$L left-handed derivatives leading to unphysical long-time limits---a point to which we return later. This protocol, however, leads to some ambiguity: which memory kernel element should one choose to test this? One might also imagine that it might be difficult to implement this protocol when the memory kernels are noisy or have high-frequency noise, as observed in systems with broad spectral densities. 

\subsection{A new protocol to choose $\tau_M$} 

To avoid these ambiguities, we propose to use a noise-robust root mean squared error (RMSE) metric. RMSE metrics integrate the error over time added over all entries of the correlation function between reference dynamics, $\mathcal{C}^{\rm ref}_{jk}(t)$, and GQME dynamics, $\mathcal{C}^{\rm GQME}_{jk}(t; \tau_M)$, obtained subject to a proposed choice of cutoff time, $\tau_M$, 
\begin{equation}
    {\rm RMSE}(\tau_M) = \sqrt{\frac{1}{T}\int_0^T {\rm d}t \sum_{jk} |\mathcal{C}^{\rm ref}_{jk}(t) - \mathcal{C}^{\rm GQME}_{jk}(t; \tau_M)|^2}.
\end{equation}
These metrics have recently been used to choose non-Markovian generator lifetimes for conformational dynamics in biophysics~\cite{dominic2023memory, dominic2023building}, charge transport in solids \cite{bhattacharyya2024mori, bhattacharyya2024anomalous, bhattacharyya2025space}, and even linear and nonlinear spectroscopic responses \cite{PhysRevA.92.032113,Sayer2024}. For SC-GQMEs, the natural reference would be either the numerically exact result or the choice of SC theory in the problem. However, access to numerically exact solutions would defeat the purpose of resorting to SC dynamics, and choosing the SC dynamics as a reference can introduce ambiguities. For example, choosing $\tau_M$ when the RMSE built with respect to the bare SC dynamics becomes minimal would bias the GQME to recover the original SC approximation, which is what we want to improve in the first place. Alternatively, while SC dynamics built on left-handed derivatives offer greater accuracy at short times, choosing these as the reference could bias the GQME to recreate unphysical instabilities at long times when the parameter regime is challenging to SC theory. We propose a middle ground. Since the bare SC dynamics remains physical throughout, we use it as the reference for the RMSE from which we choose when to cut off the memory kernel obtained from SC dynamics built on left-handed derivatives, i.e., $\mathcal{C}^{\rm ref}(t) \equiv [\mathcal{C}(t)]_{\rm LSC}$. 

We expect the RMSE for the bare SC dynamics $\mathcal{K}_{\rm LSC}^{(0L)}(t)$ to decrease monotonically with increasing $\tau_M$ (gray curves in Figs.~\ref{fig:fig7} (C) and (D)). In contrast, we expect the RMSE for $\mathcal{K}_{\rm LSC}^{(1L)}(t)$ and $\mathcal{K}_{\rm LSC}^{(2L)}(t)$ to decrease for a short time before deviating from the RMSE for $\mathcal{K}_{\rm LSC}^{(0L)}(t)$. Like for the memory kernels, these deviations should occur at $\tau_{0,1}$ and $\tau_{1,2}$. Figures~\ref{fig:fig7} (C) and (D) show that the RMSE lines for $\mathcal{K}_{\rm LSC}^{(1L)}(t)$ and $\mathcal{K}_{\rm LSC}^{(2L)}(t)$ exhibit a minimum in the RMSE before starting to increase again. Choosing this value of $\tau_M$ produces GQME dynamics that closely track $[\mathcal{C}(t)]_{\rm LSC}$ (see SI Fig.~S7 in SI Sec.~VI). One may also be tempted to choose the next intersection point between the RMSE curves for $\mathcal{K}_{\rm LSC}^{(0L)}(t)$ and $\mathcal{K}_{\rm LSC}^{(1L)}(t)$ and $\mathcal{K}_{\rm LSC}^{(2L)}(t)$. While this choice yields improved dynamics relative to $[\mathcal{C}(t)]_{\rm LSC}$, it is still suboptimal. We then expect to observe the RMSE curves for $\mathcal{K}_{\rm LSC}^{(1L)}(t)$ and $\mathcal{K}_{\rm LSC}^{(2L)}(t)$ to start deviating from each other when the short-time accuracy of $\mathcal{K}_{\rm LSC}^{(1L)}(t)$ wanes, at $\tau_{1,2}$ (dashed blue line in Figs.~\ref{fig:fig7} (C) and (D), which agree with the dashed blue lines in (A) and (B)). This intersection offers an unambiguous protocol for identifying the point in time when $\mathcal{K}_{\rm LSC}^{(1L)}(t)$ and $\mathcal{K}_{\rm LSC}^{(2L)}(t)$ should offer improved GQME dynamics relative to $\mathcal{K}_{\rm LSC}^{(0L)}(t)$ while remaining physical. Indeed, Figs.~\ref{fig:fig7} (E) and (F) confirm this expectation, with the memory-truncated $\mathcal{K}_{{\rm LSC}}^{(1L)}$ dynamics (dashed blue line) exhibiting greater accuracy than the bare LSC dynamics (gray line) while avoiding the unphysical behavior of the $\mathcal{K}_{{\rm LSC}}^{(1L)}$ dynamics subject to no cutoff (dashed red lines).

Yet, we argue that the performance of our truncated GQME dynamics can be better. Effective triangulation of the optimal $\tau_M$ requires kernels that maintain short-time accuracy and long-time stability. Although the second left-handed derivative can provide additional short-time accuracy, its propensity for long-time instability (see Figs.~\ref{fig:fig7} (A) and (B)) makes it a double-edged sword. To extend the stability of the RMSE curves---potentially allowing the curves to overlap for longer times---we propose using kernels that have at most one left-handed derivative, such as $\mathcal{K}^{(1L)}_{\rm LSC}$. One must then introduce a second construction for comparison to triangulate $\tau_M$. 
While one might be tempted to use $\mathcal{K}^{(0L)}_{\rm LSC}$, it has less short-time accuracy relative to $\mathcal{K}^{(1L)}_{\rm LSC}$ and the RMSE curves from each construction deviate at very short times (see Fig.~\ref{fig:fig7} (C) and (D)). Instead, we turn to \textit{mixed-accuracy} constructions. Specifically, we invoke the \textit{traditional} SC-GQME memory kernel construction, whose auxiliary kernels are given by Eq.~\eqref{aux_all}. The only difference between $\mathcal{K}^{(1L)}_{\rm LSC}$ and the traditional SC-GQME kernel lies in the construction of both auxiliary kernels . That is, for the former, one constructs both kernels using only the first left-handed derivative, whereas for the latter, one uses both the original SC approximation, $\big[ \mathcal{C}(t)\big]_{\rm LSC}$, and the first left-handed derivative, $\big[ \mathcal{C}^L(t)\big]_{\rm LSC}$. 

This mixing generally endows the SC-GQME memory kernel with long-time stability as well as different levels of short-time accuracy compared to $\mathcal{K}^{(1L)}_{\rm LSC}(t)$. Thus, we propose to triangulate $\tau_M$ using a \textit{single-accuracy} kernel, $\mathcal{K}^{(1L)}_{\rm LSC}(t)$ and a \textit{mixed-accuracy} kernel, $\mathcal{K}(t)$, from the SC-GQME.  
Thus, we expect both kernels to maintain long-time stability, but with different degrees of short-time accuracy, which we confirm in Fig.~\ref{fig:fig8} (A) and (B). 

Consistent with our proposal, the truncated dynamics from $\mathcal{K}^{(1L)}_{\rm LSC}(t)$ (dashed light-blue lines) in Figs.~\ref{fig:fig8} (E) and (F) exhibit greater accuracy than those in Figs.~\ref{fig:fig7} (E) and (F), including a better estimate of the long-time limit. Thus this protocol performs well in difficult parameter regimes, such as those characterized by high system-bath coupling, high energy bias, or broad spectral densities (e.g., Debye), which we examine in Fig.~S8 and S9 in SI Sec.~VI. Thus, this RMSE-based protocol leveraging the single-accuracy $\mathcal{K}^{(1L)}_{\rm LSC}(t)$ and mixed-accuracy SC-GQME kernel gives both an unambiguous path to identifying the kernel cutoff time, $\tau_M$, and significantly improves the accuracy of SC dynamics across parameter regimes and systems, even those that had proved previously inaccessible for improvement through the SC-GQME.

\subsection{Summary}

    %%%%%%%%%%%%%%%%%%%%Figure  9  begin %%%%%%%%%%%%%%%%%%%%%%%
\begin{figure*}[t]
\centering
\includegraphics[width=\linewidth]{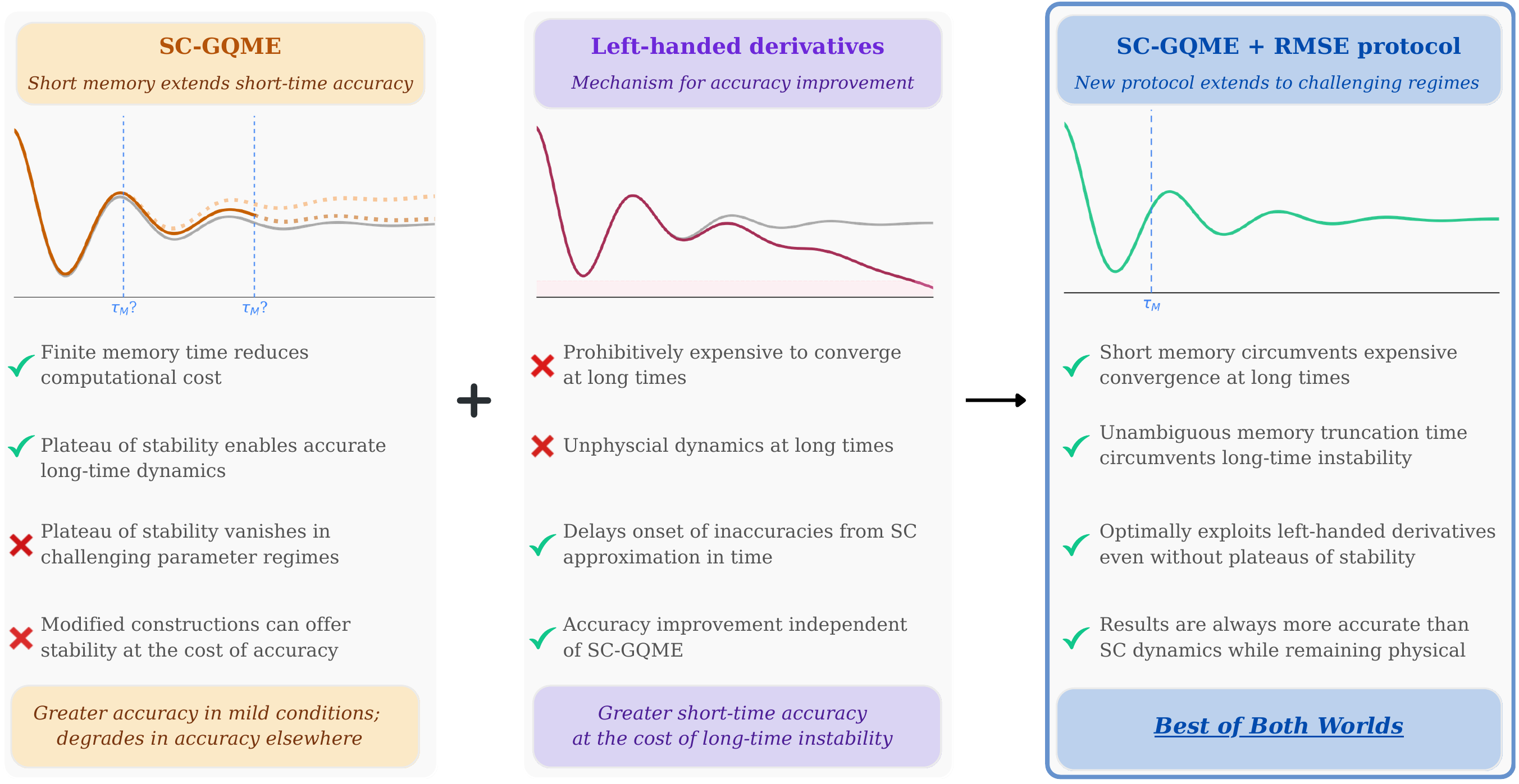}
\caption{ \textbf{Summary of mechanisms that can improve the accuracy and efficiency of SC dynamics.}  \textit{With SC-GQMEs:} For favorable parameter regimes, the SC-GQME's short-lived and decaying kernels offer the ability to reduce computational costs and unambiguous memory truncation time that leads to an improvement in accuracy relative to the bare SC dynamics. For challenging parameter regimes, the plateau of stability disappears, and an unambiguous truncation point can only be identified at the expense of accuracy gains. \textit{Without GQMEs:} Left-handed derivatives can postpone the onset of inaccuracy when the Liouville-rotated initial condition maintains a comparable or superior order of accuracy in the correlation function relative to the SC-evolved operator. This approach requires no projection operator---only the ability to compute the correlation functions generated by the rotated initial condition. In challenging parameter regimes, however, the short-time accuracy gains come at the cost of unstable long-time dynamics. \textit{Our protocol:} Our memory truncation protocol offers an unambiguous approach to identifying kernel lifetimes, even in challenging parameter regimes where previous proposals fail. This allows one to leverage the additional short-time accuracy from left-handed derivatives and reliably truncate the memory kernel to predict accurate and efficient SC dynamics.} 
\label{fig:flow}
\end{figure*}
%%%%%%%%%%%%%%%%%%%%Figure 9 end %%%%%%%%%%%%%%%%%%%%%%%

The SC-GQME \textit{does} provide a tangible benefit over integrating left-handed time derivatives, and our analysis reveals \textit{when} it is most advantageous. Specifically, when the memory kernel decays faster or even on similar timescales with the onset of inaccuracy---or even instability---in the left-handed derivatives, one can benefit from the latter's enhanced short-time accuracy and generate accurate GQME dynamics for arbitrarily long times, even in parameter regimes where previous studies had encountered difficulties. We achieved this by reframing how to choose the memory cutoff, $\tau_M$: instead of seeking a plateau of stability, which can become ambiguous in difficult parameter regimes, we shift to an RMSE-based analysis referenced to accessible SC dynamics to obtain a practical protocol for identifying a $\tau_M$. This protocol addresses \textbf{\hyperref[qs]{Q3}}: it is physically meaningful and yields predictions that are more accurate than the bare SC dynamics in parameter regimes where the traditional plateau does not exist. This approach further demonstrates that higher-order left-handed derivatives enhance short-time accuracy at the expense of long-time stability. This balance of short-time accuracy and long-time stability motivates the comparison of single- and mixed-accuracy kernel constructions that balance the competing effects, effectively extending the window of reliable short-time dynamics. Importantly, the protocol is pathway-agnostic by design and can in principle be applied to any combination of memory kernel constructions that differ in their level of short-time accuracy (see SI Sec.~VI C). These ideas demonstrate that the SC-GQME is more than the sum of its constituent SC approximations: when paired with a principled cutoff strategy, it offers a robust framework for leveraging the short-time accuracy of SC dynamics to make reliable and physically meaningful predictions of long-time dynamics.
\section{Conclusion and Outlook}

In this work, we have addressed a fundamental and long-standing challenge in SC-GQMEs---how and when the SC-GQME improves the accuracy of semiclassical dynamics---through an extensive and systematic study. To resolve this main challenge, we addressed three main questions, \textbf{\hyperref[qs]{Q1}} -- \textbf{\hyperref[qs]{Q3}}, delineated in our introduction. We began by discovering that analytically acting with a (left-handed) time derivative on the initial condition of a SC correlation can improve its accuracy, even in the absence of the GQME. This finding allowed us to answer the first question (\textbf{\hyperref[qs]{Q1}}) by demonstrating that the improvement could be understood through the short-time expansions of the exact and LSC-approximated correlation functions. However, we also showed that while left-handed derivatives can progressively improve short-time accuracy, they can become more inaccurate, and even unphysical, at long times, especially in parameter regimes that are challenging to SC theory. What is more, their integrated dynamics did not exactly match the predictions from the SC-GQME, especially at long times. This latter discrepancy between integrated and SC-GQME dynamics prompted us to address the second question (\textbf{\hyperref[qs]{Q2}}) and show how the self-consistent SC-GQME formulation can (i) appear to preserve physical constraints, such as conservation of population, and (ii) alter the predicted dynamics by mixing time derivatives with various levels of accuracy. Having resolved these apparent discrepancies, we turned to the third question (\textbf{\hyperref[qs]{Q3}}) by leveraging these novel insights with the finite lifetime of the SC-GQME's memory kernel to reap the benefits of short-time accuracy from these derivatives while avoiding the dangers of their long-time instability. We achieved this by introducing a new protocol that employs an RMSE metric evaluated with respect to the LSC dynamics to unambiguously identify a beneficial kernel cutoff. Our protocol avoids the difficulties of previous approaches relying on a plateau of stability, which is known to disappear in difficult parameter regimes. Instead, we leverage the self-consistent structure of the memory kernel to construct single-accuracy and mixed-accuracy kernels to triangulate the latest point at which the resulting memory kernels exhibit short-time accuracy. We proposed this point as a proxy for the kernel cutoff time, and demonstrated that it gives truncated dynamics with greater accuracy than the bare LSC approximation and of the integrated left-handed derivatives even in parameter regimes where previous protocols have failed. For a schematic summary of the key findings in our work, we refer the reader to Fig.~\ref{fig:flow}.

These insights open the door to applying the SC-GQME as a principled and reliable tool, and generalizing it to problems beyond model systems, even when one does not have access to exact benchmarks. By providing a clear framework for balancing the short-time accuracy of left-handed derivatives with the long-time stability enabled by memory truncation, our work enables the SC-GQME to deliver dynamics that are more accurate than the direct SC approximation. This approach is directly applicable to a wide variety of problems, including charge and energy transport and linear and non-linear spectroscopies where both transient and long-time accuracy is imperative. More broadly, our findings suggest that the SC-GQME can be systematically leveraged to capture the quantum dynamics of increasingly complex and atomistic environments, even ones with non-Gaussian statistics. 

\section*{Supplemental Material}
See the supplementary material for additional details supporting this work, including the generality of the conclusions,  computational details, derivations of short-time analysis and population conservation, and additional details on the RMSE protocol. 

\section*{Acknowledgments}

This work was supported by the National Science Foundation Early Career Award in the directorate for Mathematical and Physical Sciences under Award No.~2443961. A.M.C.~acknowledges the support from a David and Lucile Packard Fellowship for Science and Engineering. This work utilized the Alpine high-performance computing resource at the University of Colorado Boulder. Alpine is jointly funded by the University of Colorado Boulder, the University of Colorado Anschutz, Colorado State University, and the National Science Foundation (award 2201538). We thank Tianchu Li for sharing his HEOM code with us. We thank Aaron Kelly for sharing his FBTS code with us. M. R. L. thanks Ethan H. Fink for insightful discussions. We collectively thank Anthony J. Dominic  III, Zachary R. Wiethorn, and Pranay Venkatesh for comments on the manuscript.

\section*{Author Declarations}
\subsection*{Conflict of Interest}
The authors have no conflicts to disclose.
\subsection*{Data Availability}
The data that support the findings of this study are available
from the corresponding author upon reasonable request.

\section*{References}

\bibliography{references}

\end{document}

\setcounter{page}{1}
\renewcommand{\thepage}{S\arabic{page}}

\title{Supplemental material for ``How to improve the accuracy of semiclassical and quasiclassical dynamics with and without generalized quantum master equations"}

\author{Matthew Laskowski}
\affiliation{Department of Chemistry, University of Colorado Boulder, Boulder, CO 80309, USA}

\author{Srijan Bhattacharyya}
\affiliation{Department of Chemistry, University of Colorado Boulder, Boulder, CO 80309, USA}

\author{Andr\'{e}s Montoya-Castillo}
\homepage{Andres.MontoyaCastillo@colorado.edu}
\affiliation{Department of Chemistry, University of Colorado Boulder, Boulder, CO 80309, USA} 

\date{\today}

\maketitle
 
\tableofcontents

\newpage

\setcounter{figure}{0}
\renewcommand{\thefigure}{S\arabic{figure}}

\setcounter{table}{0}
\renewcommand{\thetable}{S\arabic{table}}

\setcounter{enumiv}{0}
\renewcommand{\theenumiv}{S\arabic{enumiv}}

\setcounter{equation}{0}
\renewcommand{\theequation}{A\arabic{equation}}

\section{Generality of Findings }

In this section, we emphasize the generality of several conclusions presented in the main text. In particular, we show that our conclusions and protocols remain valid (A) when focusing on coherence dynamics in the spin-boson model, (B) when using alternative SC and QC theories, and (C) when applying the SC-GQME to other models. We set $\hbar = 1$ for the remainder of this supplementary information (SI).

\subsection{Coherence dynamics}

In the main text, we focus on the ability of left-handed derivatives and the SC-GQME to improve the accuracy of population dynamics. However, observables of interest in optical spectroscopies and quantum information science also require electronic coherences, which correspond to the off-diagonal elements of the reduced electronic density matrix. Here, we demonstrate that evaluating $\big[\mathcal{C}^L(t)\big]_{\rm LSC}$ and applying our RMSE protocol can improve the accuracy of these coherences relative to the bare LSC dynamics. 
%%%%%%%%%%%%%%%%%%%%Figure  s5  begin %%%%%%%%%%%%%%%%%%%%%%%
\begin{figure*}
\centering
\includegraphics[width=\linewidth]{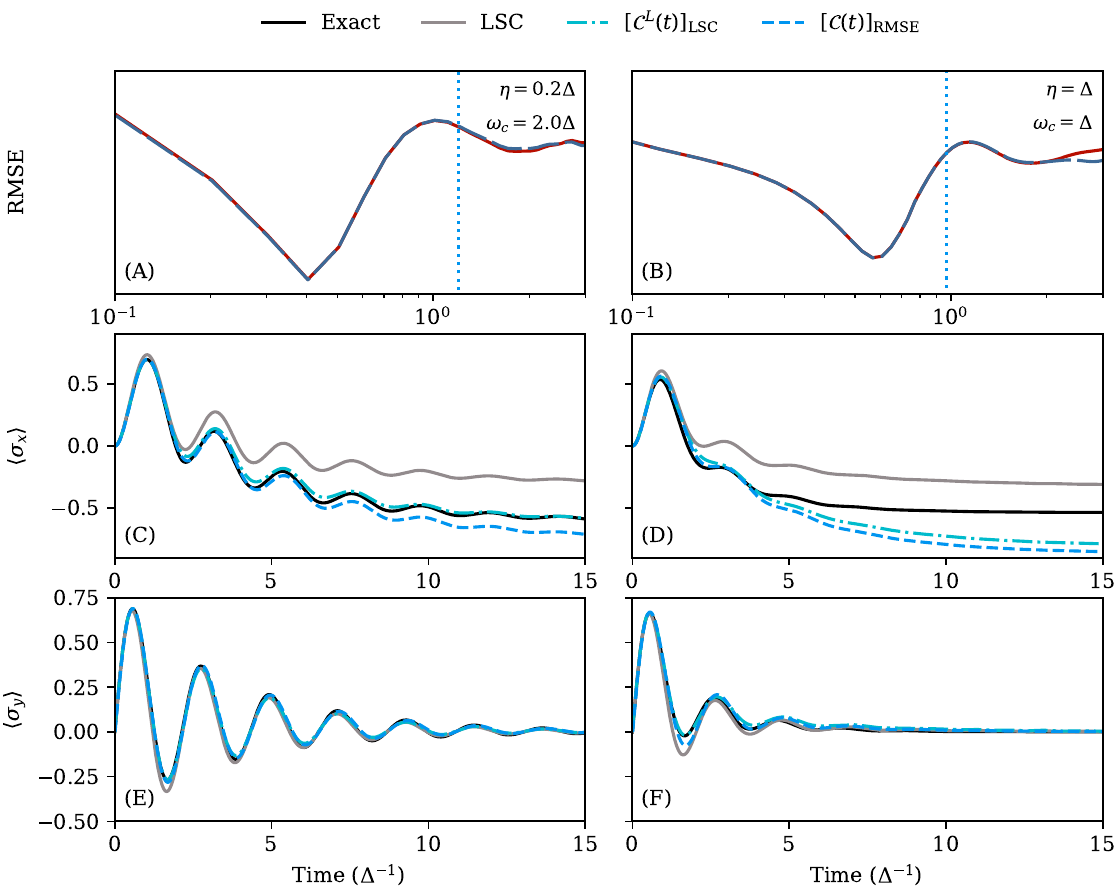}
\caption{\label{setup} \textbf{Left-handed derivatives and our RMSE protocol improve the accuracy of coherences.} Nonequilibrium coherence dynamics in the spin–boson model subject to initial condition $\rho(0) = \rho_B\ket{1}\bra{1}$ with parameters $\epsilon = \Delta$, $\beta = 5.0\Delta^{-1}$, $\omega_c = 2.0\Delta$, $\eta = 0.2\Delta$ ((A),(C), and (E)) and  $\epsilon = \Delta$, $\beta = 5.0\Delta^{-1}$, $\omega_c = \Delta$, $\eta = \Delta$ ((B), (D), and (F)) with an Ohmic spectral density. (A-B) RMSE relative to the LSC dynamics of predictions from \textit{single-accuracy} and \textit{mixed-accuracy} constructions of the memory kernels. That is, we construct $\mathcal{K}^{(1L)}_{\rm LSC}(t)$ (red line) from single-accuracy auxiliary kernels and $\mathcal{K}(t)$ (dashed dark blue line) from the SC-GQME using mixed-accuracy auxiliary kernels. (C-D) Comparison of exact (black), LSC (gray), $\big[\mathcal{C}^L(t)\big]_{\rm LSC}$ (dotted teal)  and $\big[\mathcal{C}(t)\big]_{\rm RMSE}$ (dashed blue) dynamics for $\langle \sigma_x (t) \rangle$. At low $\eta$, both methods improve the transient behavior and long-time limit. When $\eta$ increases, both $\big[\mathcal{C}^L(t)\big]_{\rm LSC}$ and $\big[\mathcal{C}(t)\big]_{\rm RMSE}$ provide greater short-time accuracy, but the incorrect long-time limit. (E-F) The same comparison for the observable $\langle \sigma_y (t) \rangle$: both $\big[\mathcal{C}^L(t)\big]_{\rm LSC}$ and $\big[\mathcal{C}(t)\big]_{\rm RMSE}$ improve the transient behavior for both low and high $\eta$ and correctly recover the long-time limit.\\ %
\fbox{%
        \parbox{0.98\textwidth}{%
            \justifying \noindent \underline{Alt text}: Six-panel figure. Top row shows RMSE curves comparing single- and mixed-accuracy memory kernels relative to LSC dynamics for two parameter regimes. Middle and bottom rows compare coherence dynamics from exact, LSC, LSC-approximated left-handed derivative, and RMSE protocol methods for the off-diagonal components of the reduced density. LSC-approximated left-handed derivative and RMSE protocol are always more accurate for the expectation value of the magnetization along the y-direction, but only match the exact dynamics in low system-bath coupling for the magnetization along the x-direction.} } %
} 
\label{fig:figs5}
\end{figure*}

%%%%%%%%%%%%%%
%%%%%%Figure s5 end %%%%%%%%%%%%%%%%%%

Figure~\ref{fig:figs5} presents the coherences, $\langle \sigma_x (t) \rangle $ and $\langle \sigma_y(t) \rangle$, in two different parameter regimes.  In a system with low system-bath coupling (see Fig.~\ref{fig:figs5} (C) and (E)), the LSC estimates of $\langle \sigma_x \rangle$ and $\langle \sigma_y \rangle$ are accurate at short times, but miss transient behavior at longer times, and even converge to the incorrect long-time limit for $\langle \sigma_x \rangle$. By evaluating $\big[\mathcal{C}^L(t)\big]_{\rm LSC}$ (dotted teal line), one better captures the transient behavior and obtains the correct long-time limit for both $\langle \sigma_x \rangle$ and $\langle \sigma_y \rangle$. The predictions from the RMSE protocol, which we refer to as $\big[\mathcal{C}(t)\big]_{\rm RMSE}$, use $\tau_M = 1.20\Delta^{-1}$ for the case of $\eta = 0.2 \Delta$, consistent with our analysis in Fig.~\ref{fig:figs5} (A). Using this cutoff, $\big[\mathcal{C}(t)\big]_{\rm RMSE}$ (dashed light blue line) also correctly captures the transient behavior of both observables and yields the correct long-time limits. For the case of high system-bath coupling (parameters from Figs.~7 and 8 in the main text), the LSC dynamics remains accurate at short times.It does not faithfully reproduce the transient evolution of $\langle \sigma_x(t) \rangle $ and $\langle \sigma_y(t) \rangle $, and fails to recover the long-time limit of $\langle \sigma_x (t)\rangle $ (see Fig.~\ref{fig:figs5} (D) and (F)). For $\langle \sigma_x(t) \rangle$, both $\big[\mathcal{C}^L(t)\big]_{\rm LSC}$ and $\big[\mathcal{C}(t)\big]_{\rm RMSE}$ --- with $\tau_M = 0.97\Delta^{-1} $ consistent with our analysis of Fig.~\ref{fig:figs5} (B) in the main text --- provide dynamics that match the exact benchmark for a longer period of time. However, although both methods improve short-time accuracy, they ultimately converge to an incorrect long-time limit that also differs from the original LSC result (see Fig.~\ref{fig:figs5} (D)). On the other hand, both methods provide excellent agreement with the exact benchmark when measuring $\langle \sigma_y \rangle$ even in the system with high system-bath coupling (see Fig.~\ref{fig:figs5} (F)). One may attribute this to persistent difficulty in correctly capturing detailed balance when the equilibrium value of the observable deviates from zero. Overall, this analysis suggests that the protocols described in this paper can improve the accuracy of LSC predictions for coherence dynamics.

\subsection{Applicability to other semiclassical and quasiclassical theories}

In the main text, the SC theory of choice is LSC. While some derivations in this work explicitly demonstrate why left-handed derivatives improve short-time accuracy, we illustrate these ideas specifically using LSC dynamics (see SI Sec.~IV). Here, we show that the same ideas for the enhanced accuracy of left-handed derivatives holds in other SC and QC theories. We numerically verify that left-handed derivatives can also enhance the accuracy of other methods and show that the proposed RMSE protocol is applicable to a broader class of SC (LSC, PBME, FBTS) and QC (Ehrenfest) theories.

We begin with the Poisson bracket mapping equation (PBME)\cite{kim2008quantum, kelly2011mixed}, which is sometimes referred to in the literature as LSC(I)\cite{Gao2020, liu2024combining}. PBME is equivalent to LSC in all regards except one: the form of the operators measured at the final time of the correlation function, $t$. To evaluate a correlation function within this formalism, one comes upon the same form as our LSC expression (see Eq.~\eqref{lsc-apporx}), but to compute subsystem observables at $t \geq 0$ one instead evaluates the Wigner-transformed bosonic creation and annihilation operators (see Eq.~\eqref{project}) propagated to time $t$:
\begin{equation} \label{measurement2}
    e^{i\mathcal{L}^Wt}\big(\ket{n}\bra{m}\big)^W = \sigma_{nm}^W(\mathbf{X}(t), \mathbf{P}(t)).
\end{equation}
Here, $\mathcal{L}^W$ denotes the Poisson bracket and $(\mathbf{X}(t), \mathbf{P}(t))$ are the corresponding time-evolved mapping variables. To compute PBME dynamics, we used the same algorithm as described in SI Sec.~II, but evaluated observables using Eq.~\eqref{measurement2}.

%%%%%%%%%%%%%%%%%%%%Figure  s6  begin %%%%%%%%%%%%%%%%%%%%%%%
\begin{figure*}[t!]
\centering
\includegraphics[width=\linewidth]{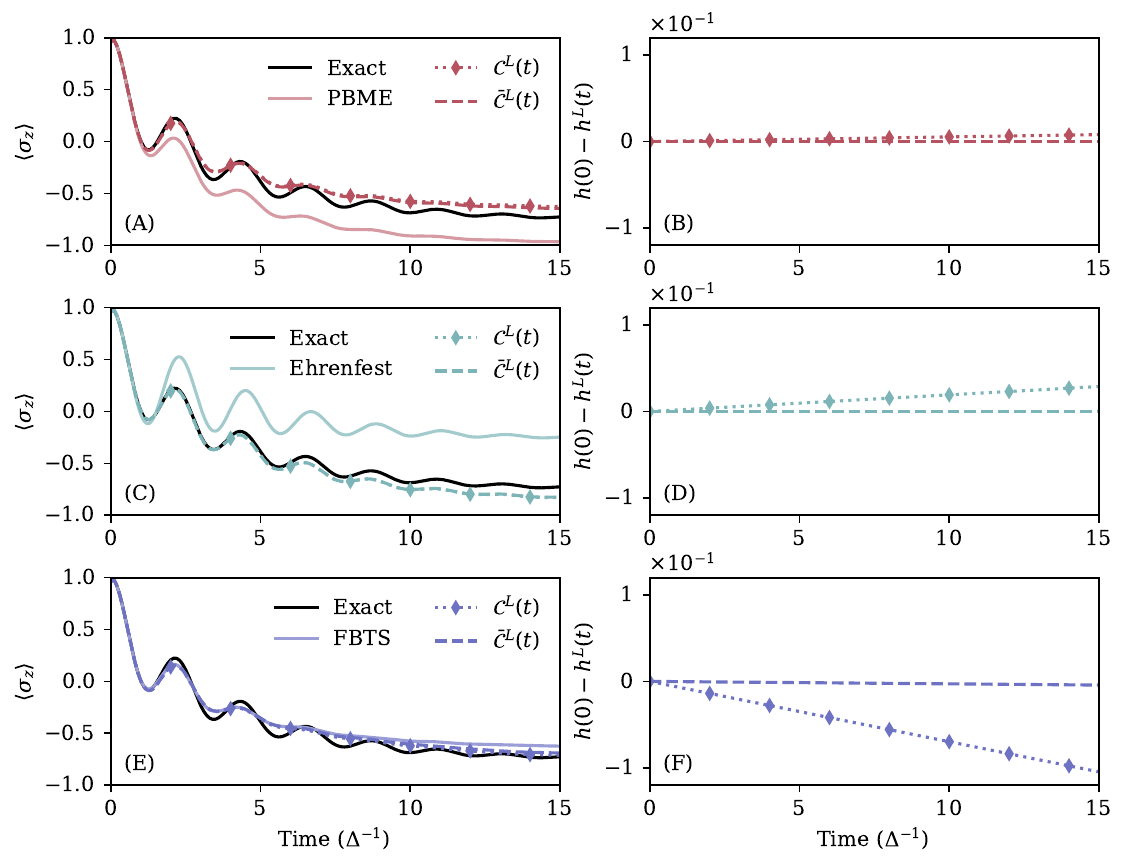}
\caption{\label{setup} \textbf{Left-handed derivatives improve the accuracy of other SC and QC theories.} Nonequilibrium population dynamics in the spin–boson model subject to initial condition $\rho(0) = \rho_B\ket{1}\bra{1}$ with parameters $\epsilon = \Delta$, $\beta = 5.0\Delta^{-1}$, $\omega_c = 2.0\Delta$, $\eta = 0.2\Delta$ with an Ohmic spectral density. Each row corresponds to a different SC or QC method: (A-B) PBME, (C-D) Ehrenfest, (E-F) FBTS. \textit{Left:} $\mathcal{C}^L(t)$ provides transient behavior and long-time limits that are more similar to exact (black) dynamics than the original SC dynamics. $\bar{\mathcal{C}}^L(t)$ provides nearly identical dynamics. \textit{Right:} Violation of total population conservation arising from integrating $\dot{\mathcal{C}}^L(t)$ for each SC and QC method in the absence of the static shift of Eq.~29 in the main manuscript. When the static shift is applied, integrating $\bar{\mathcal{C}}^L(t)$ conserves total population. \\ %
\fbox{%
        \parbox{0.98\textwidth}{%
            \justifying \noindent \underline{Alt text}: Six-panel figure whose left column shows population dynamics for three SC/QC methods---PBME, Ehrenfest, and FBTS---comparing bare dynamics against unshifted and shifted left-handed derivatives. The right column shows the population flux when integrating the unshifted and shifted left-handed derivatives for all three methods. Without the static shift, population is not conserved; applying the shift restores conservation.} } %
            }
\label{fig:figs6}
\end{figure*}

%%%%%%%%%%%%%%
%%%%%%Figure s6 end %%%%%%%%%%%%%%%%%%

The second method we utilize is the quasiclassical Ehrenfest mean-field theory \cite{mclachlan1964variational, ehrenstock1995semiclassical}. In Ehrenfest dynamics, one treats the electronic subsystem quantum mechanically while treating the nuclear bath degrees of freedom classically. Ehrenfest neglects system-bath correlations, such that the nuclei evolve on an average potential energy surface determined by the electronic wavefunction\cite{tully1998mixed, grunwald2009quantum}. To implement Ehrenfest, we follow the algorithm delineated in Refs.~\onlinecite{noneq1} and ~\onlinecite{pfalzgraff2019efficient}. Using this algorithm, we admit a timestep of $0.01 \Delta^{-1}$ when using an Ohmic spectral density.

The last method we utilize is the Forward-Backward Trajectory Solution of the quantum-classical Liouville equation (FBTS)\cite{hsieh2012nonadiabatic}. In FBTS, one propagates the subsystem using forward and backward trajectories of quantum coherent state variables while the bath degrees of freedom evolve classically under a force generated by the average of both trajectories. This approach approximately captures quantum coherence and system-bath correlations by coupling the forward and backward mapping variables through a shared bath trajectory. For implementation details, we refer the reader to Ref.~\onlinecite{hsieh2012nonadiabatic}. To obtain dynamics using this method, we used a timestep of $1.0 \times 10^{-3} \Delta^{-1}$.

%%%%%%%%%%%%%%%%%%%%Figure  s7  begin %%%%%%%%%%%%%%%%%%%%%%%
\begin{figure*}
\centering
\includegraphics[width=\linewidth]{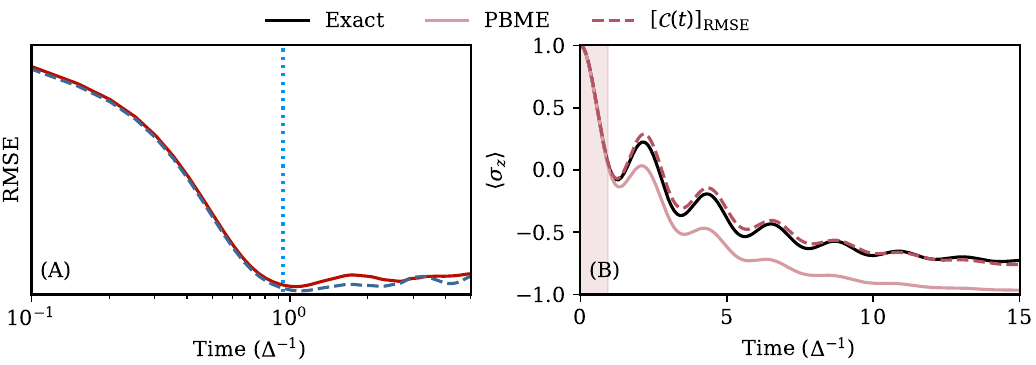}
\caption{\label{setup} \textbf{Our RMSE protocol predicts reliable $\tau_M$ using PBME.} Nonequilibrium population dynamics in the spin–boson model subject to initial condition $\rho(0) = \rho_B\ket{1}\bra{1}$ with parameters $\epsilon = \Delta$, $\beta = 5.0\Delta^{-1}$, $\omega_c = 2.0\Delta$, $\eta = 0.2\Delta$ with an Ohmic spectral density. (A) RMSE relative to the LSC dynamics of predictions from \textit{single-accuracy} and \textit{mixed-accuracy} constructions of the memory kernels. That is, we construct $\mathcal{K}^{(1L)}_{\rm LSC}(t)$ (red line) from single-accuracy auxiliary kernels and $\mathcal{K}(t)$ (dashed dark blue line) from the SC-GQME using mixed-accuracy auxiliary kernels. (B) By selecting when the RMSE curves diverge, $\tau_M = 0.94\Delta^{-1}$, the resulting nonequilibrium population dynamics from the SC-GQME have higher relative accuracy to the original PBME dynamics and remain physical at all times. The shaded red region indicates the time up to $\tau_M$. \\ %
\fbox{%
        \parbox{0.98\textwidth}{%
            \justifying \noindent \underline{Alt text}: Two-panel figure whose left panel shows the RMSE from single-accuracy and mixed-accuracy memory kernels relative to PBME dynamics. The right panel shows that the corresponding truncated PBME population dynamics yield results very similar to the exact dynamics. } } %
            }
\label{fig:figs7}
\end{figure*}
%%%%%%%%%%%%%%
%%%%%%Figure s7 end %%%%%%%%%%%%%%%%%%

Figure~\ref{fig:figs6} compares the dynamics from each of the three SC/QC methods with those obtained from $\mathcal{C}^L(t)$ and $\bar{\mathcal{C}}^L(t)$. On their own, the SC/QC methods predict dynamics that are short-time accurate, but ultimately decay to the incorrect long-time limit (see Fig.~\ref{fig:figs6} (A), (C), and (E)). Additionally, each method improves the short-time accuracy of the integrated dynamics relative to the original SC/QC dynamics when one evaluates $\mathcal{C}^L(t)$ and $\bar{\mathcal{C}}^L(t)$ (see Fig.~\ref{fig:figs6} (A), (C) and (E)). In addition, all of the integrated dynamics produce long-time limits that are closer to the exact result. Figure~\ref{fig:figs6} (B), (D) and (F) also highlights that $\mathcal{C}^L(t)$ does not preserve total population, but do so after implementing the shift introduced in Eq.~(29) of the main text. In addition to this observation, all predicted dynamics from $\mathcal{C}^L(t)$ lose or gain the total population linearly, which matches the prediction from Eq.~(28) in the main text. Once one removes the static bias at $t=0$, the subsequent dynamics of $\mathcal{C}^L(t)$ now preserve population (see Fig.~\ref{fig:figs6} (B), (D) and (F) ). Together, Fig.~\ref{fig:figs6} demonstrates that our analysis of left-handed derivatives is not unique to LSC dynamics.

In addition to extending the analysis of $\mathcal{C}^L(t)$ to other SC methods, the proposed RMSE protocol can also be applied more broadly across different SC approaches. Specifically, in Fig.~\ref{fig:figs7} we apply the RMSE protocol to PBME dynamics. Figure~\ref{fig:figs7} (A) shows that the RMSE curves have a generally different shape when using PBME in the spin-boson model compared to LSC in the main text. Specifically, both RMSE curves start at the same point and decay together. Unlike the LSC RMSE curves, the minima in the PBME RMSE curves do not correspond to reproducing the original PBME dynamics. Even with this differing shape, using our RMSE protocol, we select $\tau_M$ when the two RMSE curves diverge from each other at $\tau_M = 0.94 \Delta^{-1}$. Although the truncated dynamics do not exactly reproduce the benchmark dynamics, Fig.~\ref{fig:figs7} (B) highlights that our protocol can still select a reliable $\tau_M$ that yields improved dynamics when using other SC methods. 

\subsection{Applicability to other models: Frenkel Exciton Model}

In the main text, our analysis centered on the spin-boson model. However, this choice of model also does not limit our conclusions. To highlight this, we demonstrate the applicability of our conclusions in the Frenkel exciton model. 

The Frenkel exciton model is a paradigmatic model for describing excitation energy transport in molecular and condensed-phase systems \cite{ishizaki2009theoretical, berkelbach2012reduced2, Gao2020}. The Hamiltonian consists of local excitation states, each connected to a local Gaussian thermal environment,
\begin{equation}
    \hat{H} = \sum_i \epsilon_i \ket{i}\bra{i} + \sum_{\langle i,j \rangle} \Delta_{ij} \ket{i}\bra{j} + \frac{1}{2}\sum_i \sum_n\left( \hat{p}^2_{i,n} + \omega_{i,n}^2\hat{x}^2_{i,n} \right) + \sum_{i}\hat{V}_B^i \ket{i}\bra{i},
\end{equation}
where $\epsilon_i$ is the local excitation energy, $\Delta_{ij}$ is the electronic coupling (hopping integral) between neighboring excitations, and $\hat{V}_B^i = \sum_{n} c_{i,n}\hat{x}_{i,n}$ is the bath part of the system-bath coupling that leads to local excitation-induced polarization of the environment, with $c_{i,n}$ being the coupling of the $n^{\rm th}$ oscillator in the bath to the excitation on site $i$. 

%%%%%%%%%%%%%%%%%%%%Figure  s8  begin %%%%%%%%%%%%%%%%%%%%%%%
\begin{figure*}[t]
\centering
\includegraphics[width=\linewidth]{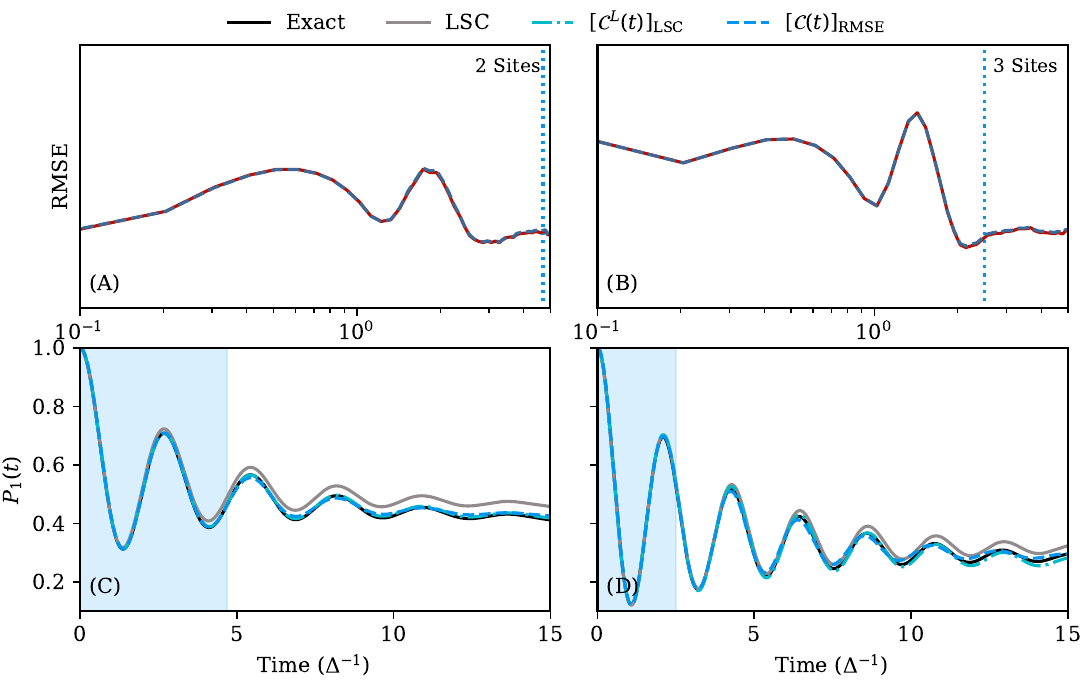}
\caption{\label{setup} \textbf{Our RMSE protocol predicts reliable $\tau_M$ in the Frenkel exciton model.} Nonequilibrium population dynamics subject to initial condition $\rho(0) = \rho_B\ket{1}\bra{1}$ with parameters $\epsilon_1 = 0.5\Delta$, $\epsilon_2 = -0.5\Delta$, for the two-site model and $\epsilon_1 = 0.5\Delta$, $\epsilon_2 = 0$, $\epsilon_3 = -0.5\Delta$ for the three-site model with $\beta = 0.48\Delta^{-1}$, $\omega_c = 0.53\Delta$, $\lambda = 0.2\Delta$ and a Debye spectral density. (A-B) RMSE relative to the LSC dynamics of predictions from \textit{single-accuracy} and \textit{mixed-accuracy} constructions of the memory kernels. (C-D) Choosing $\tau_M$ in the regions where the RMSE curves remain stable, $\tau_M = 4.70\Delta^{-1}$, for the two-site model and $\tau_M = 2.50\Delta^{-1}$ for the three-site model, the resulting SC-GQME dynamics have higher relative accuracy to the original LSC dynamics and remain physical at all times. The shaded blue region indicates the time up to $\tau_M$. \\ %
\fbox{%
        \parbox{0.98\textwidth}{%
            \justifying \noindent  \underline{Alt text}: Four-panel figure that shows the RMSE protocol applied to a 2-site Frenkel exciton model (top row) and a 3-site Frenkel exciton model (bottom row). The left column shows that the RMSE protocol yields reliable memory cutoffs that give improved accuracy for transient dynamics and long-time limits over bare LSC dynamics for both systems (right column).  } } %
            }
\label{fig:figs8}
\end{figure*}

%%%%%%%%%%%%%%
%%%%%%Figure s8 end %%%%%%%%%%%%%%%%%%

To evaluate the left-handed derivative and auxiliary kernels for the SC-GQME in the Frenkel exciton model, we use the following expressions,
\begin{subequations}
\begin{align}
    &\dot{\mathcal{C}}^L(t) = -i\mathcal{H}q^{(0,0)}(t) +  \frac{i}{2} \sum_j \left(\mathcal{Z}^{(j)-}q^{(1s,0)}_{(j,0)} - \mathcal{Z}^{(j)+}q^{(1a,0)}_{(j,0)}\right), \\
    &\mathcal{K}^{(3b)}(t) =\frac{i}{2} \sum_j \left(\mathcal{Z}^{(j)-}q^{(1s,0)}_{(j,0)} - \mathcal{Z}^{(j)+}q^{(1a,0)}_{(j,0)}\right), \\
    &\mathcal{K}^{(1)}(t) = \frac{1}{2}\sum_j \sum_k \left[ \mathcal{Z}^{(j)-}q^{(1s,1)}_{(j,k)}(t) + \mathcal{Z}^{(j)+}q^{(1a,1)}_{(j,k)}(t) \right]\mathcal{Z}^{(k)-},
\end{align}
\end{subequations}
where $q$ denotes a dynamical matrix defined in Eq.~\ref{q-approx}; its subscript indicates the site index at which the bath operator is evaluated. We have also introduced static matrices of the form,
\begin{subequations}
\begin{align}
    &\mathcal{Z}_{jk}^{(n)+} ={\rm Tr}_{\rm sys}\{ \{ \ket{n}\bra{n}, A_j^{\dagger}\} A_k\}, \\
    &\mathcal{Z}_{jk}^{(n)-} = -{\rm Tr}_{\rm sys}\{ [\ket{n}\bra{n}, A_j^{\dagger}] A_k\}, \\ 
    &\mathcal{H}_{jk} = -{\rm Tr}_{\rm sys}\{ [\hat{H}_{\rm S}, A_j^{\dagger}] A_k\},
\end{align}
\end{subequations}
with $\hat{H}_{S} = \sum_i \epsilon_i \ket{i}\bra{i} + \sum_{\langle i,j \rangle} \Delta_{ij} \ket{i}\bra{j}$, where $\langle i,j\rangle$ denotes nearest neighbor coupling.  

Using these expressions, we evaluate the accuracy of $\big[\mathcal{C}^L(t)\big]_{\rm LSC}$ and our RMSE protocol in two- and three-site Frenkel exciton models with all-to-all electronic couplings of $\Delta$ under commonly used parameter regimes\cite{ishizaki2009theoretical, berkelbach2012reduced2, Gao2020} with a Debye spectral density (see Fig.~\ref{fig:figs8}). Figure ~\ref{fig:figs8} (C) and (D) demonstrate that $\big[\mathcal{C}^L(t)\big]_{\rm LSC}$ still provides an improvement in accuracy in describing the total population on site $1$ at $t=0$ relative to the LSC dynamics in both the two- and three-site Frenkel exciton models. Additionally, Fig.~\ref{fig:figs8} (A) and (B) show that the RMSE plots relative to the LSC dynamics, for both memory kernel constructions, exhibit similar behavior at short times. In fact, the RMSE curves remain close to one another over the entire time range without significant divergence. As a result, we chose $\tau_M$ in the region where the RMSE curves remained stable. This resulted in the selection of $\tau_M = 4.70\Delta^{-1}$ for the two-site system and $\tau_M = 2.50\Delta^{-1}$ for the three-site system. By truncating the memory kernels at these $\tau_M$, Fig.~\ref{fig:figs8} (C) and (D) demonstrate the $ \big[\mathcal{C}(t)\big]_{\rm RMSE}$ gives more accurate transient behavior and long-time limits relative to the LSC dynamics. From these results, we conclude that $\big[\mathcal{C}^L(t)\big]_{\rm LSC}$ provides improved accuracy relative to the LSC dynamics, and that our proposed RMSE protocol gives a reliable $\tau_M$ in the Frenkel exciton model.

\setcounter{subsection}{0}
\setcounter{equation}{0}
\renewcommand{\theequation}{B\arabic{equation}}

\section{Computational details}
\label{spin-boson}

In this section, we describe the details of all simulations employed in this work. We set $\hbar = 1$ for the remainder of this SI. 

\subsection{Exact dynamics}

For all exact benchmark calculations, we perform time-evolving matrix product operator (TEMPO)\cite{strathearn2018efficient} calculations using the open-source Open Quantum Systems in Python (OQuPy) \cite{Fux2024} and PATHSUM \cite{kundu2023pathsum} packages. All simulations were run with a timestep of $0.01 \Delta^{-1}$, a memory length of $250$ timesteps, and precision of $10^{-7}$. 

\subsection{LSC dynamics}

We performed all our semiclassical dynamics with an in-house implementation of the standard LSC approach \cite{ wang1998semiclassical, wang1999semiclassical, miller2001semiclassical, shi2003relationship}. The basic idea of LSC is to approximate the quantum dynamics of a system via its classical dynamics subject to quantum mechanical sampling of the initial conditions and measurement of the operators at initial, $t=0$, and final times, $t$. Because classical dynamics is defined with respect to conjugate variables, generally position and canonical momenta, one needs to map discrete degrees of freedom (DOFs), like outer product $\{ \ket{j}\bra{k}\} $, onto continuous DOF and then evaluate the dynamics using the classical limit of the semiclassical propagator. Here, we provide the full details of our implementation.

\noindent \textbf{Continuous mapping of discrete states:} The mapping connecting discrete and continuous DOFs is generally based on representations of spin operators by bosonic operators, although alternative approaches based on action-angle variables are also possible. We employ the mapping initially introduced by Meyer and Miller\cite{miller1979classical} and later rigorously derived by Stock and Thoss \cite{stock1997semiclassical}---the (MMST) map---that translates all outer products of an $N$-level system to $N$ harmonic oscillators,
\begin{equation}
    \ket{j}\bra{k} \rightarrow \hat{a}_j^{\dagger}\hat{a}_k,
\end{equation}
where $\hat{a}_j^{\dagger}$ ($\hat{a}_j^{\dagger}$) is the creation (annihilation) operator for the $j$th boson. This mapping preserves the commutation relationships of spin operators when restricted to the first excitation subspace, but the bosons can, in principle, access an infinite-dimensional Hilbert space when employing approximate dynamics. This map is exact when it is restricted to the oscillator subspace with a single excitation, $\ket{1_j} \in \left\{ \ket{1_1} = \ket{1, 0, ..., 0} , ..., \ket{1_N} = \ket{0, 0, ..., 1}\right]$, and holds appeal because these bosonic operators can be written in terms of continuous Cartesian coordinates, 
\begin{subequations}
\begin{align} 
\label{anhil}
    \hat{a}_j = (\hat{X}_j + i\hat{P}_j)/\sqrt{2},\\
    \label{create}
    \hat{a}_j^{\dagger} = (\hat{X}_j - i\hat{P}_j)/\sqrt{2},
\end{align}
\end{subequations}
and thus have a direct classical analog. Here, $\mathbf{X}$ and $\mathbf{P}$ are the mapped coordinate and momentum operators for the mapped subsystem states. 

\noindent \textbf{Wigner phase space:} Because the outer products of many-body harmonic oscillator states in the ground or first excited state have a well-known spatial representation, this mapping allows one to move to continuous phase space through, for example, a Wigner transformation \cite{imre1967wigner}. In Wigner phase space, a correlation function takes the form
\begin{align} 
    \mathcal{C}(t) = \mathrm{Tr}&\left\{ \hat{\rho}_B \ket{j}\bra{k} e^{i\mathcal{L}t} \ket{n}\bra{m } \right\} \\
    \label{phase-space}
    &= \left(2\pi\right)^{-f}\int {\rm d}\mathbf{\Gamma} \; \Big( \hat{\rho}_B \ket{j}\bra{k}\Big)^W \Bigg(e^{i\mathcal{L}t} \ket{n}\bra{m}\Bigg)^W. 
\end{align}
Here, $f$ describes the total number of DOFs (electronic and nuclear), $\mathcal{L}(\cdot) \equiv [ \hat{H}, \cdot ]$ is the quantum mechanical Liouville operator, and the $W$ superscript denotes a Wigner transformation of the operators\cite{hillery1984distribution},
\begin{subequations}
\begin{align} \label{wigner}
    A^W\left(\mathbf{Q, \Pi}\right) &= \int {\rm d}\mathbf{Z} \; e^{-i\mathbf{Z\Pi}} \big\langle \mathbf{Q} + \frac{\mathbf{Z}}{2}\big|\hat{A}\big|\mathbf{Q} - \frac{\mathbf{Z}}{2}\big\rangle,
\end{align}
\end{subequations}
where $\mathbf{Q} \equiv (\mathbf{X}, \mathbf{x})$ and $\mathbf{\Pi}\equiv (\mathbf{P}, \mathbf{p})$ contain the phase space variables for the system and bath. The Wigner transform of a product of operators takes the form,
\begin{equation}
\label{product}
    \left(AB\right)^W = A^W\left(\mathbf{Q,\Pi}\right) e^{\Lambda/2i} B^W\left(\mathbf{Q,\Pi}\right),
\end{equation}
where $e^{\Lambda/2i}$ is referred to as the Moyal product and 
\begin{equation} \label{Lambda-operator}
    \Lambda = \nabla_{\Pi} \nabla_Q - \nabla_Q \nabla_{\Pi}
\end{equation}
is the generator of the Poisson bracket when acting to the left on Wigner-transformed Hamiltonian, $A^W \Lambda H^W = \{H^W, A^W\}_{PB}$, where $\{A^W, B^W\}_{PB} =  \sum_{i=1}^N \left( \frac{\partial A^W}{\partial Q_i}\frac{\partial B^W}{\partial \Pi_i} - \frac{\partial A^W}{\partial \Pi_i}\frac{\partial B^W}{\partial  Q_i}\right)$ is the Poisson bracket. 

\noindent \textbf{LSC approximation:} One can recover the LSC approximation by keeping the lowest order contribution of $\Lambda$ in the expansion of the propagator  \cite{ wang1998semiclassical, miller2001semiclassical, shi2003relationship, kapral2015quantum},
\begin{equation}
\label{lsc-apporx}
    \big[\mathcal{C}(t)\big]_{\rm LSC} = \left(2\pi\right)^{-f}\int {\rm d}\mathbf{\Gamma} \; \Big( \rho_B \ket{j}\bra{k}\Big)^W e^{i\mathcal{L}^Wt} \Big(\ket{n}\bra{m}\Big)^W,
\end{equation}
where  $\mathcal{L}^W(\cdot) \equiv i\{ H^W, \cdot \}_{PB}$. Equation~\eqref{lsc-apporx} highlights that one can approximate $\mathcal{C}(t)$ by generating classical dynamics using a phase-space representation of the Hamiltonian subject to properly quantized (Wigner transformed) initial and final conditions.

\noindent \textbf{Wigner transformation of Hamiltonian:} To bring the spin-boson Hamiltonian to Wigner phase space, one applies a Wigner transformation to the bosonic creation and annihilation operators defined in Eqs.~\eqref{anhil} and ~\eqref{create},
\begin{subequations}
\begin{align} 
\label{anhil-w}
    (a_j)^W = (X_j + iP_j)/\sqrt{2},\\
    \label{create-w}
    (a_j^{\dagger})^W = (X_j - iP_j)/\sqrt{2}.
\end{align}
\end{subequations}
These definitions of the Wigner-transformed bosonic creation and annihilation operators allow one to express the spin–boson Hamiltonian (see Eq.~(1) in the main text of the manuscript) as,
\begin{equation} \label{wigner-ham}
    H^W =  (\varepsilon + V_B^W) \sigma_z^W + \Delta \sigma_x^W + \frac{1}{2}\sum_n \big( p_n^2 + \omega_n^2 x_n^2\big), 
\end{equation}
where $\{ \sigma_i^W \}$, $i \in \{ x,y,z\}$, are Wigner-transformed Pauli matrices constructed from Wigner-transformed bosonic creation and annihilation operators,
\begin{equation} \label{project}
    \begin{split}
        \sigma_z^W &= \frac{1}{2}\big( X_1^2 + P_1^2 - X_2^2 - P_2^2 \big) \\
        \sigma_x^W &= X_1X_2 + P_1P_2, \\
        \sigma_y^W &= X_1P_2 - X_2P_1.
    \end{split}
\end{equation}

\noindent \textbf{Wigner transformation of initial and final conditions:} To quantize the initial and final conditions in Eq.~\eqref{lsc-apporx}, we Wigner-transform the outer-product states mapped with MMST. Here the spin-state, $\ket{j}\bra{k} \rightarrow \ket{1_j, 0_{\neq j}}\bra{1_j, 0_{\neq k}}$, whose corresponding Wigner transform, defined in Eq.~\eqref{wigner} is,
\begin{equation} \label{w-outer-product}
\begin{split}
    (\ket{j}\bra{k})^W(\mathbf{X}, \mathbf{P}) &= \int {\rm d}\mathbf{Z} \; e^{-i\mathbf{ZP}} \big\langle \mathbf{X} + \frac{\mathbf{Z}}{2}\ket{1_j, 0_{\neq j}}\bra{1_k, 0_{\neq k}}\mathbf{X} - \frac{\mathbf{Z}}{2}\big\rangle, \\
    &=\phi\left(\mathbf{X,P}\right)\big[ (X_j - iP_j)(X_k + iP_k) - \frac{1}{2}\delta_{jk} \big]\\
    &\equiv 2\phi(\mathbf{X}, \mathbf{P})\sigma_{jk}^W,
    \end{split}
\end{equation}
where $\phi\left(\mathbf{X,P}\right) = 2^{N+1} {\rm exp}\left[ - \sum_i (X_i^2 + P_i^2) \right]$. This definition of the Wigner-transformed MMST outer-product states in Eq.~\eqref{w-outer-product} yields Pauli matrices, $\{\tilde{\sigma}_i^{ W}\}$, $i \in \{ x,y,z\}$, that differ from those obtained by Wigner-transforming the corresponding bosonic creation and annihilation operators given in Eq.~\eqref{project},
\begin{equation}
    \begin{split}
        \tilde{\sigma}_z^W &= 2 \phi\sigma_z^W  \\
        \tilde{\sigma}_x^W &=  2 \phi\sigma_x^W, \\
        \tilde{\sigma}_y^W &= 2 \phi\sigma_y^W.
    \end{split}
\end{equation}

\noindent \textbf{Initial sampling and measurement of subsystem operators:} 
We initialize subsystem variables by stochastically sampling from $\phi$, which corresponds to the Wigner distribution of the collective $N$-oscillator ground state, consisting of a Gaussian distribution with a mean and standard deviation,
\begin{subequations}
    \begin{align}
        &\Bar{X} = 0 , &\Delta X = \frac{1}{\sqrt{2}}, \\
        &\Bar{P} = 0 , &\Delta P = \frac{1}{\sqrt{2}}. 
    \end{align}
\end{subequations}
To bias the sampling and account for the bra ($j$) and ket ($k$) oscillators in their first excited state, we record the value of $\big[ (X_j - iP_j)(X_k + iP_k) - \frac{1}{2}\delta_{jk} \big]$ at $t=0$ and multiply it by the appropriate value of the operator measured at time $t$.

To measure subsystem operators at times $t\geq 0$, we evaluate the Wigner transformed value of the MMST mapped outer-product state $(\ket{n}\bra{m})^W$ in Eq.~\eqref{w-outer-product} at time $t$:
\begin{equation} \label{measurement}
    e^{i\mathcal{L}^Wt}\big(\ket{n}\bra{m}\big)^W = 2\phi(\mathbf{X}(t),  \mathbf{P}(t))\sigma_{nm}^W(\mathbf{X}(t), \mathbf{P}(t)).
\end{equation}
For systems, like the spin-boson model and many others, where the total electronic density is conserved, $\phi(t) = \phi$ is constant.

\noindent \textbf{Bath discretization and initial sampling:} For each trajectory, we must also initialize the phase-space variables for the bath. We discretize the bath into $n_{osc}=300$, employing the following relations for the frequency, $\omega_n$, and coupling constant, $c_n$ for the $n^{\rm th}$ oscillator obeying an Ohmic spectral density \cite{craig2005chemical},
\begin{subequations}
\begin{align}
    \omega_n &= \omega_c \ln \left(\frac{n-\frac{1}{2}}{n_{osc}}\right), \\
    c_n &= \omega_n \sqrt{\frac{\eta \omega_c}{n_{osc}}},
\end{align}  
\end{subequations}
where $\omega_c$ is the cutoff frequency and $\lambda$ is the reorganization energy. One can then sample the positions and momenta of each bath oscillator from its Wigner-transformed canonical density. For harmonic oscillators, the Wigner-transformed canonical density takes the form,
\begin{equation} \label{wb}
    \rho_B^W = \prod_n^{n_{osc}} \frac{\tanh(\beta \omega_n/2) }{\pi}{\rm exp}\Bigg[ \frac{-2\tanh(\beta \omega_n /2)}{\omega_n}\Big( \frac{1}{2}p_n^2 + \frac{1}{2}\omega_n^2 x_n^2\Big)\Bigg].
\end{equation}
Equation~\eqref{wb} shows that the phase-space variables of the bath may be sampled from a Gaussian distribution with mean and standard deviation,
\begin{subequations}
    \begin{align}
        &\Bar{x}_n = 0 , \; \; \; \; \Delta x_n = \left[2\omega_n \tanh\left(\frac{\beta \omega_n}{2}\right)\right]^{-1/2}, \\
        &\Bar{p}_n = 0 , \; \; \; \; \Delta p_n = \left[ \frac{\omega_n}{2\tanh\left(\frac{\beta \omega_n}{2}\right)} \right]^{1/2} . 
    \end{align}
\end{subequations}

\noindent \textbf{Algorithm for classical evolution:} We propagate the classical trajectory using a split-operator scheme\cite{strang1968construction} in which the system and bath contributions to the Liouvillian are evolved sequentially across a timestep. Consistent with the split-operator scheme, the subsystem variables are held constant during a bath evolution step and vice versa. The evolution of the entire system, $(\mathbf{Q}(t), \mathbf{\Pi}(t))  \rightarrow (\mathbf{Q}(t + \Delta t), \mathbf{\Pi}(t+ \Delta t))$, over one timestep, $\Delta t$, thus involves three steps: 
\begin{enumerate}
    \item Evolve bath variables over $\Delta t/2$, $(\mathbf{x}(t), \mathbf{p}(t))  \rightarrow (\mathbf{x}(t + \Delta t/2), \mathbf{p}(t+ \Delta t/2))$, subject to the subsystem variables being frozen at $t$.

    \item Evolve system variables over $\Delta t$, $(\mathbf{X}(t), \mathbf{P}(t))  \rightarrow (\mathbf{X}(t + \Delta t), \mathbf{P}(t+ \Delta t/2))$, subject to the bath variables being frozen at $t + \Delta t / 2$.

    \item Evolve bath variables over $\Delta t/2$, $(\mathbf{x}(t + \Delta t/2), \mathbf{p}(t + \Delta t/2))  \rightarrow (\mathbf{x}(t + \Delta t), \mathbf{p}(t+ \Delta t))$, subject to the subsystem variables being frozen at $t + \Delta t$.
\end{enumerate}

\noindent \textit{Half timestep evolution of the bath:} We illustrate this half timestep evolution for the first half timestep. The evolution over the second half timestep is done according to item 3 above. During the evolution of the bath during the first half timestep, $\hat{\sigma}_z(t) \mapsto \frac{1}{2}\big( X_1^2 + P_1^2 - X_2^2 - P_2^2\big)$ is fixed. This permits an analytical update of the position and momentum of each bath oscillator over the half-timestep,
\begin{equation} \label{ceom}
\begin{split}
    x_n\big(t + \Delta t/2 \big) &=  \Big[x_n(t) + \gamma_n(t)\Big]\cos(\omega_n \Delta t/2) + \frac{p_n(t)}{\omega_n}\sin(\omega_n \Delta t/2)-\gamma_n(t),  \\
    p_n\big(t + \Delta t/2 \big) &= p_n(t)\cos(\omega_n \Delta t/2) - \Big[x_n(t) + \omega_n\gamma_n(t)\Big] \sin(\omega_n \Delta t/2),
\end{split}
\end{equation}
where $\gamma_n(t) = c_n/\omega_n^2 \sigma_z^W(t)$.

\noindent \textit{Full timestep evolution of the subsystem:} The instantaneous subsystem Hamiltonian defined from the Wigner-transformed bosonic creation, $(a^{\dagger})^W$, and annihilation, $(a)^W$, operators is,
\begin{equation}
    H_S^W(t) = \frac{1}{2}\big[\varepsilon + V_B^W(t) \big] \big( X_1^2(t) + P_1^2(t) - X_2^2(t) - P_2^2(t)\big) + \Delta \big( X_1(t)X_2(t) - P_1(t)P_2(t)\big),
\end{equation}
where the classical bath provides a fluctuating contribution to
the bias, $V_B^W(t) = \sum_n c_n x_n(t)$.  The quadratic nature of the Hamiltonian enables us to reexpress it as,
\begin{equation}
    H_S^W(t) = \frac{1}{2}\mathbf{z}^T(t)\mathbf{M}(t)\mathbf{z}(t),
\end{equation}
where,
\begin{equation}
    \mathbf{z}(t) = \left( \begin{matrix}
        X_1(t)\\ X_2(t) \\ P_1(t) \\ P_2(t)
    \end{matrix}\right),
\end{equation}
and,
\begin{subequations}
    \begin{align}
         \mathbf{M}(t) &= \left(\begin{matrix}
        H_S^W(t) & 0 \\
        0 & H_S^W(t) 
        \end{matrix}\right), \qquad H_S = \left(\begin{matrix}
            \varepsilon + V_B^W(t) & \Delta \\
            \Delta  & - \varepsilon - V_B^W(t)
        \end{matrix}\right).
    \end{align}
\end{subequations}
By diagonalizing $H_S(t) = U(t)\Omega(t)U^{-1}(t)$, we can represent the time-dependent Hamiltonian in the basis of eigenvectors,
\begin{equation} \label{diag}
    H_S^W(t) = \tilde{\mathbf{z}}^T(t)\mathbf{\Omega}(t)\tilde{\mathbf{z}}(t),
\end{equation}
with $\tilde{\mathbf{z}}(t) = \mathbf{U}^{-1}(t)\mathbf{z}(t) $, $\tilde{\mathbf{z}}^T(t) = \mathbf{z}^T(t)\mathbf{U}(t) $, and 
\begin{equation}
     \mathbf{U}(t) = \left(\begin{matrix}
        U(t) & 0 \\
        0 & U(t) 
        \end{matrix}\right), \qquad \mathbf{\Omega}(t) = \left(\begin{matrix}
        \Omega(t) & 0 \\
        0 & \Omega(t) 
        \end{matrix}\right).
\end{equation}
In this eigenbasis, we can write $H_S^W(t)$ using rotated positions and momenta, $\tilde{X}$ and $\tilde{P}$,
\begin{equation}
    H_S^W(t) = \sum_{k=1}^2\frac{1}{2}\Omega_k(t)\big( \tilde{X}_k^2(t) + \tilde{P}_k^2(t) \big),
\end{equation}
where $\Omega_k(t)$ is the $k^{\rm th}$ eigenvalue of $H_S^W(t)$. After this canonical transformation, the classical equations of motion can still be derived from Hamilton's equations of motion,
\begin{equation} \label{subs}
    \begin{split}
        \tilde{X}_k(t + \Delta t) &= \tilde{X}_k(t)\cos(\Omega_k(t) t) + \tilde{P}_k(t)\sin(\Omega_k(t) t), \\
        \tilde{P}_k(t + \Delta t) &= \tilde{P}_k(t)\cos(\Omega_k(t) t) - \tilde{X}_k(t)\sin(\Omega_k(t) t).
    \end{split}
\end{equation}
To complete the evolution of the subsystem over the timestep, we transform the rotated variables back to the original phase space using the inverse transformation $\mathbf{z}(t+\Delta t) = \mathbf{U}(t)\tilde{\mathbf{z}}(t+\Delta t)$. 

After the subsystem is evolved over the full timestep, we evolve the bath according to item 3 of the algorithm using Eq.~\eqref{ceom}. We employ a timestep of $\Delta t = 0.01 \Delta^{-1}$ for all simulations in this paper unless otherwise specified. Convergence of the LSC approximated correlation functions requires between $10^5$ to $10^6$ trajectories. Convergence of time-derivatives takes more trajectories with a single-time derivative taking between $10^6$ - $10^7$ trajectories and a second-time derivative taking $ > 1.0\times 10^7$ trajectories.

\setcounter{subsection}{0}
\setcounter{equation}{0}
\renewcommand{\theequation}{C\arabic{equation}}
\section{Expressions for auxiliary kernels and time derivatives}

Here, we provide explicit expressions for the auxiliary kernels needed to construct the SC-GQME. Our expressions in this section closely follow the notation set in Ref.~\onlinecite{noneq1, montoya2017approximate,  pfalzgraff2019efficient}. Specifically, we articulate the kernel construction problem in terms of static and dynamic matrices. 

The dimensions of the static and dynamic matrices, as well as the correlation function $\mathcal{C}(t)$, memory kernel, and auxiliary kernels, depends on the choice of projection operator. We use the Argyres-Kelley (AK) projector \cite{argyres1964theory}, which spans subsystem populations and coherence, $A_j \in \{\ket{j}\bra{k} \}$, with $\rho_B = e^{-\beta \hat{H}_B}/\mathrm{Tr}\{e^{-\beta \hat{H}_B}\}$. However, we again note that the same conclusions arise when using the population projector \cite{sparpaglione1988dielectric}, which only spans the populations, $A_j \in \{\ket{j}\bra{j} \}$, but we remark that this choice of projector and the parameter regime of the system under study determine the memory kernel lifetimes. We order this basis of outer product states thus: $A_1 = \ket{1}\bra{1}, A_2 = \ket{1}\bra{2}, A_3 = \ket{2}\bra{1}, A_4 = \ket{2}\bra{2}$, where $A_2 = A_3^*$. 

\subsection{Static matrices}
\label{static-mat}
We define the static matrices to describe subsystem rotations in the spin-boson model,
\begin{equation}
   \mathcal{Z}_{jk}^+ ={\rm Tr}_{\rm sys}\{ \{\hat{\sigma}_z, A_j^{\dagger}\} A_k\}, \quad \mathcal{Z}_{jk}^- = -{\rm Tr}_{\rm sys}\{ [\hat{\sigma}_z, A_j^{\dagger}] A_k\}, \quad \mathcal{X}_{jk}^- = -{\rm Tr}_{\rm sys}\{ [\hat{\sigma}_x, A_j^{\dagger}] A_k\}.
\end{equation}
The two leftmost matrices were first introduced in Ref.~\onlinecite{noneq1}, and we introduce the third, which is necessary for the analysis in this paper. In the basis of outer products defined by the AK projector, these matrices take the following forms,
\begin{align}
\mathcal{Z}^+ =
\left(\begin{matrix}
    2 & 0 & 0 & 0 \\
    0 & 0 & 0 & 0 \\
    0 & 0 & 0 & 0 \\
    0 & 0 & 0 & -2
\end{matrix}\right),
\qquad
\mathcal{Z}^- =
\left(\begin{matrix}
    0 & 0 & 0 & 0 \\
    0 & -2 & 0 & 0 \\
    0 & 0 & 2 & 0 \\
    0 & 0 & 0 & 0
\end{matrix}\right),
\qquad
    \mathcal{X}^- =
\left(\begin{matrix}
    0 & 1 & -1 & 0 \\
    1 & 0 & 0 & -1 \\
    -1 & 0 & 0 & 1 \\
    0 & -1 & 1 & 0
\end{matrix}\right).
\end{align}

\subsection{Dynamical matrices}
\label{dyn-mat}
To construct correlation functions and all auxiliary kernels necessary for our analysis, we use the following dynamical matrices,
\begin{equation} \label{q-exact}
    \begin{split}
        q^{(0,0)}(t) &= {\rm Tr} \{ \hat{\rho}_B A_j^{\dagger} A_k(t)\} \\
        q^{(0, 1)}(t) &= {\rm Tr} \{ \hat{\rho}_B A_j^{\dagger} A_k(t) \hat{V}_B(t)\}, \\
        q^{(1s, 0)}(t) &= {\rm Tr} \{ \{\hat{V}_B, \hat{\rho}_B \} A_j^{\dagger} A_k(t)\}, \\
        q^{(1a, 0)}(t) &= {\rm Tr} \{ [\hat{V}_B, \hat{\rho}_B ] A_j^{\dagger} A_k(t)\}, \\
        q^{(1s, 1)}(t) &= {\rm Tr} \{ \{\hat{V}_B, \hat{\rho}_B \} A_j^{\dagger} A_k(t)\hat{V}_B(t)\}, \\
        q^{(1a, 1)}(t) &= {\rm Tr} \{ [\hat{V}_B, \hat{\rho}_B ] A_j^{\dagger} A_k(t)\hat{V}_B(t)\},\\
        q^{(2s, 0)}(t) &= {\rm Tr} \{ \{\hat{V}_B, \{\hat{V}_B, \hat{\rho}_B \} \} A_j^{\dagger} A_k(t)\}, \\
        q^{(2a, 0)}(t) &= {\rm Tr} \{ [\hat{V}_B, [\hat{V}_B, \hat{\rho}_B ] ]A_j^{\dagger} A_k(t)\}, \\ 
        q^{(1b1s, 0)}(t) &= {\rm Tr} \{ [\hat{H}_B, \{\hat{V}_B, \hat{\rho}_B \} ] A_j^{\dagger} A_k(t)\}, \\
        q^{(1b1a, 0)}(t) &= {\rm Tr} \{ [\hat{H}_B, [\hat{V}_B, \hat{\rho}_B ] ]A_j^{\dagger} A_k(t)\}.
    \end{split}
\end{equation}
The first six dynamical matrices have been introduced in Ref.~\onlinecite{noneq1}, and we introduce the last four dynamical matrices. The expressions for the dynamical matrices in Eq.~\eqref{q-exact} are exact. For our analysis, we need their LSC approximations, which take the forms,
\begin{equation} \label{q-approx}
    \begin{split}
            &q^{(0,0)}_{jk}(t) \approx \int {\rm d}\boldsymbol{\Gamma}\; \rho_B^W  A_j^{\dagger W} A_k^W(t), \\
            &q^{(0,1)}_{jk}(t) \approx \int {\rm d}\boldsymbol{\Gamma}\; \rho_B^W  A_j^{\dagger W} A_k^W(t) V_B^W(t) \\
            &q^{(1s,0)}_{jk}(t) \approx 2 \int {\rm d}\mathbf{\Gamma} \;\rho_B^W V_B^W  A_j^{\dagger W} A_k^W(t), \\
            &q^{(1a,0)}_{jk}(t) \approx 2i\int {\rm d}\mathbf{\Gamma} \;\rho_B^W \xi^W  A_j^{\dagger W} A_k^W(t), \\
            &q^{(1s, 1)}_{jk}(t) \approx 2\int {\rm d}\mathbf{\Gamma} \; \rho_B^W V_B^W  A_j^{\dagger W} A_k^W(t) V_B^W(t), \\
            &q^{(1a, 1)}_{jk}(t) \approx 2i \int {\rm d}\mathbf{\Gamma} \; \rho_B^W \xi^W  A_j^{\dagger W} A_k^W(t) V_B^W(t), \\
            &q^{(1b1s,0)}_{jk}(t) \approx 2i\int {\rm d}\mathbf{\Gamma} \; \rho_B^W \theta^W  A_j^{\dagger W} A_k^W(t), \\
            &q^{(1b1a,0)}_{jk}(t) \approx 2\int {\rm d}\mathbf{\Gamma} \; \rho_B^W \varphi^W  A_j^{\dagger W} A_k^W(t), \\
            &q^{(2s,0)}_{jk}(t) \approx 4\int {\rm d}\mathbf{\Gamma} \; \rho_B^W (V_B^W)^{ 2}  A_j^{\dagger W} A_k^W(t), \\
            &q^{(2a,0)}_{jk}(t) \approx -2\int {\rm d}\mathbf{\Gamma} \; \rho_B^W \zeta^W  A_j^{\dagger W} A_k^W(t).
    \end{split}
\end{equation}
Here, we have invoked the following Wigner-transformed bath operators,
\begin{equation}
    \begin{split}
    &\Big(\{\hat{V}_B, \hat{\rho}_B\}\Big)^W = 2V_B^W\rho_B^W \\
    &\Big(\{\hat{V}_B,\{\hat{V}_B, \hat{\rho}_B\}\}\Big)^W = 4(V_B^W)^2\rho_B^W, \\
    &V_B^W = \sum_n c_n x_n,
\end{split}
\end{equation}

\begin{equation}
\begin{split}
    &\Big([\hat{V}_B, \hat{\rho}_B] \Big)^W = 2i \xi^W\rho_B^W, \\
    &\xi^W = -\sum_n c_n p_n\frac{ \tanh\left(\beta \omega_n/2\right)}{\omega_n}, \\
    &\Big([\hat{V}_B,[\hat{V}_B, \hat{\rho}_B]]\Big)^W = -2\zeta^W\rho_B^W \\
    &\zeta^W =  2 (\xi^W)^2 - \sum_n c_n^2 \frac{ \tanh\left(\beta \omega_n/2\right)}{\omega_n},
\end{split}
\end{equation}

\begin{equation}
    \begin{split}
        &\Big([\hat{H}_B, \{\hat{V}_B, \hat{\rho}_B\}]\Big)^W = 2\theta^W\rho_B^W, \\
        &\theta^W = -i\sum_n c_n p_n, 
    \end{split}
\end{equation}   
\begin{equation}
    \begin{split}
        &\Big([\hat{H}_B, [\hat{V}_B, \hat{\rho}_B]]\Big)^W = 2\varphi^W\rho_B^W,\\
        &\varphi^W = \sum_n c_n \omega_n \tanh(\beta\omega_n/2) x_n, 
    \end{split}
\end{equation}

\subsection{Auxiliary Kernels}
Using the static and dynamical matrices defined in Secs.~\ref{static-mat} and ~\ref{dyn-mat}, the auxiliary kernels for the SC-GQME (see Eqs.~6a, 6b, and 8 in the main manuscript) take the forms,
\begin{equation}
\begin{split}
    \mathcal{K}^{(3b)}(t) &=  \frac{i}{2}\mathcal{Z}^- q^{(1s,0)}(t) -   \frac{i}{2}\mathcal{Z}^+ q^{(1a,0)}(t), \\
    \mathcal{K}^{(3f)}(t) &= i q^{(0,1)}(t)  \mathcal{Z}^{-}, \\
    \mathcal{K}^{(1)}(t) &= \frac{1}{2}\Big[\mathcal{Z}^{-} q^{(1s, 1)}(t) + \mathcal{Z}^{+} q^{(1a, 1)}(t) \Big]\mathcal{Z}^-.
\end{split}
\end{equation}

\subsection{Time Derivatives}

We can also express time derivatives using the static and dynamical matrices defined in Secs.~\ref{static-mat} and ~\ref{dyn-mat}. We start with expressions for the first left- and right-handed time derivatives, which correspond to the action of a single Liouvillian acting on either the initial condition, $\dot{\mathcal{C}}^L(t)$, or the final measurement, $\dot{\mathcal{C}}^R(t)$,
\begin{subequations}
    \begin{align}
        \dot{\mathcal{C}}^L(t) &= -i\Big( \varepsilon\mathcal{Z}^- + \Delta\mathcal{X}^-\Big)q^{(0,0)}(t) - \frac{i}{2}\mathcal{Z}^- q^{(1s,0)}(t) -   \frac{i}{2}\mathcal{Z}^+ q^{(1a,0)}(t), \\
        \dot{\mathcal{C}}^R(t) &= -iq^{(0,0)}(t) \Big( \varepsilon\mathcal{Z}^- + \Delta\mathcal{X}^-\Big) - i q^{(0,1)}(t)    \hat{\mathcal{Z}}^{-}.
    \end{align}
\end{subequations}
The second left-handed derivative requires many more terms,
\begin{equation}
    \begin{split}
        \ddot{\mathcal{C}}^{2L}(t) = &-\Big( \varepsilon^2 \mathcal{Z}^-\mathcal{Z}^- + \varepsilon \Delta \mathcal{X}^-\mathcal{Z}^- + \varepsilon \Delta \mathcal{Z}^-\mathcal{X}^- + \Delta^2 \mathcal{X}^-\mathcal{X}^-\Big) q^{(0,0)}(t) \\
        &-\frac{1}{2}\Big( 2\varepsilon\mathcal{Z}^-\mathcal{Z}^- + \Delta \mathcal{X}^-\mathcal{Z}^- + \Delta \mathcal{Z}^-\mathcal{X}^-\Big)q^{(1s, 0)}(t) - \frac{1}{2}\Delta \mathcal{Z}^+\mathcal{X}^- q^{(1a,0)}(t) \\
        &-\frac{1}{2}\mathcal{Z}^-q^{(1b1s, 0)}(t) - \frac{1}{2}\mathcal{Z}^+q^{(1b1a, 0)}(t) - \frac{1}{4}\mathcal{Z}^-\mathcal{Z}^-q^{(2s,0)}(t) - \frac{1}{4}\mathcal{Z}^+\mathcal{Z}^+q^{(2a,0)}(t).
    \end{split}
\end{equation}
We note that sequential derivatives incur a greater computational cost partly because they require converging the linear combination of ever more correlation functions, placing stricter convergence requirements on their individual constituents. 

\setcounter{equation}{0}
\renewcommand{\theequation}{D\arabic{equation}}
\setcounter{subsection}{0}
\section{Not all derivatives are created equal}

In this section, we interrogate the differences between numerical, right-handed (those where the Liouvillian acts on the final measurement), and left-handed time derivatives (those where the Liouvillian acts on the initial condition). 

\subsection{Low-order equivalence of numerical and right-handed time derivatives}

Numerical and right-handed derivatives are \textit{not} generally equivalent. Here, we show when they start to differ. We further illustrate that, as one might expect, the numerical integration of right-handed derivatives recovers the original LSC approximation for the low orders over which numerical and right-handed derivatives agree. We start by illustrating how to determine the orders at which numerical and right-handed derivatives agree. 

The difference between the $n$th numerical derivative, which arises from applying the time derivative to an already SC-approximated correlation function, 
\begin{equation}
    \label{derivs-num}
    \partial_t^n[\mathcal{C}_{jk}(t)]_{\rm LSC} = \int {\rm d}\boldsymbol{\Gamma} \; \left(\rho_B A_j^{\dagger}\right)^W e^{i\mathcal{L}^Wt} (i\mathcal{L}^W)^n {A_k}^W,
\end{equation}
and the $n$th right-handed derivative,
\begin{equation} \label{derivs-r}
    [\mathcal{C}^{nR}(t)]_{\rm LSC} = \int {\rm d}\boldsymbol{\Gamma} \; \left(\rho_B A_j^{\dagger}\right)^W e^{i\mathcal{L}^Wt} \left( (i\mathcal{L})^n A_k\right)^W,
\end{equation}
is that the latter requires one to perform the Wigner transformation on the $n$-times Liouvillian-rotated quantum mechanical operator being evolved. Here, the $nR$ superscript denotes the $n$th right-handed derivative. To understand this difference more concretely, we consider the action of the Liouvillian on the final measurement in the numerical and right-handed derivative given in Eq.~\eqref{derivs-num} and \eqref{derivs-r} when $n=1$,
\begin{subequations}
    \begin{align}
    \label{pb1}
    (i\mathcal{L}^W) {A_k}^W &= -\left\{ H^W, A_k^W\right\}_{PB}, \\
    \label{star1}
    \left( (i\mathcal{L}) A_k\right)^W &= i\big(H^W e^{\Lambda / 2i} A_k^W - A_k^W e^{\Lambda / 2i}H^W\big).
    \end{align}
\end{subequations}
The expressions in the equations above differ due to their functional dependence on $\Lambda$. Equation~\eqref{pb1} has a linear dependence on $\Lambda$, $\{A^W, B^W\}_{PB} = -(A^W \Lambda B^W)$, while Eq.~\eqref{star1} contains higher-orders of $\Lambda$ via $e^{-\Lambda/2i} = \sum_m^{\infty} \frac{1}{m!}\big(\frac{\Lambda}{2i}\big)^m$. Hence, Eqs.~\eqref{pb1} and \eqref{star1} show that the numerical and right-handed derivatives generally differ.

However, when $n=1$, each protocol yields equivalent results for the spin-boson Hamiltonian and a final measurement that is defined solely by mapped outer-product states. This equivalence follows from the fact that, under these conditions, the treatment of the final measurement in the right-handed derivative simplifies to the Poisson bracket. One can see this by starting with Eq.~\eqref{star1} and utilizing the antisymmetry of the Moyal product $A^W e^{\Lambda / 2i} B^W = B^W e^{-\Lambda / 2i} A^W$ which allows one to write, 
\begin{equation}\label{c3}
    \begin{split}
        \left( (i\mathcal{L}) A_k\right)^W &= 2i\Big[H^W \sinh\Big(\frac{\Lambda}{2i}\Big)A_k^W \Big], \\
        &= 2i\Big[H^W \Big( \Big( \frac{\Lambda}{2i}\Big) + \frac{1}{3!}\Big( \frac{\Lambda}{2i}\Big)^3 \Big) + ...\Big) A_k^W \Big].
    \end{split}
\end{equation}
In the spin-boson model (and related models, like the Frenkel exciton model), one only needs to consider the first term of this expansion when acting on a subsystem-only operator like $A_k^W$. This behavior arises from the structure of the Wigner transformed MMST-mapped outer products of discrete states, $(\ket{j}\bra{k})^W = \phi(\mathbf{X}, \mathbf{P}) f(\mathbf{X}, \mathbf{P})$ (see Eq.~\eqref{w-outer-product}). Here, one only needs to consider the action of $\Lambda$ on $f(\mathbf{X}, \mathbf{P})$,
\begin{equation}
\begin{split}
    H^W \Lambda \Big(\phi f \Big) &= \phi \Big(H^W \Lambda  f \Big) + \Big( H^W \Lambda \phi\Big) f \\
    &=   \phi \Big(H^W \Lambda f \Big), 
    \end{split}
\end{equation}
since $ H^W \Lambda \phi = 0$. From the definition of the Wigner transformed MMST-mapped outer products (see Eq.~\eqref{w-outer-product}), $f(\mathbf{X}, \mathbf{P})$ can only be quadratic in each of subsystem phase-space variables. Thus, 
\begin{equation} \label{simple}
\begin{split}
    H^W \Big(\frac{\Lambda}{2i}\Big)^m A_k^W = 0, \qquad m > 2, \qquad A_k \in \{\ket{j}\bra{k}\} .
\end{split}
\end{equation}
Equation~\eqref{simple} allows one to further simplify the expansion in Eq.~\eqref{c3},
\begin{equation}
    \begin{split}\label{conc}
        \left( (i\mathcal{L}) A_k\right)^W &= 2i\Big[H^W  \Big( \frac{\Lambda}{2i}\Big) A^W \Big], \\
        &= H^W \Lambda A_k^W, \\
        &= -\{ H^W , A_k^W \}_{PB} = (i\mathcal{L}^W) A_k^W.
    \end{split}
\end{equation}
Hence, in the spin-boson model, for subsystem operators alone, and when $n=1$, the numerical and right-handed derivatives are equivalent to each other. Thus, upon integration, the LSC-approximated right-handed time derivative returns the original LSC-approximated correlation function, explaining the results of Fig.~1D in the main manuscript.

\subsection{Higher-order numerical and right-handed derivatives generally differ} 
Although the lower-order numerical and right-handed derivatives are equivalent, there is a critical order, $n^*$, where they begin to differ. This critical order, $n^*$, depends on the choice of observable whose derivatives one is interested in calculating. We illustrate this by focusing on the numerical and right-handed derivatives of $\hat{\sigma}_z$ and $\hat{\sigma}_x$. For simplicity, we consider a bath containing a single oscillator that has a position, $x$, momentum, $p$, coupling strength, $c$, and frequency, $\omega$. To generalize this to a many-oscillator bath, one may use the following replacements: $c \rightarrow \sum_n c_n$, $c^2 \rightarrow \sum_n c_n^2$, and $cx \rightarrow \sum_n c_nx_n$. 

We begin with the first four numerical derivatives of $\tilde{\sigma}_z^W$,
\begin{equation}\label{c6}
    \begin{split}
        \partial^1_t \tilde{\sigma}_z^W &= 2\Delta \tilde{\sigma}_y^W, \\
        \partial^2_t\tilde{\sigma}_z^W &= 4\left( \varepsilon + cx\right) \Delta \tilde{\sigma}_x^W   -4\Delta^2  \tilde{\sigma}_z^W, \\
        \partial^3_t\tilde{\sigma}_z^W &= -8 \left( \varepsilon^2 \Delta  + \Delta^3 \right) \tilde{\sigma}_y^W -16 \varepsilon \Delta c x \tilde{\sigma}_y^W + 4\Delta c p \tilde{\sigma}_x^W -8 \Delta c^2  x^2  \tilde{\sigma}_y^W,\\
        \partial^4_t \tilde{\sigma}_z^W&= 16\Delta^2 \left( \left(\varepsilon + cx\right)^2 + \Delta^2 \right) \tilde{\sigma}_z^W - 16\Delta\left(\varepsilon + cx\right)\left( \left(\varepsilon + cx\right)^2 + \Delta^2 \right) \tilde{\sigma}_x^W \\
        &- 24 \Delta c^2 xp \tilde{\sigma}_y^W- 4\Delta \omega^2 cx\tilde{\sigma}_x^W {\color{red} - 4\Delta c^2 \left(\sigma_z^W \tilde{\sigma}_x^W \right)}.
    \end{split}
\end{equation}
While the first three right-handed derivatives are equivalent to the first three numerical derivatives of $\hat{\sigma}_z$ in the LSC treatment, i.e., $\partial^n_t \tilde{\sigma}_z^W =  \big(\tilde{\sigma}_z^{nR}\big)^W$ for $n \leq 3$, the fourth numerical derivative produces an extra term (in red, above):
\begin{equation} \label{c8}
\begin{split}
    \big(\tilde{\sigma}_z^{4R}\big)^W&= \Big[16\Delta^2 \left( \left(\varepsilon + cx\right)^2 + \Delta^2 \right) \tilde{\sigma}_z^W - 16\Delta\left(\varepsilon + cx\right)\left( \left(\varepsilon + cx\right)^2 + \Delta^2 \right) \tilde{\sigma}_x^W \\
    &- 24 \Delta c^2 xp \tilde{\sigma}_y^W- 4\Delta \omega^2 cx\tilde{\sigma}_x^W\Big].
\end{split}
\end{equation}
This additional term emerges from the non-commutativity between bath operators and $H_{SB}$, causing terms beyond first order in $\Lambda$ to survive in the Moyal product expansion (see Eq.~\eqref{c3}). To see this, consider the numerical and right-handed time derivatives acting on $4\Delta cp \tilde{\sigma}_x^W$, which is present in the third order of Eqs.~\eqref{c6} and \eqref{c8}. Starting with the system-bath contribution to the numerical time derivative,
\begin{equation} \label{pb}
\begin{split}
    (i\mathcal{L}_{SB}^W)4\Delta cp \tilde{\sigma}_x^W &= -4\Delta c \{ cx \sigma_z^W , p \tilde{\sigma}_x^W\}_{PB}, \\
    &= -8\Delta \phi c^2 \Big( \{\sigma_z^W, \sigma_x^W\}_{PB}xp + \{x, p\}_{PB} \sigma_z^W\sigma_x^W\Big), \\
    &= -4\Delta c^2 \Big(2 xp \tilde{\sigma}_y^W + \sigma_z^W\tilde{\sigma}_x^W\Big).
    \end{split}
\end{equation}
In contrast, consider the full Moyal product for the system-bath contribution to the right-handed derivative,
\begin{equation}
    \begin{split}
       ( (i\mathcal{L}_{SB})4\Delta c \hat{p} \hat{\sigma}_x)^W &= 4i\Delta c \Big( cx \sigma_z^W e^{\Lambda/2i}p \tilde{\sigma}_x^W - p \tilde{\sigma}_x^We^{\Lambda/2i} cx \sigma_z^W \Big), \\
       &= 8i\Delta \phi c^2 \Big( (x e^{\Lambda_B / 2i} p)(\sigma_z^W e^{\Lambda_S / 2i} \sigma_x^W) - (p e^{\Lambda_B / 2i} x)(\sigma_x^W e^{\Lambda_S / 2i}\sigma_z^W ) \Big), \\
       &= -8\Delta \phi c^2 \Big( (xp + \frac{i}{2})\sigma_y^W + (px - \frac{i}{2})\sigma_y^W \Big), \\
       &= - 4\Delta c^2 \Big( 2xp \tilde{\sigma}_y^W \Big),
    \end{split}
\end{equation}
where we have used the fact that subsystem and bath variables commute, which allows for the factorization, $e^{\Lambda / 2i} = e^{(\Lambda_S + \Lambda_B)/2i} = e^{\Lambda_S/2i}e^{\Lambda_B/2i}$. Thus, the critical order at which the numerical and right-handed derivatives acting on observable $\hat{\sigma}_z$ in the spin-boson model treated within the LSC approximation start to differ is $n^* = 4$. 

Turning to $\hat{\sigma}_x$, we again start with its numerical time derivatives up to $3^{\rm rd}$ order,
\begin{equation}\label{c10}
    \begin{split}
        \partial^1_t \tilde{\sigma}_x^W  &= -2\varepsilon \tilde{\sigma}_y^W - 2cx \tilde{\sigma}_y^W \, \\
        \partial^2_t \tilde{\sigma}_x^W &= 
        -4\big(\varepsilon + cx)^2 \tilde{\sigma}_x^W + 4\Delta\big(\varepsilon + cx)\tilde{\sigma}_z^W - 2cp \tilde{\sigma}_y^W , \\
        \partial^3_t\tilde{\sigma}_x^W &=  8 \big( \varepsilon + cx)^3 + 8\Delta^2 \big( \varepsilon + cx) + 2c\omega^2] \tilde{\sigma}_y^W - 12 \varepsilon cp \tilde{\sigma}_x^W + 8 \Delta cp \tilde{\sigma}_z^W \textcolor{red}{+2c^2 \Big( \sigma_z^W \tilde{\sigma}_y^W \Big) }  .
    \end{split}
\end{equation}
For $\hat{\sigma}_x$, the right-handed derivatives only begin to differ at $n= 3$,
\begin{equation} \label{c11}
    \begin{split}
        \big(\tilde{\sigma}_x^{3R}\big)^W &=  8 [\big( \varepsilon + cx)^3 + 8\Delta^2 \big( \varepsilon + cx) + 2c\omega^2] \tilde{\sigma}_y^W - 12 \varepsilon cp \tilde{\sigma}_x^W + 8 \Delta cp \tilde{\sigma}_z^W .
    \end{split}
\end{equation}
Thus, the degree of equivalence between of numerical and right-handed derivatives are equivalent depends on the system and the observable of interest in the final measurement. 

\subsection{Left-handed derivatives differ from numerical time derivatives}

More so than right-handed derivatives, left-handed derivatives (which also depend on the initial condition in the correlation function or nonequilibrium average of interest) differ from numerical derivatives within LSC dynamics. We establish this difference by showing that left-handed and numerical time derivatives differ from each other at $\textit{all}$ orders. 

We begin by comparing expressions for the numerical and left-handed derivatives,
\begin{subequations} \label{derivl}
\begin{align}
    \partial_t[\mathcal{C}_{jk}(t)]_{\rm LSC} &= \int {\rm d}\boldsymbol{\Gamma} \; \left(\rho_B A_j^{\dagger}\right)^W e^{i\mathcal{L}^Wt} (i\mathcal{L}^W) {A_k}^W, \\
            \label{num-ic}
    &= \int {\rm d}\boldsymbol{\Gamma} \; \Big(-i\mathcal{L}^W\big(\rho_B A_j^{\dagger}\big)^W\Big) e^{i\mathcal{L}^Wt} {A_k}^W,
\end{align}
\end{subequations}
and 
\begin{equation}\label{left-ic}
    [\dot{\mathcal{C}}^{L}(t)]_{\rm LSC} = \int {\rm d}\boldsymbol{\Gamma} \; \left(-i\mathcal{L}\rho_B A_j^{\dagger}\right)^W e^{i\mathcal{L}^Wt}  A_k^W.
\end{equation}
The expressions in Eqs.~\eqref{derivl} and \eqref{left-ic} differ from each other due to their treatment of the initial condition. Unlike numerical and right-handed derivatives, whose treatments of the final measurement did not lead to any differences until $3^{\rm rd}$ or $4^{\rm th}$ order (depending on the observable of interest), the treatment of the initial condition from a left-handed time derivative differs from a numerical derivative even at $1^{\rm st}$ order within the spin-boson model. To see this, consider the treatment of the initial conditions in Eqs.~\eqref{derivl} and ~\eqref{left-ic},
\begin{subequations}
\begin{align}
    \label{ic-pb}
    &(-i\mathcal{L}^W)\Big(\rho_B A_j^{\dagger}\Big)^W = \{ H^W, \rho_B^WA_j^{\dagger W}\}_{PB} \\
        \label{ic-star}
    &\Big(-i\mathcal{L}\rho_B A_j^{\dagger}\Big)^W = -i \Big( H^W e^{\Lambda/2i}\rho_B^WA_j^{\dagger W} + \rho_B^WA_j^{\dagger W} e^{\Lambda/2i} H^W \Big).
\end{align}
\end{subequations}
Even superficially, Eqs.~\eqref{ic-pb} and ~\eqref{ic-star} are different because the former is linear in $\Lambda$, while the latter contains higher orders of $\Lambda$. Even in the spin-boson model, these higher orders of $\Lambda$ are non-zero when acting on the system-bath initial condition:
\begin{equation} \label{c15}
      \Big(-i\mathcal{L}\rho_B A_j^{\dagger}\Big)^W  = -2i\Big[H^W \Big( \Big( \frac{\Lambda}{2i}\Big) - \frac{1}{3!}\Big( \frac{\Lambda}{2i}\Big)^3 \Big) + ...\Big) \rho_B^WA_j^{\dagger W} \Big].
\end{equation}
In this situation, the observable of interest contains contributions from subsystem and bath variables. Because $H^W_{SB}$ in the spin-boson Hamiltonian is quadratic in subsystem variables and linear in bath variables, it is possible for the second term in the expansion (third order in $\Lambda$) to be non-zero if the observable of interest is also at least second order in subsystem variables and linear in bath variables. Since the subsystem initial condition is effectively quadratic (remember that $\phi$ does not contribute) and $\rho_B^W$ features an exponential dependence on the bath variables, one must consider the second term in the expansion of the Moyal product in Eq.~\eqref{c15},
\begin{equation}\label{conc}
\begin{split}
    \Big(-i\mathcal{L}\rho_B A_j^{\dagger}\Big)^W &= -2i\Big[H^W \Big( \Big( \frac{\Lambda}{2i}\Big) - \frac{1}{3!}\Big( \frac{\Lambda}{2i}\Big)^3 \Big) \rho_B^WA_j^{\dagger W} \Big] \\
    &= (-i\mathcal{L}^W)\Big(\rho_B A_j^{\dagger}\Big)^W + \frac{2i}{3!}H^W \Big( \frac{\Lambda}{2i}\Big)^3 \rho_B^WA_j^{\dagger W}.\\
\end{split}
\end{equation}
Equation~\eqref{conc} highlights that the treatment of the initial condition in numerical and left-handed time derivatives differs even at $1^{\rm st}$ order in the spin-boson model. Hence, if one evaluates an LSC approximated left-handed derivative (see Eq.~\eqref{left-ic}), one should \textit{not} obtain the numerical time derivative of the LSC approximated correlation function (see Eq.~\eqref{num-ic}). In this scenario, upon integration, the left-handed derivative must produce a different prediction for the correlation function than the bare LSC dynamics.

\subsection{Short-time analysis of correlation functions }
While the previous analysis identifies when right- and left-handed derivatives match the result of numerical derivatives, it does not explain why integrating left-handed derivatives can lead to improved LSC dynamics. To address this question, here, we analyze the short-time accuracy of various LSC-approximated correlation functions relative to exact quantum mechanical expressions. This analysis reveals \textit{how} a LSC-approximated left-handed derivative can produce a correlation function with greater accuracy.  

A straightforward way to describe short time behavior is through a Taylor expansion around $t=0$. The coefficients (static moments) from these expansions offer a direct means to assess the relative short-time accuracy of a given correlation function. We start by considering the Taylor expansion of a correlation function, 
\begin{equation}
    \label{corr_exp}
    \mathcal{C}_{jk}(t) = \sum_{n=0}^{\infty} \frac{\mathcal{C}^{(n)}_{jk}(0)}{n!} t^n,
\end{equation}
where $\mathcal{C}^{(n)}_{jk} \equiv \partial^n_t \mathcal{C}_{jk}(t)\big\rvert_{t=0}$. To keep our analysis as transparent as possible, we focus on $\mathcal{C}(t) = {\rm Tr}\{ \hat{\rho}_B  \ket{1}\bra{1} e^{i\mathcal{L}t}\hat{\sigma}_z\}$, where $\hat{\rho}_B$ is the canonical density of the bath. This correlation function corresponds to initializing an excitation in state $\ket{1}\bra{1}$ and measuring the difference in population on each site as a function of time. To calculate the static moments, one computes the time derivative and evaluates at $t=0$. 

We start with the first $6$ exact static moments of the correlation function,
\begin{equation}\label{exactexp}
    \begin{split}
        &\partial_t[\mathcal{C}(t)]_{{\rm QM}}\big\rvert_{t=0} = 0, \\
       &\partial^2_t [\mathcal{C}(t)]_{{\rm QM}}\big\rvert_{t=0} = -4\Delta^2, \\
       &\partial^3_t [\mathcal{C}(0)]_{{\rm QM}}\big\rvert_{t=0} = 0 ,\\
        &\partial^4_t[\mathcal{C}(0)]_{{\rm QM}}\big\rvert_{t=0} = 16\Delta^2\left(\varepsilon^2 + \Delta^2 + c^2 \langle \hat{x}^2 \rangle \right), \\    
        &\partial^5_t[\mathcal{C}(0)]_{{\rm QM}}\big\rvert_{t=0} = 0, \\
        &\partial^6_t[\mathcal{C}(0)]_{{\rm QM}}\big\rvert_{t=0}=  -16\Delta^2 \Big( 4\left(\varepsilon^2 + \Delta^2 \right)^2 - 8c^2 \langle \hat{p}^2 \rangle + 24 \epsilon^2 c^2 \langle \hat{x}^2 \rangle 
        \\
        &\qquad \qquad \qquad \qquad \qquad \qquad \quad + 4c^4 \langle \hat{x}^4 \rangle+ 8\Delta^2 c^2 \langle \hat{x}^2 \rangle  + 9 \omega^2c^2 \langle \hat{x}^2 \rangle {\color{blue} + 5\varepsilon  c^2 }  \Big).
    \end{split}
\end{equation}
In the last line, we have highlighted in blue the term that the LSC-approximated Taylor expansion captures incorrectly.

We now consider the LSC-approximated correlation function (see Eq.~\eqref{lsc-apporx}),
\begin{equation}
\begin{split}
  [\mathcal{C}(t)]_{\rm LSC} &=\left(2\pi\right)^{-f}\int {\rm d}\mathbf{\Gamma} \;  \rho_B^W  \tilde{\sigma}_{11}^W  e^{i\mathcal{L}^Wt} \tilde{\sigma}_z^W, \\
    &=\left(2\pi\right)^{-f}\int {\rm d}\mathbf{\Gamma} \;  \rho_B^W  (4\phi^2) \sigma_{11}^W  e^{i\mathcal{L}^Wt} \sigma_z^W,
  \end{split}
\end{equation}
where, we remind the reader, the MMST mapping of the outer product states gives the following expression for the initial condition $(\ket{1}\bra{1})^W \equiv \tilde{\sigma}_{11}^W = 2\phi \sigma_{11}^W =  \phi(X_1^2 + P_1^2 - \frac{1}{2})$, from.
We find that the moments of the LSC-approximated correlation function, $\partial_t^n[\mathcal{C}(t)]_{\rm LSC}\big\rvert_{t=0}$, only begin to differ from the quantum mechanical moments at 6th order, i.e., $\partial^n_t [\mathcal{C}(t)]_{\rm QM}\big\rvert_{t=0} = \partial_t^n[\mathcal{C}(t)]_{\rm LSC}\big\rvert_{t=0}$ for $n < 6$, with
\begin{equation} \label{crexp}
\begin{split}
    &\partial_t^6[\mathcal{C}(t)]_{\rm LSC}\big\rvert_{t=0} = \\
    & -16\Delta^2 \Big( 4\left(\varepsilon^2 + \Delta^2 \right)^2 - 8c^2 \langle p^2 \rangle + 24 \epsilon^2 c^2 \langle x^2 \rangle + 4c^4\langle x^4 \rangle + 8\Delta^2 c^2 
    \langle x^2 \rangle 
    + 9 \omega^2c^2 \langle x^2 \rangle  \Big)  \\
    & \qquad {\color{red}- \frac{16\varepsilon \Delta^2 c^2}{(2\pi)^2}  \int {\rm d}\mathbf{\Gamma} \; \rho_B^W  \tilde{\sigma}_{11}^W \Big[     \left(\sigma_x^W\tilde{\sigma}_x^W\right) 
    - 5\left(\sigma_y^W\tilde{\sigma}_y^W\right) + 9 \left(\sigma_z^W\tilde{\sigma}_z^W\right) \Big]}. 
\end{split}
\end{equation}
We note that differences between the $6^{\rm th}$ static moments in Eqs.~\eqref{exactexp} and ~\eqref{crexp} arise from the non-zero contributions of terms containing products of Wigner transformed Pauli matrices consistent with previous analysis indicating that such contributions first appear at $6^{\rm th}$ order for LSC-approximated population dynamics \cite{golosov2001classical}. Thus, through this short-time analysis, we have determined that the LSC-approximated correlation function is accurate up to $\mathcal{O}(t^6)$. We note, however, that this order, $\mathcal{O}(t^6)$, differs from the critical order we identified for the operator for the population difference, $\hat{\sigma}_z$, i.e., $\mathcal{O}(t^4)$ (see Eq.~\eqref{c6} and \eqref{c8}), indicating that one can benefit from error cancellation against an initial condition resulting in some terms averaging out to $0$. That is, the additional terms in Eq.~\eqref{c8} relative to Eq.~\eqref{c6} do not contribute at $t=0$ under the selected closure,
\begin{equation} \label{no-contribute}
   \frac{1}{(2\pi)^2} \int {\rm d}\mathbf{\Gamma} \; \rho_B^W  \tilde{\sigma}_{11}^W  \big( \sigma_n^W \tilde{\sigma}_m^W) = \frac{1}{(2\pi)^2}\int {\rm d}\mathbf{\Gamma} \;  \rho_B^W (4\phi^2) \sigma_{11}^W  \big( \sigma_n^W \sigma_m^W) = \frac{1}{2}\delta_{n,m}. 
\end{equation}
Thus, the initial condition can \textit{improve} the accuracy of the LSC-approximated correlation function relative to the accuracy of the LSC-evolved operator alone. However, the \textit{choice} of initial condition can change the order of such relative improvements. 

Now, we consider what happens when we first act a left-handed derivative on the initial condition and subsequently calculate the higher-order static moments by acting numerical time derivatives. Doing this holds appeal because, as we have shown in Fig.~1c of the main manuscript, it can improve the accuracy of the semiclassically approximated correlation function. The question is: how does this happen? 

Consider the same correlation function as before, $\mathcal{C}(t)$. Here, we found that the LSC-approximated evolution of $\hat{\sigma}_z$ is accurate to $\mathcal{O}(t^4)$, which closure with the initial condition $\rho_B \ket{1}\bra{1}$ improves in accuracy to $\mathcal{O}(t^6)$. If one were to take a \textit{numerical} time derivative of this LSC-approximated correlation function, one should expect it to be accurate only to $\mathcal{O}(t^5)$ (i.e., the operator being evolved semiclassically becomes accurate only to $\mathcal{O}(t^3)$) such that, upon numerical integration that simply undoes the numerical derivative, one should recover the original LSC-approximated correlation function, which is accurate to $\mathcal{O}(t^6)$. Doing this numerical time derivative, as we have shown, is equivalent to performing the right-handed derivative. However, if one were to apply a left-handed derivative, the LSC-approximated \textit{operator} being evolved remains accurate to $\mathcal{O}(t^4)$. Then, because the choice of initial condition can increase the accuracy of the LSC-approximated correlation function relative to the LSC-approximated operator to different extents, one can \textit{hope} that the new initial condition generated by the exact Liouvillian-mediated rotation of the original initial condition can, at worst, cause the same improvement as the original initial condition. If this is the case, then the LSC-approximated time derivative of $\mathcal{C}(t)$ would be accurate to $\mathcal{O}(t^7)$. Upon numerical integration, $\mathcal{C}(t)$ would then become accurate to $\mathcal{O}(t^6)$. That is, left-handed derivatives could provide a mechanism to delay the onset of LSC-caused inaccuracy. Is this true? 

It \textit{is} true, with \textit{errors emerging only beyond $\mathcal{O}(t^6)$}. This constitutes a significant improvement as the original LSC approximation to $\mathcal{C}(t)$ was only accurate to $\mathcal{O}(t^6)$. Hence, the correlation function, $\dot{\mathcal{C}}^L(t)$, has greater short-time accuracy than the original LSC approximated correlation function.

\setcounter{equation}{0}
\renewcommand{\theequation}{E\arabic{equation}}
\setcounter{subsection}{0}
\section{Population conservation in SC-GQME}
\label{sum-rules}

Here, we interrogate how the SC-GQME conserves population. We discover that population conservation is \textit{not} guaranteed from the self-consistent structure of the SC-GQME, but rather emerges from the matrix structure of the auxiliary kernels. 

\subsection{Preservation of column linear dependence via matrix multiplication}

It is straightforward to show that \textit{right multiplication by a matrix with linearly dependent columns preserves that linear dependence}. To illustrate this property, we start by considering an $n \times n $ matrix $A$.  If $A$ has a pair of linearly dependent $k$ and $k'$ columns (where $k \neq k'$), then
\begin{equation} \label{linear-dep}
    A_{jk} = \lambda A_{jk'},
\end{equation}
where $\lambda$ is a constant. Considering an arbitrary $n \times n $ matrix, $B$, the pair of linearly dependent columns in $A$ are the same in the product $BA$,
\begin{equation} \label{column-structure}
\begin{split}
    (BA)_{jk} &= \sum_i B_{ji}A_{ik}, \\
    &= \lambda \sum_i B_{ji}A_{ik'}, \\
    &= \lambda (BA)_{jk'}.
    \end{split}
\end{equation}
This property generally does not hold when the order of multiplication is reversed (i.e., the product $AB$ does not satisfy it).

\subsection{ Population conservation in GMEs }

How does the mathematical observation above help us understand how GMEs conserve total population? In our construction of $\mathcal{C}(t)$, the rows encode the initial conditions while the columns encode the final measurements. Specifically, when we use the AK projector, which spans subsystem populations and coherences, the first column (indexed 1) refers to measuring the population on site 1 ($\ket{1}\bra{1}$) and the last column (indexed 4) refers to measuring the population on site 2 ($\ket{2}\bra{2}$). We can now describe the total population in the system as a function of time subject to the $j$th initial condition, $h_j(t)$, 
\begin{equation} \label{conserv}
    \frac{dh_j}{dt} = \dot{\mathcal{C}}_{j1}(t) + \dot{\mathcal{C}}_{j4}(t),
\end{equation}
where $j \in \{ 1,2,3,4\}$. To conserve the total population through time, one must satisfy
\begin{equation}\label{conserv}
    \dot{\mathcal{C}}_{j1}(t) = -\dot{\mathcal{C}}_{j4}(t).
\end{equation}
Equation~\eqref{conserv} implies that the GME preserves population when the first and fourth columns (or any combination of columns describing the measurement of populations) of $\dot{\mathcal{C}}(t)$ are linearly dependent with a proportionality factor $\lambda = -1$ (equal and opposite).

To calculate $\dot{\mathcal{C}}(t)$ within the GME, one uses the Mori-Nakajima-Zwanzig equation,
\begin{equation} \label{mnz2}
    \dot{\mathcal{C}}(t) = \mathcal{C}(t)\dot{\mathcal{C}}(0) - \int^t_0 {\rm d}s \; \mathcal{C}(t-s)\mathcal{K}(s).
\end{equation}
Equation~\eqref{conserv} requires the first and fourth columns of $\dot{\mathcal{C}}(t)$ to be equal and opposite for the GME to conserve population. Equation~\eqref{mnz2} then implies that, for $\dot{\mathcal{C}}(t)$ to satisfy this condition, the terms on the right-hand side collectively must obey it. The first term, $\mathcal{C}(t)\dot{\mathcal{C}}(0) $, satisfies this property if one uses the analytical form of $\dot{\mathcal{C}}(0)$. This is because in the spin-boson model $\dot{\mathcal{C}}(0)$ has equal and opposite first and fourth columns,
\begin{equation} \label{analytical}
    \dot{\mathcal{C}}(0) = (\mathbf{A}|\mathcal{L}|\mathbf{A}) = \left(\begin{matrix}
        0 & -\Delta & \Delta & 0 \\
        -\Delta & 2\epsilon & 0 & \Delta \\
        \Delta & 0 & -2\epsilon & -\Delta \\
        0 & \Delta & -\Delta & 0
    \end{matrix}\right).
\end{equation} 
One might also evaluate $\dot{\mathcal{C}}(0)$ numerically from one's choice of approximate dynamics, but $\dot{\mathcal{C}}(0)$ the resulting product may not have first and fourth columns that are equal and opposite.
However, if one uses the analytical form of $\dot{\mathcal{C}}(0)$, the first term on the right-hand side of Eq.~\eqref{mnz2} is guaranteed to have first and fourth columns that are equal and opposite. Consequently, for $\dot{\mathcal{C}}(t)$ to conserve population, the second term on the right-hand side of Eq.~\eqref{mnz2} must exhibit the same column structure.

The second term on the right-hand site of Eq.~\eqref{mnz2} contains an integral (sum) of products. In each of these products, the memory kernel, $\mathcal{K}(t)$, appears on the right.  Thus, each of these products is guaranteed to have first and fourth columns that are equal and opposite if the memory kernel also exhibits the same column structure. When the memory kernel has this column structure, the second term on the right-hand side of Eq.~\eqref{mnz2} has equal and opposite first and fourth columns and the GME satisfies the population-preservation condition in Eq.~\eqref{conserv}. Hence, a GME constructed using the AK projector is guaranteed to conserve population if the first term in Eq.~\eqref{mnz2} is constructed using the analytical expression of $\dot{\mathcal{C}}(0)$ and the memory kernel has columns that correspond to each population being equal and opposite. 

\subsection{The AK SC-GQME conserves total population}

In the standard AK SC-GQME, one uses the analytical form of $\dot{\mathcal{C}}(0)$ to evaluate $\dot{\mathcal{C}}(t)$ using Eq.~\eqref{mnz2}. Thus, to prove the AK SC-GQME conserves total population, we show the memory kernel constructed in the SC-GQME has first and fourth columns (columns that correspond to measuring populations) that are equal and opposite in sign. 

We begin by noting that the memory kernel in the SC-GQME obeys the self-consistent expansion in terms of auxiliary kernels given by Eq.~(7) in the main manuscript, 
\begin{equation}\label{construct-si}
    \mathcal{K}(t) = \mathcal{K}^{(1)}(t) + \int^t_0 {\rm d}s \; \mathcal{K}^{(3b)}(t-s)\mathcal{K}(s).
\end{equation}
One can Fourier transform both sides of this expression to obtain,
\begin{equation}
    \Tilde{K}(\omega) = \tilde{\mathcal{R}}(\omega)\tilde{K}^{(1)}(\omega).
\end{equation}
where $\tilde{\mathcal{R}}(\omega) \equiv \left[ 1 - \tilde{K}^{(3b)}(\omega) \right]^{-1}$. Taking the inverse-Fourier transform, one obtains,
\begin{equation} \label{fk1}
    \mathcal{K}(t) = \int^t_0 {\rm d}s \; \mathcal{R}(t-s)\mathcal{K}^{(1)}(s).
\end{equation}
The above expression reveals that we can write $\mathcal{K}(t)$ as the integral (sum) of products of $\mathcal{R}(t)$ and $\mathcal{K}^{(1)}(t)$. Hence, for the memory kernel to satisfy $\mathcal{K}_{j1}(t) = -\mathcal{K}_{j4}(t)$, one only needs to show that 
\begin{equation} \label{sum-k1}
    \mathcal{K}^{(1)}(t) = -\big[\ddot{\mathcal{C}}^{LR}(t)\big]_{\rm LSC} + \dot{\mathcal{C}}(0)\big[\dot{\mathcal{C}}^R(t)\big]_{\rm LSC} + \big[\dot{\mathcal{C}}^L(t)\big]_{\rm LSC}\dot{\mathcal{C}}(0) + \dot{\mathcal{C}}(0)\big[\mathcal{C}(t)\big]_{\rm LSC}\dot{\mathcal{C}}(0)
\end{equation}
satisfies the same condition.

We prove that the first and fourth columns of each term in Eq.~\eqref{sum-k1} are equal and opposite. First, $\big[\dot{\mathcal{C}}^L(t)\big]_{\rm LSC}$ does \textit{not} on its own have any columns that are linearly dependent as the finite sampling of static bath operators can introduce a static bias to individual matrix elements. However, taking a right-handed time derivative of $\big[\ddot{\mathcal{C}}^{LR}(t)\big]_{\rm LSC}$ eliminates this static bias as the derivative depends only on relative rates of change rather than absolute values. We illustrate this by explicitly showing the LSC approximated matrix elements $\big[\ddot{\mathcal{C}}^{LR}_{11}(t)\big]_{\rm LSC}$ and $\big[\ddot{\mathcal{C}}^{LR}_{14}(t)\big]_{\rm LSC}$, which correspond to the second time derivative of initializing a charge on site $\ket{1}\bra{1}$ in the spin-boson model and measuring the total population at time $t$, sum to $0$,
\begin{equation}
    \begin{split}
        \big[\ddot{\mathcal{C}}^{LR}_{11}(t)\big]_{\rm LSC} + \big[\ddot{\mathcal{C}}^{LR}_{14}(t)\big]_{\rm LSC} = &-\Delta^2\int {\rm d}\mathbf{\Gamma} \; \rho_B^W \tilde{\sigma}_y^W \tilde{\sigma}_y^W(t) 
    + 2\Delta\int {\rm d}\mathbf{\Gamma} \; \xi^W \rho_B^W \tilde{\sigma}_{11}^W \tilde{\sigma}_y^W(t)   \\ 
      &+\Delta^2\int {\rm d}\mathbf{\Gamma} \; \rho_B^W \tilde{\sigma}_y^W \tilde{\sigma}_y^W(t) 
    - 2\Delta\int {\rm d}\mathbf{\Gamma}\; \xi^W \rho_B^W \tilde{\sigma}_{11}^W \tilde{\sigma}_y^W(t)  \\
         &= 0.
            \end{split}
\end{equation}
This property holds for all rows, $j \in \{1,2,3,4\}$. Thus, the first term in Eq.~\eqref{sum-k1} has first and fourth columns that are equal and opposite. In the second term, the right side of the product is $\big[\dot{\mathcal{C}}^R(t)\big]_{\rm LSC}$. Because LSC dynamics conserve the total population\cite{wang1999semiclassical}, one can be certain that this right-handed derivative, which is equivalent to the numerical derivative, has first and last columns that are equal and opposite. Thus, because $\big[\dot{\mathcal{C}}^R(t)\big]_{\rm LSC}$ is on the right side of the product in the second term on the right-hand side of Eq.~\eqref{sum-k1}, this second term also has first and fourth columns that are equal and opposite (see Eq.~\eqref{column-structure}). Finally, the last two terms on the right-hand side of Eq.~\eqref{sum-k1} contain products with $\dot{\mathcal{C}}(0)$ on the right side. Since the analytical form of $\dot{\mathcal{C}}(0)$ has first and fourth columns that are equal and opposite (see Eq.~\eqref{analytical}), and because $\dot{\mathcal{C}}(0)$ is on the right side of each of these products, these terms also have first and fourth columns that are equal and opposite (see Eq.~\eqref{column-structure}). Thus, $\mathcal{K}^{(1)}_{j1}(t) = - \mathcal{K}^{(1)}_{j4}(t)$.

We have thus demonstrated that the memory kernel constructed from the AK SC-GQME has equal and opposite first and fourth columns (columns that correspond to measuring populations). This proof only relied on the use of the analytical form of $\dot{\mathcal{C}}(0)$ (or an approximate form that still obeys the same analytical constraints). Hence, the memory kernel constructed using the SC-GQME predicts dynamics that preserve total population (see Eq.~\eqref{conserv}).

\begin{table}[h]
\centering
\begin{tabular}{|c|c|c|c|c|}
\hline
Aux. Kernel & Term 1 & Term 2 & Term 3 & Term 4 \\
\hline
$\mathcal{K}^{(1)}(t)$ & $-\big[\ddot{\mathcal{C}}^{LR}(t)\big]_{\rm LSC}$ & $\big[\dot{\mathcal{C}}^L(t)\big]_{\rm LSC}\dot{\mathcal{C}}(0)$ &  $\dot{\mathcal{C}}(0)\big[\dot{\mathcal{C}}^R(t)\big]_{\rm LSC} $& $-\dot{\mathcal{C}}(0)\big[\mathcal{C}(t)\big]_{\rm LSC}\dot{\mathcal{C}}(0) $ \\
\hline
$\mathcal{K}^{(1)}_{M1}(t)$ & $-\big[\ddot{\mathcal{C}}^{LR}(t)\big]_{\rm LSC}$ & $\big[\dot{\mathcal{C}}^L(t)\big]_{\rm LSC}\dot{\mathcal{C}}(0)$ &  {\color{red} $\dot{\mathcal{C}}(0)\big[\dot{\mathcal{C}}^L(t)\big]_{\rm LSC} $} & $-\dot{\mathcal{C}}(0)\big[\mathcal{C}(t)\big]_{\rm LSC}\dot{\mathcal{C}}(0) $ \\
\hline
$\mathcal{K}^{(1)}_{M2}(t)$ & $-\big[\ddot{\mathcal{C}}^{LR}(t)\big]_{\rm LSC}$ & {\color{blue} $\big[\dot{\mathcal{C}}^R(t)]_{\rm LSC}\dot{\mathcal{C}}(0)$} &  $\dot{\mathcal{C}}(0)\big[\dot{\mathcal{C}}^R(t)\big]_{\rm LSC} $& $-\dot{\mathcal{C}}(0)\big[\mathcal{C}(t)\big]_{\rm LSC}\dot{\mathcal{C}}(0) $ \\
\hline
$\mathcal{K}^{(1)}_{M3}(t)$ & {\color{red} $-\big[\ddot{\mathcal{C}}^{2L}(t)\big]_{\rm LSC}$} & $\big[\dot{\mathcal{C}}^L(t)\big]_{\rm LSC}\dot{\mathcal{C}}(0)$ &  $\dot{\mathcal{C}}(0)\big[\dot{\mathcal{C}}^R(t)\big]_{\rm LSC} $& $-\dot{\mathcal{C}}(0)\big[\mathcal{C}(t)\big]_{\rm LSC}\dot{\mathcal{C}}(0) $ \\
\hline
$\mathcal{K}^{(1)}_{M4}(t)$ & $-\big[\ddot{\mathcal{C}}^{LR}(t)\big]_{\rm LSC}$ & $\big[\dot{\mathcal{C}}^L(t)\big]_{\rm LSC}\dot{\mathcal{C}}(0)$ &  $\dot{\mathcal{C}}(0)\big[\dot{\mathcal{C}}^R(t)\big]_{\rm LSC} $& {\color{blue} $-\dot{\mathcal{C}}(0)\big[\mathcal{C}^L(t)\big]_{\rm LSC}\dot{\mathcal{C}}(0) $} \\
\hline
$\mathcal{K}^{(1)}_{M5}(t)$ & $-\big[\ddot{\mathcal{C}}^{LR}(t)\big]_{\rm LSC}$ & $\big[\dot{\mathcal{C}}^L(t)\big]_{\rm LSC} {\color{red} \big[\dot{\mathcal{C}}^L(0)\big]_{\rm LSC}} $ &  $ {\color{red} \big[\dot{\mathcal{C}}^L(0)\big]_{\rm LSC}} \big[\dot{\mathcal{C}}^R(t)\big]$& $-{\color{red} \big[\dot{\mathcal{C}}^L(0)\big]_{\rm LSC}}\big[\mathcal{C}(t)\big]_{\rm LSC} {\color{red} \big[\dot{\mathcal{C}}^L(0)\big]_{\rm LSC}}$ \\
\hline
\end{tabular}
\caption{Modifications in the construction of $\mathcal{K}^{(1)}(t)$. The first row corresponds to the original construction of $\mathcal{K}^{(1)}(t)$. The following rows describe quantum mechanically equivalent auxiliary kernels to $\mathcal{K}^{(1)}(t)$ that start to differ when the LSC approximation is imposed at different levels.  \\ %
\fbox{%
        \parbox{0.98\textwidth}{%
            \justifying \noindent \underline{Alt text}: Summary of five alternative constructions of the auxiliary kernel $\mathcal{K}^{(1)}(t)$. Each row describes the components used to construct $\mathcal{K}^{(1)}(t)$, where blue entries highlight changes that preserve the column structure required for population conservation, while red entries highlight changes that do not preserve this column structure.   } } %
            }
\label{tab:4x4}
\end{table}

%%%%%%%%%%%%%%%%%%%%Figure  s1  begin %%%%%%%%%%%%%%%%%%%%%%%
\begin{figure*}
\centering
\includegraphics[width=\linewidth]{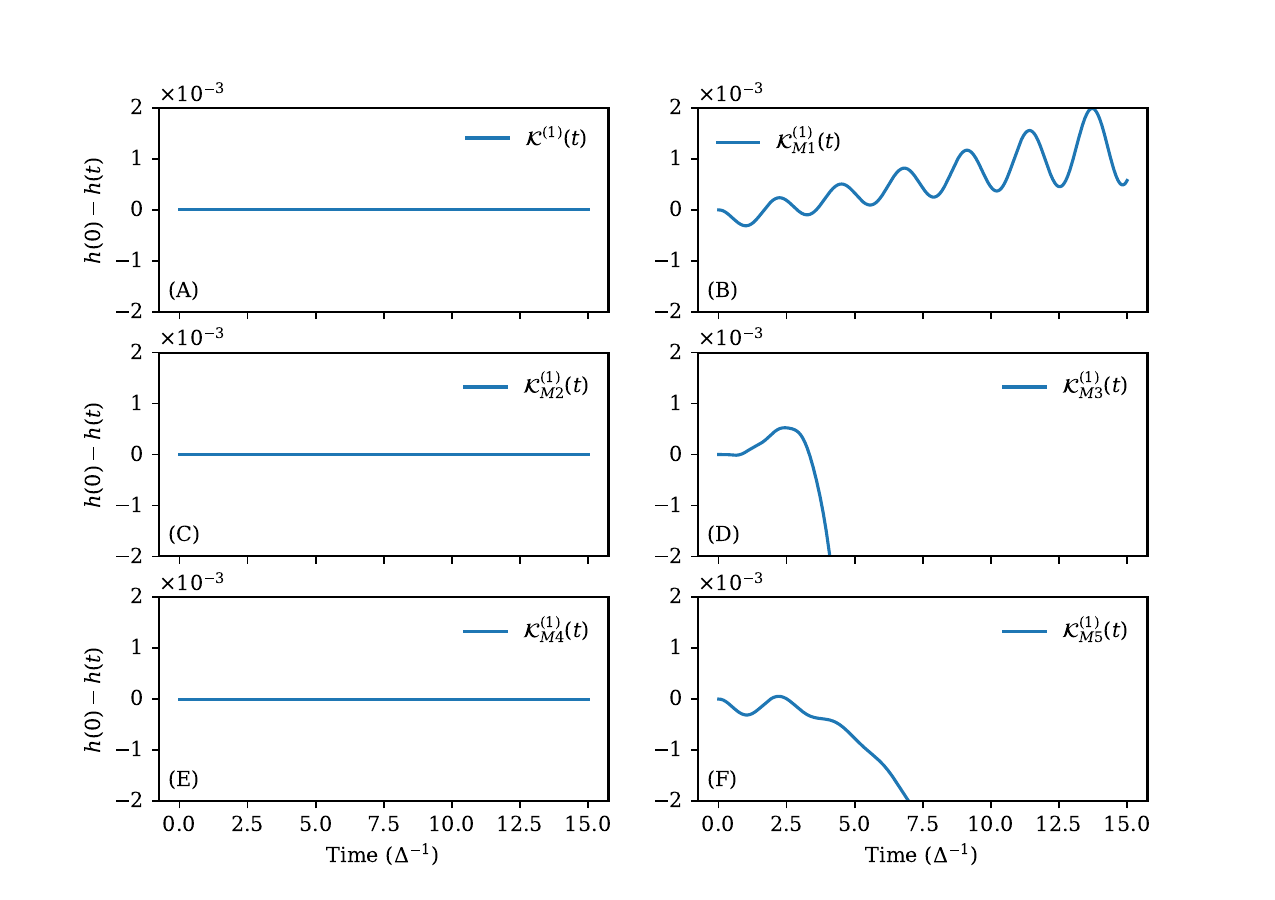}
\caption{\label{setup} \textbf{Modifying column structure of $\mathcal{K}^{(1)}(t)$ alters population conservation in SC-GQME.} Population flux, $h$, obtained from the SC-GQME dynamics of the spin-boson model with parameters $\epsilon = \Delta, \beta = 5.0\Delta^{-1}, \omega_c = 2.0\Delta, \eta = 0.2\Delta$ and an Ohmic spectral density and subject to initial condition $\rho(0) = \rho_B\ket{1}\bra{1}$. Each panel corresponds to SC-GQME dynamics predicted from a memory constructed using $\mathcal{K}^{(3b)}(t)$ and the variations of $\mathcal{K}^{(1)}(t)$ described in Table~\ref{tab:4x4}. \\ %
\fbox{%
        \parbox{0.98\textwidth}{%
            \justifying \noindent \underline{Alt text}: Six-panel figure describing population flux over time for SC-GQME dynamics using the six kernel constructions from Table S1. The left column shows the constructions that have a population flux of zero, while the right column shows constructions that have non-zero population flux. } } %
            }
\label{fig:figs1}
\end{figure*}
%%%%%%%%%%%%%%%%%%%%Figure s1 end %%%%%%%%%%%%%%%%%%

\subsection{Numerical tests}
We have analytically shown that the column structure $\mathcal{K}^{(1)}(t)$---specifically, that the sum of the columns (1 and 4, in our case) that track populations for each initial condition are equal and opposite---is what ensures population conservation in the SC-GQME. Now we illustrate this numerically by modifying the construction of $\mathcal{K}^{(1)}(t)$ in Eq.~\eqref{sum-k1}. In particular, we replace different components of this equation with analogues that, when computed exactly, would lead the GME to again recover the exact result. However, upon making the LSC approximation, these terms can change the resulting dynamics. Table~\ref{tab:4x4} summarizes five different variations of the construction of $\mathcal{K}^{(1)}(t)$, with the first line corresponding to the original SC-GQME expression for $\mathcal{K}^{(1)}(t)$ in Eq.~\eqref{sum-k1}. In {\color{blue} blue} are the replacements that yield a memory kernel can be expected to continue to conserve probability because the column structure remains unchanged. In {\color{red} red} are the replacements that can be expected to compromise population conservation. Figure~\ref{fig:figs1} illustrates the extent to which these variations of $\mathcal{K}^{(1)}(t)$ conserve population, confirming the veracity of our analysis.

\subsection{Single-population SC-GQME}

\begin{figure*}
\centering
\includegraphics[width=\linewidth]{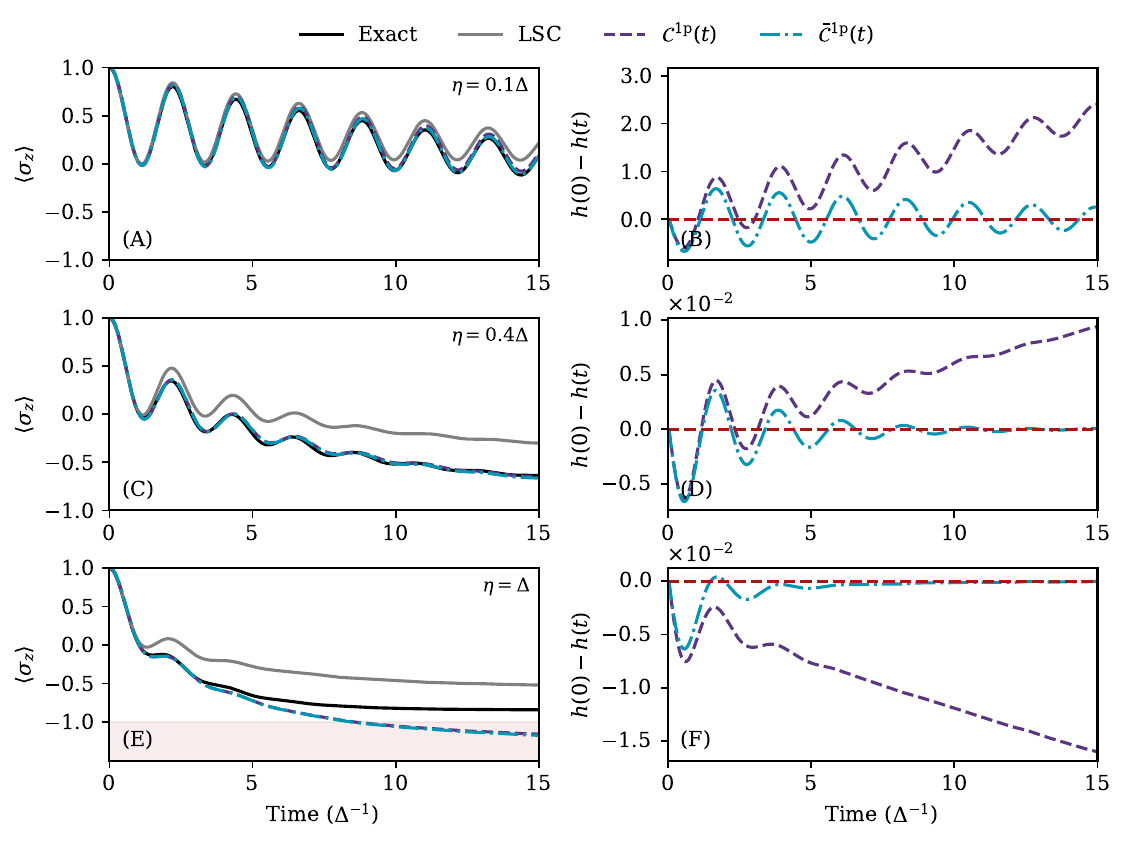}
\caption{\label{setup} \textbf{Single population SC-GQME preserves total population after removing static bias.} Nonequilibrium population dynamics in the spin–boson model subject to initial condition $\rho(0) = \rho_B\ket{1}\bra{1}$ with parameters $\epsilon = \Delta$, $\beta = 5.0\Delta^{-1}$, $\omega_c = 2.0\Delta$, and various levels of system-bath coupling, $\eta$, with an Ohmic spectral density. Each row corresponds to a different parameter regime. \textit{Left:} Comparison of exact (black) and LSC (gray) population dynamics to the single-population SC-GQME before \big( $\mathcal{C}^{\rm 1p}(t)$, dashed purple\big) and after \big( $\mathcal{\bar{C}}^{\rm 1p}(t)$, dotted blue\big) removing static bias at $t=0$. \textit{Right:} Change in total population, $h$, plotted as a function of time for the corresponding dynamics. The unshifted single-population SC-GQME, $\mathcal{C}^{\rm 1p}(t)$, exhibits population dynamics with a bias that linearly grows in time and can be expected to diverge at long times. In contrast, the shifted single-population SC-GQME, $\mathcal{\bar{C}}^{\rm 1p}(t)$, has a well-behaved population in all parameter regimes considered. The dotted red line denotes $0$. Red shaded regions denote unphysical negative populations. \\ %
\fbox{%
        \parbox{0.98\textwidth}{%
            \justifying \noindent \underline{Alt text}: Six-panel figure whose left column compares exact, LSC, single-population SC-GQME, and a shifted single-population SC-GQME population dynamics at three system-bath coupling strengths. The right column highlights the population flux for the single-population SC-GQME, which is non-zero, and the shifted single-population SC-GQME, which oscillates around zero. } } %
            } 
\label{fig:fignew}
\end{figure*}

Here, we examine how projecting the SC-GQME onto fewer states than the AK projector---specifically onto a single population---affects total population conservation. To simulate the single-population SC-GQME, we employ the expressions derived in Ref.~\onlinecite{Mulvihill2022}. We present the corresponding nonequilibrium population dynamics obtained from $1.0\times 10^{5}$ LSC trajectories that we use to construct the single-population SC-GQME in Fig.~\ref{fig:fignew} (A), (C), and (E). By only projecting onto a single population (a scalar), it is impossible for the auxiliary kernels to have equal and opposite columns corresponding to each population in the kernel matrix. Thus, based on the analysis in the previous subsections, one would expect the single-population SC-GQME to be unable to preserve the total population. To test this hypothesis, we plot the total population of the system as a function of time in Fig.~\ref{fig:fignew} (B), (D), and (F). In all parameter regimes considered here, the total population of the system diverges as a function of time (see Fig.~\ref{fig:fignew} (B), (D), (F)). Just as shifting the static bias at $t=0$ enables  $\big[\mathcal{\bar{C}}^L(t)\big]_{\rm LSC}$ to preserve population, applying the same shift proposed ensures a non-divergent total population in the single-population SC-GQME. While this shift does not generally yield a significant improvement in accuracy for the parameters considered here (see Fig.~\ref{fig:fignew} (A), (C), (D)), we nonetheless recommend it as a matter of best practice, since a divergent total population is undesirable, especially when propagating dynamics over arbitrarily long times.

%%%%%%%%%%%%%%%%%%%%Figure  s4  begin %%%%%%%%%%%%%%%%%%%%%%%
%%%%%%%%%%%%%%%%%%%%Figure s4 end %%%%%%%%%%%%%%%%%%
%\newpage

\setcounter{equation}{0}
\renewcommand{\theequation}{F\arabic{equation}}

\section{Memory cutoffs}

Here, we provide additional details on the effectiveness of the memory cutoff protocols proposed in the main text. In particular, we describe the relative accuracy of alternative memory cutoffs when using the root mean squared error (RMSE) metric. Moreover, we demonstrate that our proposed protocol yields a reliable $\tau_M$, even in cases that have proven challenging to the SC-GQME. Finally, we show the RMSE protocol can predict reliable dynamics when using other mixed-accuracy constructions of the memory kernel.

\subsection{Alternative memory cutoffs}

In this section, we provide the SC-GQME dynamics obtained by applying memory cutoffs selected from different points along the RMSE curves. In Fig.~7 of the main text, we demonstrate that our protocol for choosing $\tau_M$ by computing the RMSE relative to LSC dynamics from kernels constructed with varying levels of left-handed derivatives provides truncated dynamics with higher accuracy than LSC dynamics. One may also wonder about the quality of dynamics one could obtain when truncating the memory kernel at other notable times in the RMSE plots. Here, we use the same RMSE plots as in the main text (Fig.~7) corresponding to systems with strong system-bath coupling (see Fig.~\ref{fig:figs2} (A) and (B)). Additionally, because both RMSE plots in Fig.~\ref{fig:figs2} (A) and (B) share a similar structure, we choose four new memory truncation times corresponding to distinct structural features. Using each of these memory truncation times, we subsequently compare the corresponding nonequilibrium population dynamics from the kernels to the exact and LSC dynamics. At short times, all RMSE curves decrease from a shared initial value before reaching a minimum. Applying a cutoff in this region (gray-purple line), produces dynamics with lower accuracy than the original LSC approximation, (see Fig.~\ref{fig:figs2} (C) and (D)). Thus, if one aims to improve the accuracy of LSC dynamics, one should never apply a memory cutoff before the minimum in the RMSE curve. Alternatively, one can apply a cutoff at the minimum itself (orange dashed lines). Because the RMSE is relative to the LSC dynamics, this cutoff produces dynamics that roughly recover the original LSC dynamics and obtain little to no improvement (see Fig.~\ref{fig:figs2} (E) and (F)). Beyond their minimum, all RMSE curves increase again. While additional improvement is expected over some time interval in this regime, the optimal choice of memory cutoff becomes unclear in the absence of a well-defined protocol. One distinct point in this region is the intersection between the RMSE curves for $\mathcal{K}_{\rm LSC}^{(0L)}(t)$ and $\mathcal{K}_{\rm LSC}^{(1L)}(t)$ and $\mathcal{K}_{\rm LSC}^{(2L)}(t)$ (dashed blue line). By selecting $\tau_M$ at this intersection, the kernel predicts nonequilibrium population dynamics with higher relative accuracy than the original LSC dynamics, but, in general, this point cannot recover the exact long-time limit (see Fig.~\ref{fig:figs2} (G) and (H)). Finally, if one disregards all structurally distinct points and chooses a memory cutoff beyond the RMSE minimum, there is no a priori guarantee that the dynamics from the kernel remain physical. We illustrate this point by arbitrarily selecting $\tau_M = 2.0\Delta^{-1}$ (dashed green line). In some parameter regimes, this selection of $\tau_M$ allows the kernel to predict dynamics within graphical accuracy of the exact benchmark (see Fig.~\ref{fig:figs2} (I)), although this level of agreement is largely accidental. Thus, this choice of $\tau_M$ is not always reliable, as it can predict unphysical dynamics in challenging parameter regimes (see Fig.~\ref{fig:figs2} (J)).

%%%%%%%%%%%%%%%%%%%%Figure  s2  begin %%%%%%%%%%%%%%%%%%%%%%%
\begin{figure*}
\centering
\includegraphics[trim={0 0.1cm 0 0.1cm},width=\textwidth,height=\textheight,keepaspectratio]{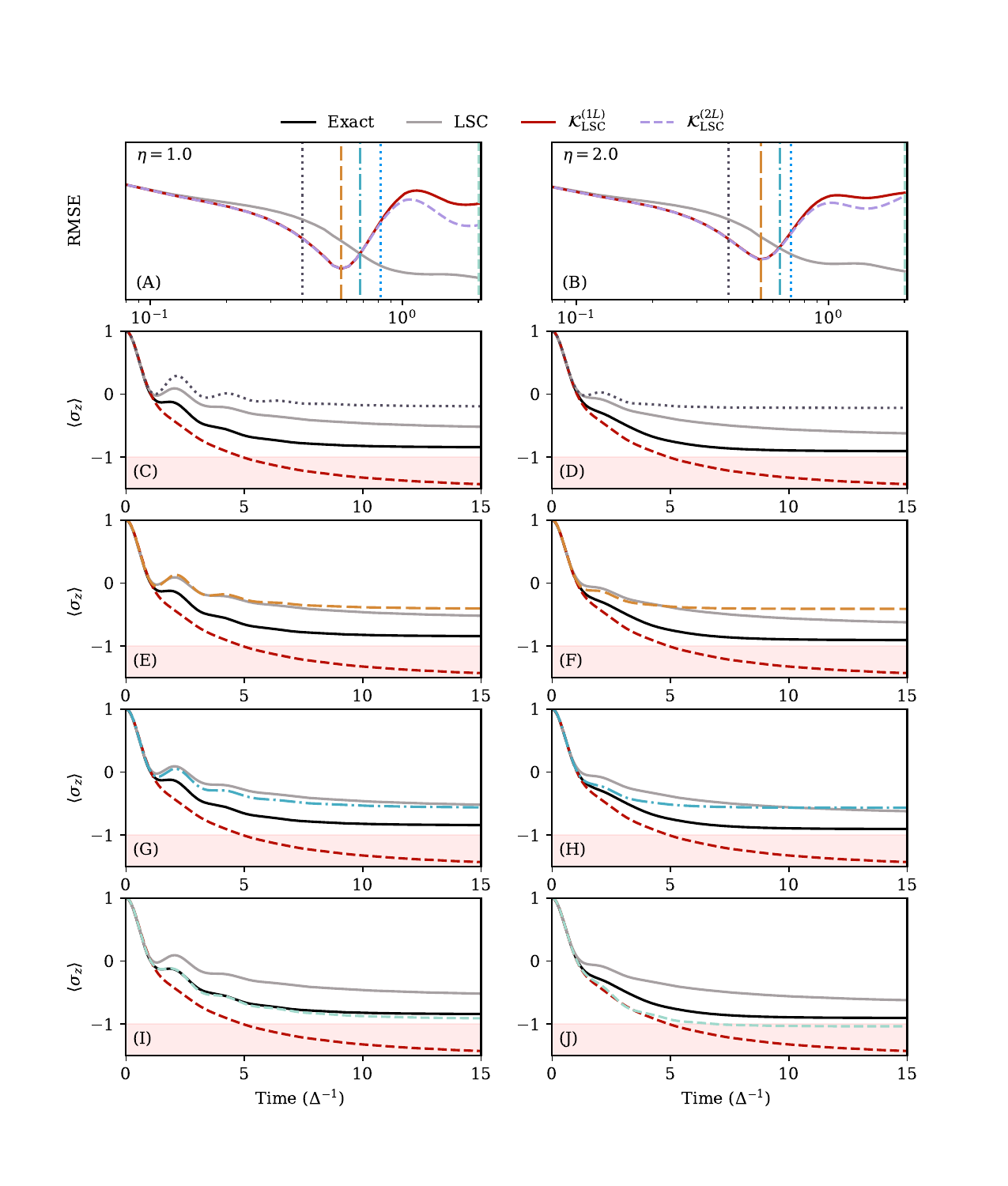}
\thispagestyle{empty} % optional: removes header/footer for full-page figure
\end{figure*}
\clearpage
\begin{figure*}

\caption{\label{fig:figs2} \textbf{Alternative memory time cutoffs provide various levels of accuracy}. Alternative choices for $\tau_M$ determined from RMSE curves relative to the LSC dynamics for the spin-boson model with an Ohmic spectral density and $\epsilon = \Delta$, $\beta = 5.0\Delta^{-1}$, $\omega_c = \Delta$ at various system-bath couplings, $\eta$, subject to initial condition $\rho(0) = \rho_B\ket{1}\bra{1}$. We generated these predictions using single-accuracy kernels constructed with varying numbers of left-handed derivatives (see Fig~7 in main text). (A-B): Four alternative values of $\tau_M$. (C-D): Lower-than-LSC-accuracy population dynamics obtained when choosing $\tau_M$ before the minimum in the RMSE plot (dotted line). (E-F): LSC-level accuracy population dynamics obtained when choosing $\tau_M$ at the minimum (orange dashed line). (G-H) Improved-accuracy population dynamics obtained when choosing $\tau_M$ as the intersection between the RMSE curves for $\mathcal{K}_{\rm LSC}^{(0L)}(t)$ and $\mathcal{K}_{\rm LSC}^{(1L)}(t)$ and $\mathcal{K}_{\rm LSC}^{(2L)}(t)$ (dashed blue line). Note that while these resulting dynamics are always physical, they fail to recover the exact long-time limit. (I-J): Benchmark-accuracy population dynamics obtained when choosing a cutoff later in time, such as $\tau_M = 2.0\Delta^{-1}$ (dashed green line). We remark that this agreement is accidental and may become much worse, even unphysical, in challenging parameter regimes. \\ %
\fbox{%
        \parbox{0.98\textwidth}{%
            \justifying \noindent \underline{Alt text}: Ten-panel figure whose top column shows truncation times on RMSE plots from Fig. 8 of the main text. The bottom four rows show the corresponding population dynamics from the selected memory truncation times in the top row. Cutoffs before or at the RMSE minimum recover LSC-level accuracy at best, while later principled choices yield improved dynamics, and arbitrary late cutoffs can produce unphysical results.  } } %
            }
\end{figure*}
%%%%%%%%%%%%%%%%%%%%Figure  s2  end %%%%%%%%%%%%%%%%%%%%%%%

\subsection{Previously inaccessible parameter regimes}
%%%%%%%%%%%%%%%%%%%%Figure  s3  begin %%%%%%%%%%%%%%%%%%%%%%%
\begin{figure*}
\centering
\includegraphics[width=\linewidth]{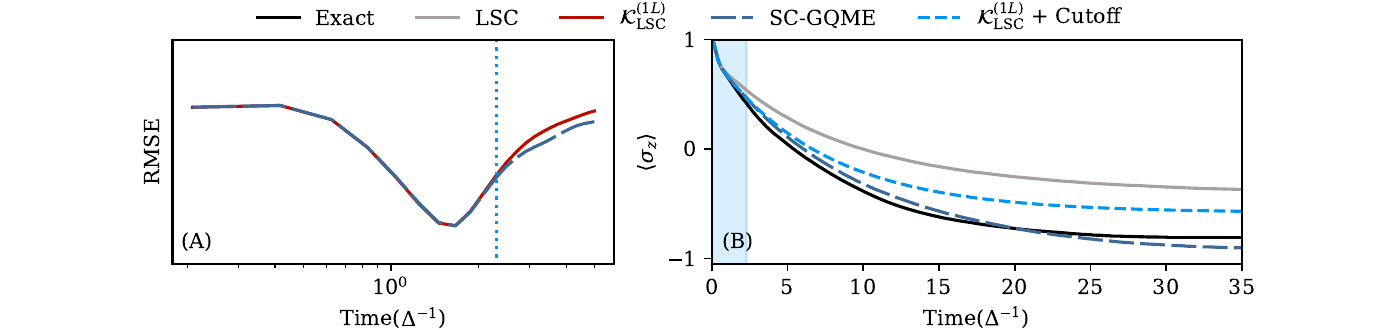}
\caption{\label{setup}  \textbf{Our RMSE protocol predicts reliable $\tau_M$ in system with large energy bias.}
RMSE protocol to determine $\tau_M$, in the spin-boson model with an Ohmic spectral density and $\epsilon = 3.0\Delta$, $\beta = 0.3\Delta^{-1}$, $\omega_c = \Delta$, $\eta = \Delta$, subject to initial condition $\rho(0) = \rho_B\ket{1}\bra{1}$ (A) RMSE relative to the LSC dynamics of predictions from \textit{single-accuracy} and \textit{mixed-accuracy} constructions of the memory kernels. That is, we construct $\mathcal{K}^{(1L)}_{\rm LSC}(t)$ from single-accuracy auxiliary kernels and $\mathcal{K}(t)$ from the SC-GQME using mixed-accuracy auxiliary kernels. (B) By selecting when the RMSE curves diverge, $\tau_M = 2.30\Delta^{-1}$, the resulting nonequilibrium population dynamics from the SC-GQME have higher relative accuracy to the original LSC dynamics and remain physical at all times. The shaded blue region indicates the time up to $\tau_M$. \\ %
\fbox{%
        \parbox{0.98\textwidth}{%
            \justifying \noindent  \underline{Alt text}: Two-panel figure showing RMSE curves to determine memory truncation time and corresponding population dynamics in the spin-boson model with large energy bias. The RMSE protocol yields population dynamics that outperform bare LSC dynamics and remain physical for all times.   } } %
            }
\label{fig:figs3}
\end{figure*}
%%%%%%%%%%%%%%%%%%%%Figure s3 end %%%%%%%%%%%%%%%%%%

Here, we illustrate the effectiveness of our RMSE protocol leveraging short-time accurate and long-time stable kernels to triangulate a reliable $\tau_M$ across traditionally challenging scenarios. Consistent with our final suggestion for the protocol that reproducibly yields predictions of greatest accuracy in the main text (see Fig.~8), the kernels we use are the \textit{single-accuracy} $\mathcal{K}^{(1L)}_{\rm LSC}(t)$ and \textit{mixed-accuracy} $\mathcal{K}(t)$, from the SC-GQME. Specifically, we consider systems that are well beyond the accuracy limits of SC dynamics, such as those with high energy bias between electronic states and systems with a Debye spectral density. 
\\
\noindent \textbf{High energy bias}: SC dynamics generally struggle to achieve accurate results in systems with a strong energy bias \cite{tully1998mixed, golosov2001classical, grunwald2009quantum}. Even worse, previous work has observed that no plateau of stability exists when using the SC-GQME in systems with increasing energy bias,\cite{Amati2022} limiting its applicability  increasingly. In light of this challenge, we test our protocol to triangulate $\tau_M$ in a system with high energy bias by considering a spin-boson model with parameters $\varepsilon=3.0\Delta$, $\omega_c = \Delta$, $\eta = \Delta$, $\beta = 0.3\Delta^{-1}$ with an Ohmic spectral density and a simulation timestep of, $\Delta t = 5.0 \times 10^{-3} \Delta^{-1}$. Figure~\ref{fig:figs3} (A) shows that the RMSE curves relative LSC dynamics from each kernel construction have a similar shape to those in Figs.~7 and 8 in the main text. Initially, the two RMSE curves start from the same point and decrease until they reach a minimum. After the minimum, both curves increase and then eventually begin to separate. We select $\tau_M$ as the point where the RMSE curve of $\mathcal{K}^{(1L)}(t)$ first diverges from that of the SC-GQME, indicated by the dotted blue line at $\tau_M = 2.30 \Delta^{-1}$. Because this $\tau_M$ lies beyond the RMSE minimum in Fig.~\ref{fig:figs3} (A), we expect it to outperform LSC dynamics, which we confirm in Fig.~\ref{fig:figs3} (B). Although the truncated $\mathcal{K}^{(1L)}_{\rm LSC}(t)$ dynamics do not exactly reproduce the benchmark long-time limit, Fig.~\ref{fig:figs3} (B) highlights that our protocol can still select a reliable $\tau_M$ that yields more accurate dynamics than the direct LSC approximation, even in parameter regimes with large energy biases.

%%%%%%%%%%%%%%%%%%%%Figure  s4  begin %%%%%%%%%%%%%%%%%%%%%%%
\begin{figure*}[t]
\centering
\includegraphics[width=\linewidth]{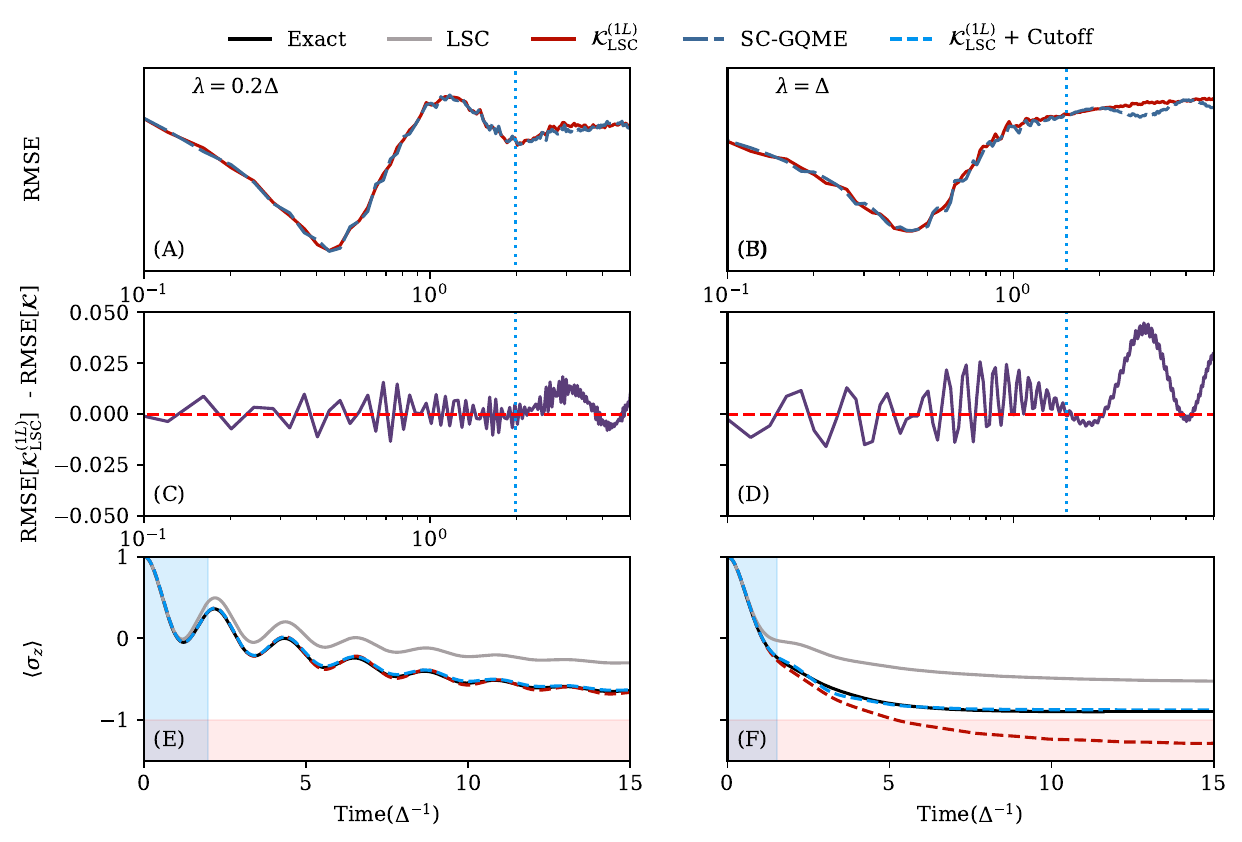}
\caption{\label{setup} \textbf{Our RMSE protocol predicts reliable $\tau_M$ in system with Debye spectral densities.}
RMSE protocol to determine $\tau_M$, in the spin-boson model with an Debye spectral density and $\epsilon = \Delta$, $\beta = 5.0\Delta^{-1}$, $\omega_c = \Delta$, and various amounts of system-bath coupling, $\eta$, subject to initial condition $\rho(0) = \rho_B\ket{1}\bra{1}$ (A-B) RMSE relative to the LSC dynamics of predictions from \textit{single-accuracy} and \textit{mixed-accuracy} constructions of the memory kernels. That is, we construct $\mathcal{K}^{(1L)}_{\rm LSC}(t)$ from single-accuracy auxiliary kernels and $\mathcal{K}(t)$ from the SC-GQME using mixed-accuracy auxiliary kernels. (C-D) Because we cannot visually determine $\tau_M$ from (A-B), we plot the difference between the RMSE curve for $\mathcal{K}^{(1L)}_{\rm LSC}(t)$ and the RMSE curve for $\mathcal{K}(t)$ from the SC-GQME. $\tau_M$ (dashed blue line) is selected as the last time the oscillations encompass $0$ (dashed red line) and before a clear oscillatory structure emerges. (E-F) The choice of $\tau_M$ from (C-D) provides reliable nonequilibrium population dynamics even when the system has a Debye spectral density. \\ %
\fbox{%
        \parbox{0.98\textwidth}{%
            \justifying \noindent  \underline{Alt text}: Six-panel figure whose top row shows RMSE curves to determine memory truncation in spin-boson models with low and high system-bath coupling and a Debye spectral density. Because the RMSE curves are noisy, the middle row shows the difference between both RMSE curves, which reveals when they diverge. This disambiguates the choice of kernel truncation times. The bottom row highlights that this memory truncation yields population dynamics that improve upon bare LSC dynamics.    } } %
            }
\label{fig:figs4}
\end{figure*}
%%%%%%%%%%%%%%%%%%%%Figure s4 end %%%%%%%%%%%%%%%%%%
\noindent \textbf{Debye spectral density}: We also consider our protocol in systems with a broad Debye spectral density. The Debye spectral density, which takes the form,
\begin{equation}
    J(\omega) = \frac{2\lambda \omega \omega_c}{\omega^2 + \omega_c^2},
\end{equation}
where $\lambda$ is the reorganization energy and $\omega_c$ is the cutoff frequency of the bath, provides a more challenging case for trajectory-based dynamical methods due to the presence of bath degrees of freedom with large frequencies \cite{berkelbach2012reduced}. To address this challenge, we test our protocol by calculating the RMSE relative to the LSC dynamics in a spin-boson model with the parameters, $\epsilon =  \Delta$, $\omega_c = \Delta$, $\beta = 5.0\Delta^{-1}$ with a Debye spectral density and various strengths of system-bath coupling, $\eta$ (see Fig.~\ref{fig:figs4} (A) and (B)). The RMSE curves exhibit a similar overall shape to those in other parameter regimes. However, unlike the previously considered Ohmic spectral densities, the Debye spectral density produces RMSE curves that appear significantly more jagged. This ``noise'' makes it more difficult to visually determine when the RMSE curves in Fig.~\ref{fig:figs4} (A) and (B) deviate from each other. To overcome this problem, we plot the difference between the RMSE curves from $\mathcal{K}^{(1L)}_{\rm LSC}(t)$ and the RMSE curve from $\mathcal{K}(t)$ (SC-GQME) in Fig.~\ref{fig:figs4} (C) and (D). These panels highlight that the noise can cause differences between the RMSE curves even at short times. In the main text, we deterministically selected $\tau_M$ as the time point when the RMSE curves diverged from each other (see Fig.~7). Here, to better account for the noise in Fig.~\ref{fig:figs4} (C) and (D) while remaining consistent with this framework, we select $\tau_M$ as the time when the oscillations no longer encompass $0$ (dashed red line) and  when a clear oscillatory structure emerges. We highlight the selected memory cutoffs in Fig.~\ref{fig:figs4} (C) and (D) with dashed blue lines: for $\eta = 0.2\Delta$, $\tau_M = 1.98\Delta^{-1}$ (see Fig.~\ref{fig:figs4} (C)), and for $\eta = \Delta$, $\tau_M = 1.53\Delta^{-1}$ (see Fig.~\ref{fig:figs4} (D)).  
To benchmark the spin-boson model with a Debye spectral density, we utilize the hierarchical equations of motion (HEOM) as a dynamical solver\cite{shi2009efficient}. For the given parameter regimes, we set $K = 10$ and $L = 26$ to obtain converged dynamics. 
We utilize this benchmark to show in Fig.~\ref{fig:figs4} (E) and (F) that these choices of $\tau_M$ predict nonequilibrium population dynamics that are of higher accuracy than the LSC counterparts at both strengths of system-bath coupling. In the case of $\eta = 0.2\Delta$, the dynamics even match the exact long-time limit. Hence, by employing the combination of single-accuracy and mixed-accuracy kernels, our RMSE protocol enables us to determine a reliable $\tau_M$ that improves the LSC dynamics, even when using challenging spectral densities and high reorganization energies.  

\subsection{Alternative Memory Kernel Constructions}

\begin{figure*}
\centering
\includegraphics[width=\linewidth]{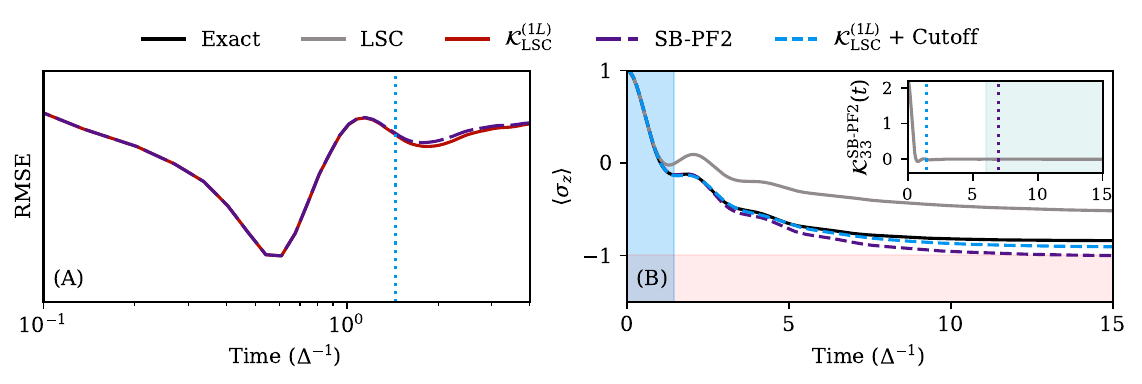}
\caption{\label{setup} \textbf{Our RMSE protocol predicts reliable $\tau_M$ using alternative mixed-accuracy closures of memory kernel.} RMSE protocol to determine $\tau_M$ for nonequilibrium population dynamics in the spin-boson model with an Ohmic spectral density and $\epsilon = \Delta$, $\beta = 5.0\Delta^{-1}$, $\omega_c = \Delta$, and $\eta = \Delta$, subject to initial condition $\rho(0) = \rho_B\ket{1}\bra{1}$. (A) RMSE relative to the LSC dynamics of predictions from \textit{single-accuracy} and \textit{mixed-accuracy} constructions of the memory kernels. That is, we construct $\mathcal{K}^{(1L)}_{\rm LSC}(t)$ from single-accuracy auxiliary kernels and $\mathcal{K}^{\rm SB-PF2}(t)$ from Eq.~7 in the main text. (B) Comparison of dynamics obtained with exact HEOM (black), LSC (gray), the SC-GQME with truncation of $\mathcal{K}^{\rm SB-PF2}(t)$ determined from plateau of stability (purple), and the SC-GQME with truncation of $\mathcal{K}^{\rm 1L}(t)$ determined from RMSE protocol (blue). Red shaded regions denote unphysical, negative populations. The shaded blue region highlights the time up to $\tau_M$ selected from the RMSE protocol. (Inset): Memory kernel element from  $\mathcal{K}^{\rm SB-PF2}(t)$, with the green shaded region highlighting the plateau of stability. To determine $\tau_M$ from the plateau of stability, we select any time within the green region, $\tau_M = 7.00 \Delta^{-1}$ (dashed purple line). The RMSE protocol identifies $\tau_M = 1.45 \Delta^{-1}$ (dashed blue line). \\ %
\fbox{%
        \parbox{0.98\textwidth}{%
            \justifying \noindent  \underline{Alt text}: Two-panel figure whose first panel shows RMSE curves constructed from a single-accuracy memory kernel and a SB-PF2 kernel (playing the role of the mixed-accuracy kernel) to determine memory truncation time. The right panel shows that the plateau-of-stability approach using the SB-PF2 memory kernel at high system-bath coupling can yield unphysical dynamics, while using the SB-PF2 in the RMSE protocol yields physically valid dynamics with improved accuracy compared to bare LSC dynamics.  } } %
            }
\label{fig:figsk3f}
\end{figure*}

In this section, we demonstrate that our proposed RMSE protocol is compatible with other constructions of the mixed-accuracy memory kernel. Thus far, we have constructed the mixed accuracy memory kernel using Eq.~5 in the main text. However, Ref.~\onlinecite{mulvihill2021road} suggests alternative kernel constructions, including a promising closure that corresponds to calculating the memory kernel using Eq.~7 in the main text, which the authors refer to as ``SB-PF2". This closure holds promise because the memory kernel has a well-defined plateau of stability across many parameter regimes. Thus, when the system-bath coupling $(\eta)$ is low, identifying the plateau of stability of the SB-PF2 gives an unambiguous kernel cutoff to predict dynamics that are in excellent agreement with the exact dynamics \cite{mulvihill2021road}. We find that the SB-PF2 closure continues to display a well-defined plateau of stability at high $\eta$ (highlighted green region, inset of Fig.~\ref{fig:figsk3f} (B)). However, any such choice of $\tau_M$ within this region yields improved short-time accuracy at the cost of unphysical long-time population dynamics. (see Fig.~\ref{fig:figsk3f} (B)).

%%%%%%%%%%%%%%%%%%%%Figure  s5  begin %%%%%%%%%%%%%%%%%%%%%%%

%%%%%%%%%%%%%%
%%%%%%Figure s5 end %%%%%%%%%%%%%%%%%%

Given this unphysical behavior that arises at high system-bath coupling, we test our RMSE protocol using the SB-PF2 closure. That is, we use $\mathcal{K}^{(1L)}$ as the single-accuracy kernel and the SB-PF2 memory kernel as the mixed accuracy kernel. Figure~\ref{fig:figsk3f} (A) shows the RMSE curves from these constructions relative to the LSC dynamics. The RMSE curves begin to diverge from each other at $\tau_M = 1.45 \Delta^{-1}$. This $\tau_M$ is slightly beyond the truncation time predicted in Fig.~8 of the main text, which was $\tau_M = 0.97 \Delta^{-1}$, and occurs earlier than the start of the plateau of stability (see inset of Fig.~\ref{fig:figsk3f} (B)). Using this new $\tau_M$, the predicted population dynamics exhibit enhanced short-time accuracy relative to the LSC dynamics and remain physical for all times (see Fig.~\ref{fig:figsk3f} (B)). We conclude that our RMSE protocol can predict a reliable $\tau_M$ in systems with high system-bath coupling where various mixed-accuracy kernels---which produce unphysical long-time behavior in the absence of truncation---nonetheless yield well-behaved dynamics when properly truncated.

\newpage
\bibliography{references}